\documentclass[useAMS,usenatbib,usegraphicx]{mn2e}
\usepackage{pslatex,color,bm,url,amsmath,amsfonts}
\def\idm#1{{\mbox{\scriptsize #1}}}
\def\vec#1{{\pmb #1}}
\def\kepler{{\em Kepler}}
\def\code#1{{\sc #1}}
\def\Y{\langle Y \rangle}
\def\norm#1{\|#1\|}
\def\mean#1{\langle{}#1{}\rangle}

\def\vec#1{{\pmb #1}}  
\def\Tr{\hbox{Tr}\,} 
\DeclareMathAlphabet{\mathsfit}{\encodingdefault}{\sfdefault}{m}{sl}
\SetMathAlphabet{\mathsfit}{bold}{\encodingdefault}{\sfdefault}{bx}{sl}
\newcommand{\m}[1]{\bm{\mathsfit{#1}}}
\def\Real{\mathbb R} 
\def\Toro{\mathbb T} 

\newcommand{\au}{\mbox{au}} 
\newcommand{\msun}{\mbox{m}_{\odot}}

\newcommand{\mE}{\mbox{m}_{\oplus}}

\newcommand{\Mmean}{\mathcal{M}}

\newcommand{\ab}{a_{\idm{b}}}
\newcommand{\ac}{a_{\idm{c}}}
\newcommand{\ad}{a_{\idm{d}}}
\newcommand{\eb}{e_{\idm{b}}}
\newcommand{\ec}{e_{\idm{c}}}

\newcount\exno

\definecolor{myred}{rgb}{0.89, 0.0, 0.13}
\definecolor{myblue}{rgb}{0.1,0.0,0.6} 
\definecolor{mybrown}{rgb}{0.9,0.4,0.3}
\newcommand\ednote[1]{{\global\advance\exno by 1\
    \color{myblue}$\spadesuit$({\bfseries\the\exno}).
\color{myblue}\bfseries\em #1}}

\newcommand\hide[1]{}

\def\kepler{{\sc Kepler}}

\def\idm#1{{\mbox{\scriptsize #1}}}

\title[The Reversibility Error Method]%
{The Reversibility Error Method (REM): 
a new, dynamical fast indicator for planetary dynamics}
\author[F.~Panichi, K.~Go\'zdziewski \&  G.~Turchetti]{
Federico Panichi$^{1}$\thanks{e-mail: federico.panichi@stud.usz.edu.pl},
Krzyszof~Go\'zdziewski$^{2}$\thanks{e-mail: krzysztof.gozdziewski@umk.pl}
\& Giorgio~Turchetti$^{3}$\thanks{e-mail: giorgio.turchetti@unibo.it}\\
$^{1}$Institute of Physics and CASA*, Faculty of Mathematics and Physics,
University of Szczecin,  Wielkopolska 15, 70-451 Szczecin, Poland\\
$^{2}$Centre for Astronomy, Faculty of Physics, Astronomy and Informatics,
Nicolaus Copernicus University, Grudziadzka 5, 87-100 Toru\'n, Poland \\
$^{3}$Department of Physics and Astronomy, 
Alma Mater Studiorum -- University of Bologna, 
Viale Berti Pichat 6/2, 40-127 Bologna, Italy \\
}
\begin{document}
%
\date{Accepted 2017 February 9; Received 2017 February 7; in original form 2016 September 24}
\pagerange{\pageref{firstpage}--\pageref{lastpage}} \pubyear{2017}
\maketitle
\label{firstpage}
\begin{abstract}
We describe the Reversibility Error Method (REM) and its applications to
planetary dynamics. REM is based on \textit{the time-reversibility analysis} of
the phase-space trajectories of conservative Hamiltonian systems. The round-off
errors break the time reversibility and the displacement from the initial
condition, occurring when we integrate it forward and backward for the same time
interval, is related to the dynamical character of the trajectory. If the motion
is chaotic, in the sense of non-zero maximal Characteristic Lyapunov Exponent
(mLCE), then REM increases exponentially with time, as $\exp \lambda t$, while
when the motion is regular (quasi-periodic) then REM increases as a power
law in time, as $t^\alpha$, where $\alpha$ and $\lambda$ are real
coefficients. We compare the REM with a variant of mLCE, the Mean Exponential
Growth factor of Nearby Orbits (MEGNO). The test set includes the restricted
three body problem and five resonant planetary systems: HD~37124, Kepler-60,
Kepler-36, Kepler-29 and Kepler-26. We found a very good agreement between
the outcomes of these algorithms. Moreover, the numerical implementation of REM is
astonishing simple, and is based on solid theoretical background. The REM
requires only a symplectic and time-reversible (symmetric) integrator of the
equations of motion. This method is also CPU efficient. It may be particularly
useful for the dynamical analysis of multiple planetary systems in the 
\kepler{} sample, characterized by low-eccentricity orbits and relatively weak
mutual interactions. As an interesting side-result, we found a possible {\em
stable chaos} occurrence in the Kepler-29 planetary system.
\end{abstract}
\begin{keywords}
methods: numerical, celestial mechanics, stars: individual: Kepler-26, stars: individual: Kepler-29, stars: individual: Kepler-36,
planetary systems
\end{keywords}

%
\section{Introduction}
%
During the past few years, the space mission \textsc{Kepler} has discovered
more than 550 multi-planet compact systems with relatively small-mass
super-Earth planets\footnote{\url{http://exoplanetarchive.ipac.caltech.edu/}}.
This has brought new understanding of the orbital architectures and the
long-term evolution of extrasolar systems. Short period exoplanets in
multi-planet systems raise a puzzling scenario of their formation and evolution.
In such near-resonant or resonant compact configurations, wide ranges of
gravitational interactions between planets are expected and chaotic dynamics due
to resonance overlap \citep{Chirikov1979,Wisdom1983,Quillen2011} may lead to
close encounters \citep{Chambers1996,Chatterjee2008} and self-disrupting systems
\citep{Chambers1999}. The mean motion resonances (MMRs) and secular resonances
are the crucial factors for the orbital evolution of compact planetary systems
and determine their long-term stability
\citep{book:Morbidelli2002,Guzzo2005,Quillen2011}.
 
A dynamical analysis of the observational data is often a challenge by itself.
Short baseline, sparse sampling and noisy measurements introduce uncertainties
and biases of the inferred orbital parameters. Uncertainties of the best-fitting
models may cover qualitatively different orbital configurations. Just to
mention a~few examples, we recall here planetary systems of Kepler-223 \citep{Mills2016}, HD~202206
\citep{Couetdic2010}, $\nu$-Octantis \citep{Ramm2016,Gozdziewski2013},
HR~8799~\citep{Marois2010,Gozdziewski2014}, HD~47366 \citep{Sato2016}. The
dynamical analysis of the best-fitting planetary models has become a standard
approach. For compact, resonant, strongly interacting systems, the optimization
of observational models may benefit from implicit constraints of the dynamical
stability \citep[i.e.,][]{Gozdziewski2008,Gozdziewski2014}.

Analysis of such problems makes use of the so called fast dynamical indicators
which are common for the dynamical systems theory. These numerical techniques
make it possible to analyse efficiently large volumes of the
phase/parameter-space. The fast indicators are developed to distinguish between
stable and unstable (regular or chaotic) motions on the basis of relatively
short arcs of phase-space trajectories of their dynamical systems. The most
common tools in this class are algorithms based on the maximal Characteristic
Lyapunov Exponent \cite[mLCE,][]{Benettin1980}, the Fast Lyapunov Indicator
\citep[FLI,][]{Froeschle1997}, the Mean Exponential Growth factor of Nearby
Orbits \citep[MEGNO,][]{Cincotta2000,Cincotta2003,Giordano2016}, the
Smaller/Generalized Alignment Index \citep[SALI and GALI,][]{book:Souchay2010},
the Orthogonal Fast Lyapunov Indicator \citep[OFLI and
OFLI2,][]{Barrio2016} as well as on a few variants of the refined Fourier frequency
analysis, like the Numerical Analysis of Fundamental Frequencies
\citep[NAFF,][]{Laskar1990,Laskar1992}, the Frequency Modified Fourier Transform
\citep[FMFT,][]{Nesvorny1996}, and the Spectral Number
\citep[SN,][]{Michtchenko2001}.

The Hamiltonian formulation of the equations of motion makes it possible to
construct symplectic integrators (SI) which preserve the geometrical properties
of the Hamiltonian flow \citep{book:Hairer2006}. Regarding the planetary
$N$-body problem, SI are CPU efficient and reliable methods for long-term
integration intervals that have brought a~breakthrough in this field
\citep[][]{Wisdom1991}. Remarkably, SI are usually time-reversible (symmetric)
schemes like the second order leapfrog \citep{Yoshida1990,book:Hairer2006}. 

A numerical breakup of the time-reversibility has been proved to be a sufficient
condition to detect chaotic trajectories in the phase-space
\citep{Aarseth1994,Lehto2008,Faranda2012}. Unlike regular orbits, an ergodic
motion is expected to result in large displacements of the initial condition
$\vec{x}_0$ after the forward and backward integration. Since SI are equivalent
to symplectic maps, it makes it possible to determine and rigorously prove
analytic properties of a numerical approach based on this idea developed in a
series of papers \citep{Turchetti2010a,Turchetti2010b, Faranda2012,Panichi2016}.

This relatively new dynamical fast indicator, called Reversibility Error
Method (REM from hereafter), is based on the time reversibility of the
ordinary differential equations (ODE). Rather than studying the divergence
of phase-space trajectories with the \textit{shadow orbits} algorithm or with the variational
equations of the equations of motion \citep[i.e.,][]{Benettin1980}, REM relies on integrating the same orbit
forward and backward with a time-reversible (symmetric) numerical integrator. A
phase-space orbit may be classified w.r.t. the growth rate of the global error
due to the accumulation of the round-off errors occurring in each integration
step (forward and backward). If the orbit is regular, in the sense of mLCE, the
accumulation of numerical errors develops as a power law in time, $\sim t^{\alpha}$,
while for mLCE-unstable trajectory this effect is exponentially amplified by its
chaotic nature, $\sim \exp \lambda t$, where $\alpha$ and $\lambda$ are real
coefficients.

Numerical applications of REM to low-dimensional dynamical systems has revealed
that it could be a sensitive and CPU efficient numerical fast indicator.
Given its similarity to mLCE \citep{Turchetti2010a,Faranda2012}, the
advantage is a great simplicity of numerical implementation. 

The main aim of this paper is to introduce the REM algorithm for studying
dynamical properties of compact systems of Earth-like planets discovered by the
\kepler{} mission. These systems are resonant or near-resonant, however with
orbits in small and moderate eccentricity range. We intend to show that REM is
 an effective and precise fast indicator for this class of systems as common mLCE
methods. 

The paper is structured as follows. After the Introduction, in Sect.~\ref{sec1}, we
briefly introduce the fast indicators REM,  MEGNO and FMFT as reference
tools. Next, based on the perturbation criterion for near-integrable dynamical
systems, we select a few examples to compare these indicators. Section
~\ref{sec2} is devoted to a brief presentation of these dynamical systems. We
recall a simple Hamiltonian system which exhibits the Arnold diffusion and the
restricted three body problem. The main target of our work are compact
3-planet systems, HD~37124 and Kepler-60, as
well as 2-planet low-order MMR systems, Kepler-29, Kepler-26 and Kepler-36,
which may be examples of ``typical'' near-resonant or resonant pairs of
Super-Earth planets in the \kepler{} sample. In Section~\ref{sec3} we
present the results of numerical experiments with the fast indicators. Section
~\ref{sec4} is devoted to numerical integrators, numerical accuracy and CPU
efficiency of the REM. After Conclusions (Sect.~\ref{conclusion}), Appendix A
shows a detailed theoretical background of this approach by comparing the
Lyapunov error, due to the initial displacement, with the forward and
reversibility errors due to random perturbations along the orbit. 
%
\section{Dynamical fast indicators}
\label{sec1}
%
The analysis of the long-term evolution of planetary systems is based on 
various analytic
theories and on the direct, numerical integration of the
equations of motion
\citep[e.g.,][]{Wisdom1991,Chambers1999,Laskar2001,Ito2002,Laskar2009}. Besides
these approaches, fast indicators are common tools to analyse the structure of
chaotic and quasi-periodic motions in the phase-space. Here, we briefly describe
REM and MEGNO, which may be considered as mLCE-related fast indicators, and a
variant of the spectral algorithms, FMFT.
%
\subsection{Reversibility Error Method (REM)}
\label{subs:REM}
%
The formal derivation of the REM for linear maps, its properties and 
connection with the mLCE are presented in \citep{Panichi2016}. For
Hamiltonian systems studied in this paper, which split into two individually
integrable terms, we prove analytical properties of the reversibility error and
charterise its changes for different regimes of motion. A detailed
introduction and analysis of REM for nonlinear symplectic maps, which generalize
the results in \citep{Panichi2016}, are given in Appendix ~\ref{appendix}. Here
we present only a brief and ``practical'' introduction.

Given an autonomous Hamiltonian system ${\cal H}$, the phase-space evolution of
its solutions can be defined as the symplectic map $\m{M}(\vec{x})$ which
iterates the conjugate variables $\vec{x}$,
\begin{equation}\label{eq:1} 
\vec{x}_{n} = \m{M}(\vec{x}_{n-1}), \quad n=1,\ldots,
\end{equation}
where $n$ is the iteration index, and $\vec{x}_{0}$ is the initial condition,
$\vec{x}_0\equiv\vec{x}(t=t_0)$. We introduce a perturbed map
$\m{M}_{\gamma}(\vec{x})$ where $\gamma$ is a measure of the perturbation
amplitude. For a generic Hamiltonian map, the reversibility error at iteration
$n$ is (see Appendix \ref{appendix}),
\begin{equation}\label{eq:2}
 d_n^\idm{(R)} = \sqrt{\mean{\left\| \m{M}_{\gamma}^{-n}(\m{M}_{\gamma}^{n}(\vec{x}_0)) 
 - \vec{x}_0 \right\|^2}},
\end{equation}
where ``${-n}$'' denotes the $n$-th backward iteration and ``$n$'' the $n$-th
forward iteration of $\m{M}_{\gamma}$. The kind of perturbation and its
amplitude are quite arbitrary: for Hamiltonian flows it may be the white
noise, for a symplectic map it may be a random additive perturbation or the
round-off error due to finite machine precision. 

To apply Eq.~\ref{eq:2} numerically, we must guarantee that the map is
invertible \citep{Faranda2012,Panichi2016}. For a numerical integrator affected
by a round-off error of amplitude $\gamma$, we change Eq.~\ref{eq:2} into 
\begin{equation}\label{eq:3}
d_n^{(R)} =
 \sqrt{
    \norm{
   \Phi^{\cal H}_{\gamma,-nh} \circ \Phi^{\cal H}_{\gamma,nh} (\vec{x}_0) - \vec{x}_0
         }^2
      },   
\end{equation}
where $\Phi^{\cal H}_{nh}$ denotes an SI scheme
advancing the initial condition from $t=0$ to $t=nh \equiv T$, 
where $h$ is the integration step. 
The scheme is time reversible, so that
\begin{equation}\label{eq:4}
  \Phi^{\cal H}_{-h} \circ \Phi^{\cal H}_{h} \equiv id,
\end{equation}
for one integration step $h$ \citep{book:Hairer2006}. (Symplectic
integrators may be not time-reversible integrators and vice-versa). The
reversibility condition is lost for maps with the round-off and/local errors
$\Phi^{\cal H}_{\gamma, nh}$. Note that in Eq.~\ref{eq:3}, we dropped the
average which appears in Eq.~\ref{eq:2}, since unlikely for random perturbation,
just a single realization of round-off errors is available. 

The reversibility error is therefore the norm of the displacement from a
selected initial condition in the phase-space, after integrating the equations
of motion forward and back for the same time interval $T=nh$ (the 
number of steps). 

Most symplectic integrator schemes $\Phi^{\cal H}_h$ used in practice are
symmetric by design. For instance, if the Hamiltonian may be split
into two terms, ${\cal H}={\cal H}_A +{\cal H}_B$, which are
individually integrable,  then the second order leapfrog scheme
\begin{equation}\label{eq:5} 
 \Phi^{\cal H}_{h} \equiv 
 \phi_{h/2}^{\cal A} \circ \phi_{h}^{\cal B} \circ \phi_{h/2}^{\cal A},
\end{equation} 
is composed of symmetric flows $\phi^{\cal A}_t$ and $\phi^{\cal B}_t$
for Hamiltonians ${\cal H}_A$ and ${\cal H}_B$, respectively.
This time-reversible scheme results in the local error $O(h^3)$. 

A great advantage of the leapfrog is that it may be easily generalized to higher
order schemes, as shown by \cite{Yoshida1990}. Here, we apply the 4th
order integrator of Yoshida, as well as a family of symmetric and symplectic
integrators called SABA$_n$ and SBAB$_n$ \citep{Laskar2001}.

\begin{figure}
\centering
 \vbox{
 \includegraphics[width=0.4762\textwidth]{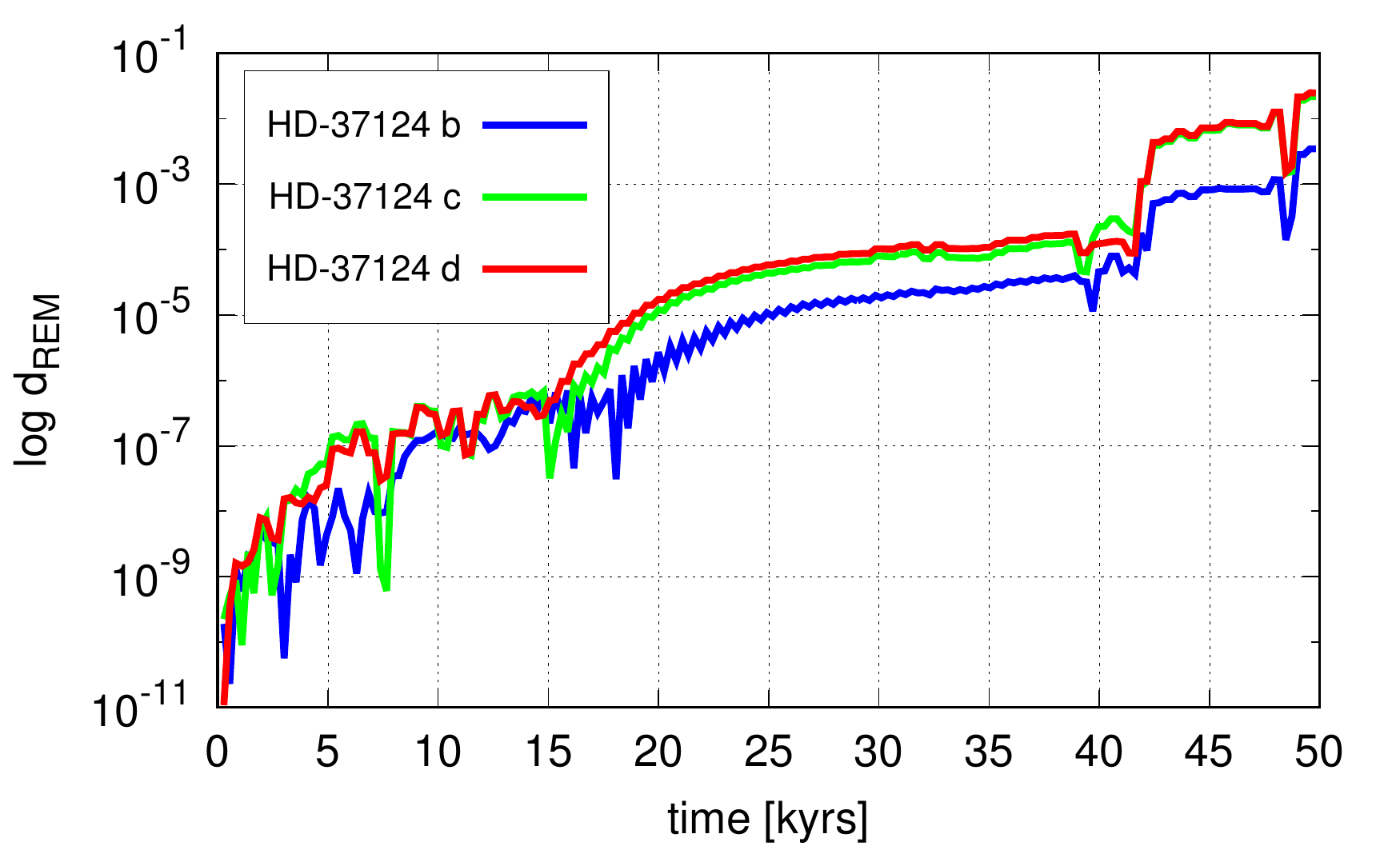}
 \includegraphics[width=0.4762\textwidth]{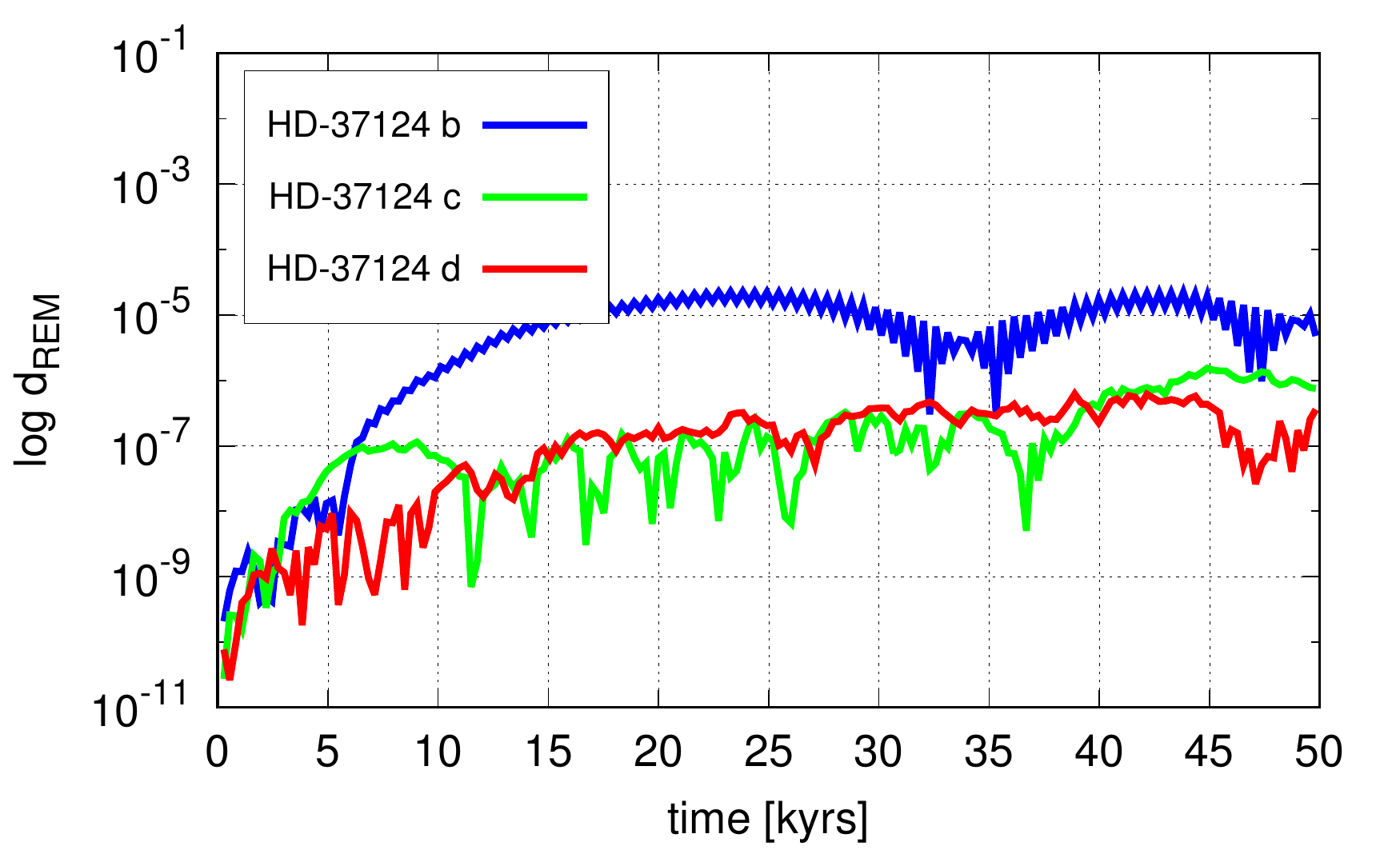}
}
\caption{
Time evolution of REM for the HD~37124 planetary system. The top panel is for
an unstable configuration, the bottom panel is for a stable, quasi-periodic
solution. The REM is computed for each orbit separately, and marked with
different colours (grey shades). The innermost planet (black, blue in the online version) appears to be most
influenced by the chaotic system, due to large value of REM ($10^{-5}$) at the
end of the total integration interval of $2\times 50$~kyrs. The second planet
(light-grey, green in the online version) and the third one (grey, red in the online version) exhibit slower increase of REM which reach
$10^{-7}$ at the end of the simulation. For the unstable configuration, the REM
components increase much faster, and they reach $0.1$, a~few orders of magnitude
larger value than for the regular model. 
}
\label{fig:figure1}
\end{figure}

A typical behaviour of REM for chaotic and regular phase-space trajectories is
illustrated in Fig.~\ref{fig:figure1}. This shows the time-evolution of the REM computed
for each individual planet in the three-planet system HD~37124 (see Sect.
\ref{subsubs:HD37124} for details). The integration has been performed for a
forward interval of 50~kyrs, and with the 4th order SABA$_4$ scheme with fixed
time-step equal to 1~day. For each planet, the REM increases following a power law
w.r.t. the integration time for a stable solution. We note that the deviation must
increase due to the accumulation of the numerical round-off and, possibly, due
to the local truncation error. We would like to note that the error with respect
to exact flow depends on both the truncation and the round-off errors and
estimates are difficult unless one of them is dominant. For the chaotic orbit,
the reversibility error increase rate has an exponential character. The crucial
point is that the final REM deviations differ by $\sim 7$ orders of magnitude,
and the orbits signatures could be easily distinguished one from each
other. 

We make use of this property in Sect.~\ref{sec3} by constructing dynamical maps
in planes of selected orbital and dynamical parameters. The REM values are
classified through their character of time-variability and relative ranges. We
note that a very similar calibration is known for the FLI
\citep{Froeschle1997} or the mLCE itself, since these indicators do not offer an
absolute measure of the instability degree in finite intervals of time.
%
\subsection{Mean Exponential Growth factor of Close Orbits}
\label{subs:MEGNO}
%
Together with the evolution of the phase-space trajectory, Eq.~\ref{eq:1}, it is
possible to propagate an initial displacement vector $\vec{\eta}$ with the
tangent map $D\m{M}$ defined as $D\m{M}_{ij}= \partial \m{M}_i/\partial
\vec{x}_j$, $i,j=1,\ldots 2N$, and $N$ is the number of the degrees of freedom,
\begin{equation}\label{eq:6} 
\vec{\eta_n} = D\m{M}(\vec{x}_{n-1})\,\vec{\eta}_{n-1}, \quad n>0.
\end{equation}
(See also Appendix~A). This discretization means solving the
Hamiltonian ODE system including the equations of motion and the variational
equations. The evolution of $\vec{\eta}(t)$ determines the maximal
Characteristic Lyapunov Exponent \cite[mLCE,][]{Benettin1980} 
\[
 \lambda \equiv \lim_{n\rightarrow\infty}
 \frac{1}{n} \log \frac{\norm{\vec{\eta}_n}}{\norm{\vec{\eta}_0}}, 
 \quad \vec{\eta}_0 \equiv {\vec{\eta}}(t_0),
\]
or its close relatives, like the Fast Lyapunov Indicator
\cite[FLI,][]{Froeschle1997} and the Mean Exponential Growth factor of Nearby Orbits
\cite[MEGNO,][]{Cincotta2000,Cincotta2003}.

Though MEGNO has been primarily defined for continuous ODEs, 
here we choose its formulation for maps,
consistent with REM formalism in other parts of this paper. It reads
\citep{Cincotta2003}
\begin{equation}\label{eq:7} 
  Y_n = \frac{2}{n}\sum_{k=1}^{n} k\, 
        \ln \frac{\norm{\vec{\eta}_k}}{\norm{\vec{\eta}_{k-1}}}, \quad
  \Y_n = \frac{1}{n}\sum_{k=1}^{n}Y_k,
\end{equation}
where $\vec{\eta_k}$ is the tangent vector at step $k$, $\vec{\eta}_0$ is random
initial vector, $\norm{\vec{\eta}_0}=1$, and $n$ is the number of steps. To
propagate the MEGNO map Eqs.~\ref{eq:7} for $N$-body planetary problem, we
implemented \citep{Gozdziewski2008} a symplectic tangent map \citep{Mikkola1999}
that solves the equations of motion and the variational equations
simultaneously.

The discrete map $\Y_n$ asymptotically tends to 
\[
  \Y_n = a n + b,
\]
with $a=0, b=2$ for a quasi-periodic orbit, $a=b=0$ for a stable, isochronous
periodic orbit, and $a=\lambda/2,\,b=0$ for a chaotic orbit, where $\lambda$ is
the mLCE approximation. Thus we can estimate the mLCE on a finite time interval
by fitting the straight line to $\Y_n$ \citep[see][for details]{Cincotta2003}.

Since MEGNO is essentially equivalent to FLI \citep{Mestre2011}, and makes it possible
to estimate the mLCE values, we consider it a well tested and a representative
fast indicator in the large family of variational algorithms \citep{Barrio2009}.

In general, the fixed step size symplectic integrators cannot be used for
configurations suffering from close encounters due to eccentric orbits. In such
cases, we use the MEGNO formulation for ODEs \citep{Cincotta2000} with
adaptive-step Bulirsch-Stoer-Gragg extrapolation method \citep[][ODEX
code]{book:Hairer2006}.
%
%
\subsection{Frequency modified Fourier Transform}
\label{subs:MFFT}
%
%
For one example system tested in this paper (Kepler-29), we used the
\citep[FMFT,][]{Nesvorny1996}, which is classified as a spectral algorithm. We
analyse the time series of heliocentric Keplerian elements ${\cal S}_i = \{
a_i(t_k) \exp(\mbox{i}\lambda_i(t_k)\}$ of planets
$i=\mbox{b},\mbox{c},\mbox{d},\ldots$ and $k=1,\ldots 2^N$, where $N$ is the
number of samples. These elements are inferred from canonical Poincar\'e
coordinates through usual two-body orbit transformation
\citep{book:Morbidelli2002}. For a near-integrable planetary system, the FMFT
transform of such series provides one of the fundamental, canonical frequencies,
namely the proper mean motion, $n_i$ associated with the largest amplitude
$a^0_i$ (the proper mean motion) of signal ${\cal S}_i$, for each of its planets.

We are interested in the diffusion of these proper mean motions, hence for each
planet we define a coefficient of the diffusion of fundamental frequencies
\citep{Robutel2001}:
\[
\sigma_{f} = \frac{n_{\Delta t\in [0, T]}}{n_{\Delta t\in [T, 2T]}}-1, 
\quad T=Nh ~, 
\]
where $h$ is the sampling step. If the frequencies for time intervals $\Delta t
\in [0,T]$ and $\Delta t \in[T,2T$] do not change, the motion is quasi-periodic,
while $\sigma_{f}$ different from zero indicates a~chaotic solution. This fast
indicator has been proved to be very sensitive for chaotic motions
\citep{Robutel2001,Nesvorny1996}.
%
\section{Between strong and weak perturbations}
\label{sec2}
%
We consider a near-integrable Hamiltonian system 
\begin{equation}
 {\cal H}(\vec{I},\vec{\theta}) = 
 {\cal H}_0(\vec{I}) + \epsilon {\cal H}_1(\vec{I},\vec{\theta}), \quad \epsilon \in [0,1),
\label{eq:8} 
\end{equation}
composed of the integrable term ${\cal H}_0(\vec{I})$ and the perturbation term $\epsilon
{\cal H}_1(\vec{I},\vec{\theta})$, w.r.t. the action--angle variables
$(\vec{I},\vec{\theta})$. We assume that ${\norm{\cal H}}_0 \simeq 
{\norm{\cal H}}_1$. The 
features determining the phase-space structure of this system are resonances
between the fundamental frequencies,
$
\dot{\vec{\omega}}_0 = {\partial\,{\cal H}_0(\vec{I})}/{\partial\,\vec{I}}.
$
They govern the long-term evolution of the phase-space trajectories. Depending
on the perturbation strength, the chaotic diffusion along these resonances
\citep{Morbidelli1995,Guzzo2002} may lead to macroscopic, geometric changes of
the phase-space trajectories. A simple measure of the complexity of a dynamical
system and chaotic diffusion is the perturbation parameter~$\epsilon$, which may
be expressed by the norm ratio of the perturbed $\norm{{\cal H}_1}$ to the
integrable $\norm{{\cal H}_0}$ term. The KAM theorem
\citep{Kolmogorov1954,Moser1958,Arnold1963} guarantees the existence of KAM
invariant tori provided that the value of the perturbation is smaller than some
threshold depending on the particular resonance. After that threshold, the KAM
tori are destroyed and the absence of topological barriers allows the chaotic
trajectories to globally diffuse \citep{Chirikov1979,Froeschle2005}. 

In this paper, we consider a few models of the form of Eq.~\ref{eq:8} and different
perturbation strengths. We focus on numerically revealing their resonant
structures with the help of the fast indicators. 

To solve the equations of motion and the variational equations associated with
model Eq.~\ref{eq:8}, required to determine MEGNO, we use a family of
symplectic, symmetric integrators SABA$_n$/SBAB$_n$ \citep{Laskar2001} which
exhibit the local error $O(\epsilon^2 h^2 + \epsilon^2 h^n)$, where $n$ is the
order of the scheme, and $h$ is the time-step. Therefore, for
splittings that provides $\epsilon$ small, as in Eq.\ref{eq:8}, these schemes
usually behave as higher order integrators without introducing negative
sub-steps \citep{Laskar2001}. Therefore even the second-order, modified
SABA$_2$/SBAB$_2$ schemes as well as the second order leapfrog with local error
$O(\epsilon h^3)$ offer sufficient accuracy and small CPU overhead. (More
technical details are presented in Sect.~\ref{sec4}).
%
\subsection{A Hamiltonian with the Arnold web presence}
\label{subs:FGL}
%
The first example for the REM and MEGNO tests is a three-dimensional dynamical
system introduced by \citep{Froeschle2000} to study qualitative features of
the resonance overlap in the phase-space of conservative Hamiltonian systems.
The Froeschl\'e--Guzzo--Lega (FGL from hereafter) Hamiltonian reads
\begin{equation}
\label{eq:9} 
    {\cal H}(\vec{I},\vec{\theta}) = 
\frac{I_1^2+I_2^2}{2} + I_3 + \frac{\epsilon}{\textrm{cos}(\theta_1)+\textrm{cos}(\theta_2)+\textrm{cos}(\theta_3)+4} .
\end{equation}
The perturbation term ${\cal H}_1(\vec{\theta})$ scaled by
$\epsilon\in[0,1)$ depends only on angles ${\cal \theta} =
[\theta_1,\theta_2,\theta_3]$. The fundamental frequencies exhibit full Fourier
spectrum. Resonances description may be reduced to the linear relation between actions
$\vec{I} = [I_1,I_2,I_3]$ through $m_1 I_1 + m_2 I_2 + 2\pi m_3 = 0$, with $m_1,
m_2, m_3$ $ \in \mathbb{Z} / 0$ \cite[see][for details]{Froeschle2000}. They
form a dense net, and their widths depend on $\epsilon$. Overlapping of these
resonances leads to fractal structures in the phase-space, interpreted as the
Arnold web. Due to the complexity of these dynamical structures and rich
long-term dynamical behaviours, which are provided by very simple
equations of motion, Hamiltonian~Eq.\ref{eq:9} is a great model to test
numerical integrators and fast indicators. This three-degrees of freedom
dynamical system exhibits all qualitative features which may be found in
multi-dimensional $N$-body systems.

%
\subsection{The circular restricted three body problem}
\label{subs:r3bp}
%
Perhaps the most attractive passage between simple dynamical systems and
planetary systems is the circular restricted three body problem (RTBP). We use
this model to demonstrate the REM algorithm and equivalence of the results when
the equations of motion are solved by relatively simple symplectic
algorithms.

The RTBP may be considered as the limit case of the $N$-body planetary problem,
when the star and a massive planet are primaries moving in a circular, Keplerian
orbit, and we investigate the motion of a mass-less particle (i.e.: an asteroid,
a comet). Any ``regular'' $2-$planet system may be transformed to the RTPB
by setting the mass of one planet to zero, and fixing a circular orbit of the
second one. Then we may solve the equations of motion with an appropriate
algorithm.

The same problem may be described in the non-inertial frame rotating with
the apsidal line of the primaries. Its dynamics is governed by the Hamiltonian
\begin{equation}
 {\cal H}(p_x,p_y,x,y) = {\cal T}(p_x,p_y,x,y) + {\cal U}(x,y) 
  \equiv {\cal H}_A + {\cal H}_B,
\label{eq:10} 
\end{equation}
where the kinetic energy ${\cal T}(p_x,p_y,x,y) \equiv {\cal H}_A(p_x,p_y,x,y)$ reads
\begin{equation}\label{eq:11} 
{\cal T}(p_x,p_y,x,y) = \frac{1}{2}
\left( x - p_y \right)^2 +  \frac{1}{2} \left( y +p_x \right)^2, 
\end{equation}
and the potential energy ${\cal U}(x,y) \equiv {\cal H}_B(x,y)$ is
\begin{equation}\label{eq:12} 
{\cal U}(x,y) = -\frac{x^2 + y^2}{2} - \frac{1-\mu}{\rho_1} - \frac{\mu}{\rho_2},
\end{equation}
where $(x,y)$ are barycentric coordinates
and momenta $(p_x,p_y)$ of the massless particle, and its
distances from primaries  
\[
\rho^2_1(x,y)=(x+\mu)^2+y^2, \quad \rho^2_2(x,y)=(x+1-\mu)^2+y^2.  
\]
Each term of Eq.~\ref{eq:10} in the absence of the others generates 
equations of motion that are solvable.

The equations of motion of the kinetic part
expressed by the gradient components of ${\cal T}$ w.r.t. $(p_x,p_y,x,y)$ canonical
coordinates,
\begin{equation}\label{eq:13}
\dot{x} = {\cal T}_{p_x}, \quad \dot{y} = {\cal T}_{p_y}, \quad
\dot{p}_x = -{\cal T}_{x}, \quad  \dot{p}_y = -{\cal T}_{y},
\end{equation}
form the linear ODE system, which has
a~well known solution \citep[e.g.,][]{Dulin2013} 
equivalent to $\phi^{\cal A}_h$,
\begin{equation}\label{eq:14}
\begin{split}
& x(h) = b_1 \sin (2h) + b_2 \cos(2h) + c_1, \\
& y(h) = b_1 \cos (2h) - b_2 \sin(2h) + c_2,\\
& p_x(h) = b_1 \cos (2h) - b_2 \sin(2h) - c_2,\\
& p_y(h) = -b_1 \sin (2h) - b_2 \cos(2h) + c_1,
\end{split}
\end{equation}
where coefficients $b_1,b_2, c_1, c_2$ are expressed through the initial
condition $(p_{x,0}, p_{y,0}, x_0, y_0)$, i.e., the momenta and coordinates at
time $t_0=0$,
\begin{equation}\label{eq:15}
\begin{split}
& b_1 = \tfrac{1}{2} \left( y_0+p_{x,0} \right), 
\quad b_2 = \tfrac{1}{2} \left( x_0-p_{y,0} \right),\\
& c_1 = \tfrac{1}{2} \left( x_0-p_{y,0} \right), 
\quad c_2 = \tfrac{1}{2} \left( y_0-p_{x,0} \right).
\end{split}
\end{equation}
The equations of motion for the potential are even more simple,
\begin{equation}\label{eq:16}
\dot{x} = 0, \quad \dot{y} = 0, \quad 
\dot{p}_x = -{\cal U}_x,\quad \dot{p}_y = -{\cal U}_y,
\end{equation}
where ${\cal U}_x$ and ${\cal U}_y$ are gradient components of the potential
${\cal U}$.
The solution to these equations, equivalent to 
$\phi^{\cal B}_h$, is essentially trivial,
\begin{equation}\label{eq:17}
\begin{split}
& x(h) = x_0, \\
& y(h) = y_0, \\
& p_x(h) = -{\cal U}_x(x_0,y_0) h + p_{x,0}, \\
& p_y(h) = -{\cal U}_y(x_0,y_0) h + p_{y,0}.
\end{split}
\end{equation}
Splitting into Hamiltonians ${\cal T}$ and ${\cal U}$ is non-natural in the sense
that the kinetic energy in a non-inertial, rotating frame depends not only on
momenta, but also on coordinates. 
%
\subsection{$N$--body planetary problem}
\label{subs:systems}
%
We define the main target of our numerical experiments, which is the $N$-body
planetary problem, w.r.t. canonical heliocentric Poincar\'e coordinates
\citep{book:Morbidelli2002}, sometimes called the democratic
heliocentric-barycentric coordinates. We apply the same formulation as in
\citep{Gozdziewski2008}. The Hamiltonian is composed of two terms ${\cal H} =
{\cal H}_0 + {\cal H}_1$. The first term reads
\begin{equation}\label{eq:18}
{\cal H}_0(\vec{p},\vec{r}) = 
\frac{1}{2}\sum_{i=1}^{N}\frac{
\vec{p}_{i}^{2}}{m_i}-k^2m_0\sum_{i=1}^{N}\frac{m_i}{r_i}, 
\end{equation} 
where $k^2$ is the Gauss gravitational constant, $\vec{p}_i=m_i \vec{v}_i$ are
the canonical (barycentric) momenta, $m_i$ the mass of the $i-th$ planet,
$\vec{v}_i$ is its barycentric velocity and $\vec{r}_i$ the heliocentric
coordinates of the planet, and $m_0$ is the stellar mass.

The second term of the Hamiltonian, which involves the perturbation of
Keplerian orbits due to the mutual interactions of the planets in the
system, is defined as
\begin{equation}\label{eq:19}
  {\cal R} \equiv \epsilon {\cal H}_1(\vec{p},\vec{r}) = \frac{1}{2~m_0}
\left ( \sum_{i=1}^{N}\vec{p}_{i} \right )^2 
-k^2\sum_{i=1}^{N}\sum_{j=i+1}^{N}\frac{m_i~m_j}{\norm{\vec{r}_i-\vec{r}_j}}.
\end{equation} 
Hamiltonian ${\cal H}$ is the direct sum of $N$ integrable
Keplerian Hamiltonians perturbed by the mutual gravitational potential of the
planets ${\cal R}$. Since  ${\cal H}_0$, and two terms of ${\cal R}$
in Eq.~\ref{eq:19} are individually integrable
\citep[for details, see, for instance,][]{Gozdziewski2008}, it leads to a natural splitting
used to construct the symplectic planetary integrators prototyped in the
remarkable paper of \cite{Wisdom1991}. Their scheme is based on splitting
the planetary Hamiltonian in Jacobi-coordinates, and may be generalized
 to other splittings, like the one we applied here.
\subsection{A characterization of tested planetary systems}
\label{subs:table}
Table~\ref{tab:tab1} displays orbital elements and masses of five resonant
planetary systems tested in the next Section.
\begin{table}
\centering
\caption{
Nominal, osculating heliocentric Keplerian elements for planetary systems tested
in this paper. The masses of parent stars are $0.78\,\msun$ for HD~37124
\citep{Vogt2005}, $0.55\,\msun$ for Kepler-26, $1.105\,\msun$
for Kepler-60 and $1.071\,\msun$ for Kepler-36, 
$1.0\,\msun$ for Kepler-29 \citep{Rowe2015}. All systems are coplanar with $I=90^{\circ}$ and
$\Omega=0^{\circ}$.
}
\begin{tabular}{l c c c r r}
\hline
\hline
System & $m\,[\mE]$ & $a\,[\au]$ & $e$ & 
$\varpi$\,[deg] & $\Mmean$\,[deg]\\
\hline
HD~37124~b  & $198$  & $0.51866$ & $0.079$ & $138.4$   & $259.0$ \\
HD~37124~d  & $180$  & $1.61117$ & $0.152$ & $268.9$   & $109.5$ \\
HD~37124~d  & $226$  & $3.14451$ & $0.297$ & $269.5$   & $124.1$ \\
\hline
Kepler-26~b & $5.1$   & $0.08534$ & $0.042$ & $9.6$     & $190.3$ \\
Kepler-26~c & $6.3$   & $0.10709$ & $0.025$ & $-18.6$   & $257.2$ \\
\hline
Kepler-29~b & $7.7$   & $0.09192$ & $0.006$  &  $23.6$ & $313.9$ \\
Kepler-29~c & $6.3$   & $0.10872$ & $0.007$  &  $-151.8$ & $29.0$  \\
\hline
Kepler-60~b & $4.6$   & $0.07497$ & $0.115$ & $-145.4$  & $-158.4$\\
Kepler-60~c & $4.9$   & $0.08700$ & $0.069$ & $-128.5$  & $-292.6$\\
Kepler-60~d & $4.8$   & $0.10558$ & $0.088$ & $-152.1$  & $-345.1$\\
\hline
Kepler-36~b & $4.2$   & $0.11541$ & $0.044$  &  $-126.5$ & $212.4$ \\
Kepler-36~c & $7.6$   & $0.12840$ & $0.020$  &  $-158.7$ & $24.0$  \\
\hline
\hline
\end{tabular}
\label{tab:tab1}
\end{table}
\begin{figure}
\centering
\includegraphics[width=0.47\textwidth]{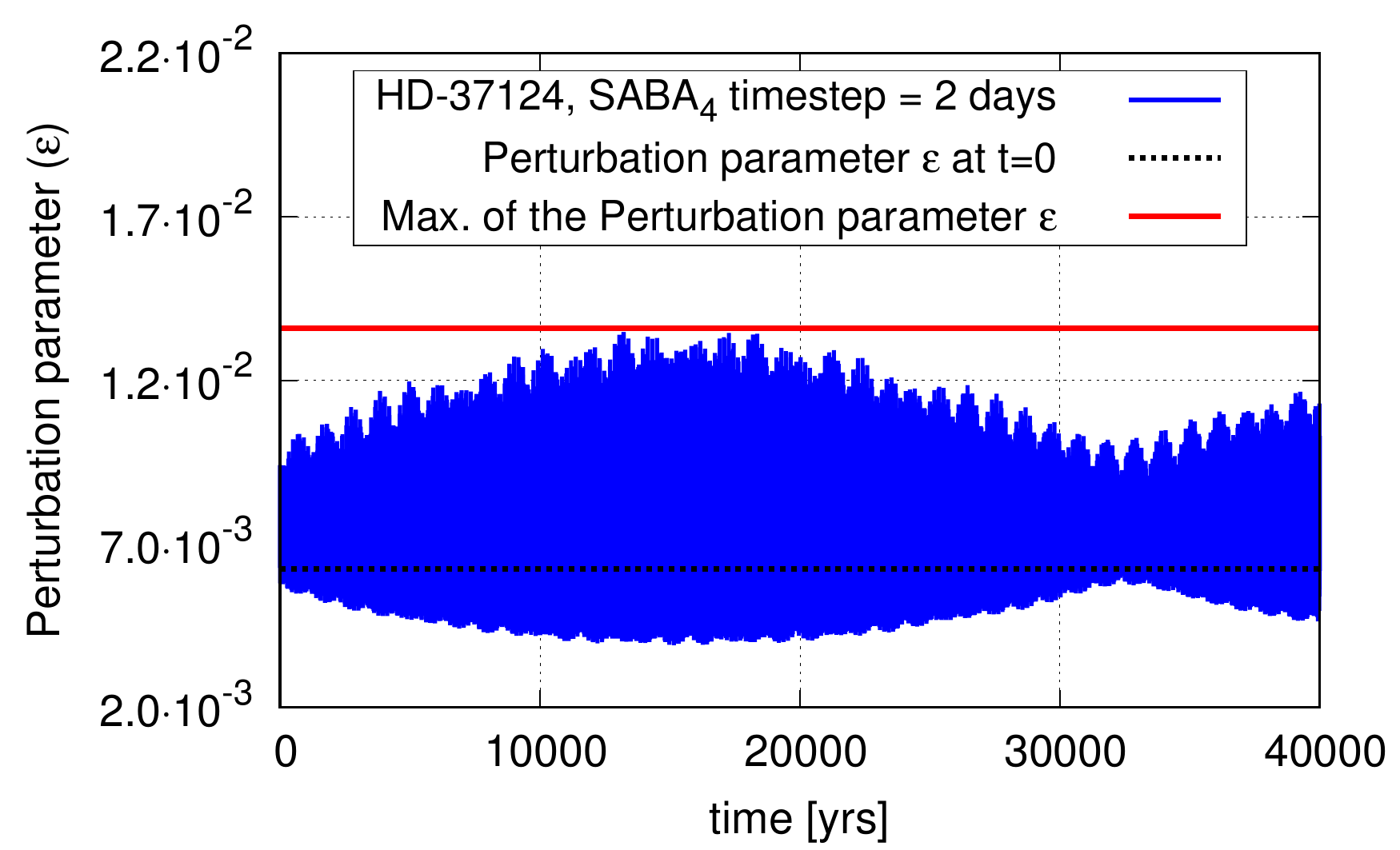}
\caption{
Variability of the perturbation parameter $\epsilon(t)$ for HD~37124 initial
condition (Tab.~\ref{tab:tab1}). The initial condition has been integrated for
40~kyrs.
}
\label{fig:figure2}
\end{figure}
Table~\ref{tab:tab2} displays estimates of the perturbation parameter
$\epsilon$, which may be the  measure of systems complexity in
Tab.~\ref{tab:tab1}. The strength of mutual perturbations affects and forces a
non-Keplerian evolution of the orbits, which we expect to be revealed in
dynamical maps obtained with the fast indicators. 

We determine this parameter for the nominal initial conditions as
$\epsilon(t=0)$, see Tab.~\ref{tab:tab2}. Obviously, $\epsilon$ is a function of
time, and, as illustrated for HD~37124 system (Fig.~\ref{fig:figure2}), it may
vary during the orbital evolution. Therefore, we integrated all systems in
Tab.~\ref{tab:tab1} for $2\times10^3$ outermost orbits, and we choose the
maximal $\epsilon$ attained during the integration as the measure of the
perturbation. We also note, that $\max\epsilon$ in Tab.~\ref{tab:tab2} is only
a~reference value for dynamical maps, which span a range of orbital elements
around the nominal parameters.
\begin{table}
\centering
\caption{
Planetary systems classified by the perturbation parameter $\epsilon \equiv
\norm{{\cal R}/{\cal H_0}}$. Units are scaled with the choice of the Gaussian
constant $k^2=1$. We consider coplanar systems, hence the number of the degrees
of freedom for each system is $4 \times N$, where $N$ is the number of planets. 
}
\label{tab:tab2}
\begin{tabular}{l c c c c}
\hline
system & $\norm{{\cal H}_0}$ & $\norm{\cal R}$ & $\epsilon(t=0)$  & $\epsilon \equiv \max\epsilon$ \\
\hline\hline
HD~37124~b,c,d  & $6\times 10^{-11}$  & $4\times 10^{-13}$ &  $6\times 10^{-3}$ & $1.3\times 10^{-2}$ \\
Kepler-26~b,c   & $9\times 10^{-12}$  & $2\times 10^{-15}$ &  $2\times 10^{-4}$ & $2.5\times 10^{-4}$ \\
Kepler-60~b,c   & $2\times 10^{-11}$  & $3\times 10^{-15}$ &  $1\times 10^{-4}$ & $1.6\times 10^{-4}$ \\
Kepler-36~b,c   & $1\times 10^{-11}$  & $2\times 10^{-15}$ &  $2\times 10^{-4}$ & $1.3\times 10^{-4}$ \\
Kepler-29~b,c   & $5\times 10^{-12}$  & $3\times 10^{-16}$ &  $5\times 10^{-5}$ & $5.3\times 10^{-5}$ \\
\hline
\end{tabular}
\end{table}
We briefly characterize the sample of planetary systems below.   
\subsubsection{HD~37124: three planets in Jovian mass range} 
\label{subsubs:HD37124}
The HD~37124 planetary system \citep{Vogt2005} is likely a compact configuration
of three massive, Jovian-like planets discovered with the Radial Velocity
technique. Its dynamics has been intensively investigated
\citep{Gozdziewski2008,Baluev2008,Wright2011}.
The perturbation parameter $\epsilon$ depends not only on the number of planets,
but also on their mutual distance and their masses. Since we intend to use
reversible SI with constant step size, even moderate
eccentricities of compact orbits may be challenging for such numerical schemes,
in the sense of accuracy and conservation of the integrals of motion. HD~37124
planetary system may be a good example of such demanding system. Its Jovian
companions are present in a region spanned by low-order 2-body and 3-body MMRs
\citep{Gozdziewski2008,Baluev2008}. Given their relatively large masses, the
expected mutual gravitational interactions between the planets are the strongest
in the sample, as shown in Tab.~\ref{tab:tab2}.
\subsubsection{Kepler-26: two planets near 7:5 MMR}
\label{subsubs:kepler26}
A resonant planetary system that exhibits complex dynamics is Kepler-26
\citep{Steffen2012}. It consist of two super-Earth planets near to the second
order 7:5 MMR. Since the orbits may appear very near one to another, the mutual
gravitational interaction may become also very strong. Kepler-26 has the largest
$\epsilon$ value among \kepler{} systems displayed in Table~\ref{tab:tab2}.
We note that actually Kepler-26 hosts four confirmed planets
\citep{JontofHutter2016} but we neglect the innermost and the outermost planet
since the available observations do not make it possible to reliably constrain
their orbits and physical properties. The two-planet configuration is selected
merely to have an example of a~particular resonant system. This is motivated
through the recent studies of this system
\citep{JontofHutter2016,Hadden2015,Deck2016}. We determined the planetary masses
through a re-analysis of the long cadence Q1-Q17 TTV dataset in \citep{Rowe2015}.
%
\subsubsection{Kepler-60: three super-Earths in the Laplace resonance}
\label{subsubs:kepler60}
%
Recently, \cite{Gozdziewski2016} analysed the Kepler-60 extrasolar system and
two resonant best-fitting solutions to the long cadence TTV measurements were
found. Both of them may be interpreted as generalized, zeroth-order three-body
mean motion Laplace resonance. The Kepler-60 is an example of an extremely
compact configuration of relatively massive planets in orbits with periods of
$\simeq 7.1$, $\simeq 8.9$ and $\simeq 11.9$~days, respectively. This resonance
could be either a ``pure'' three-body MMR with only the Laplace critical
argument $\phi_L = \lambda_b -2 \lambda_c + \lambda_d$ librating with a small
amplitude, or it may simultaneously form a chain of two-body 5:4 and 4:3
MMRs. In both cases the resonant Kepler-60 system is dynamically active and
exhibits complex dynamics, both regarding limited zones of stable motions in the
phase-space, as well as the presence of Arnold web structures. Given the close
orbits, it is also a very demanding orbital configuration for tracking the
long-term evolution and stability.
\subsubsection{Kepler-36: massive super-Earths in stable chaos?}
\label{subsubs:kepler36}
The Kepler-36 system is one of the first configurations detected with the
analysis of its clear TTV signal \citep{Deck2012}. It exhibits the smallest
$\epsilon$ in the sample shown in Tab.~\ref{tab:tab2}. This system brought our
attention due to the presence of the so called {\em stable chaos}
\citep{Deck2012}. The stable chaos means the long-term 
stable orbits in the sense of Lagrange, in spite of large mLCE. To verify this phenomenon with more recent TTV data, we did a
preliminary re-analysis of the Q1-Q17 TTV measurements with the genetic
algorithm \citep{Charbonneau1995}. We choose one of the best-fitting orbital
solutions displayed in Tab.~\ref{tab:tab1} for numerical tests of REM.
\subsubsection{Kepler-29: two super-Earths in 9:7~MMR}
\label{subsubs:kepler29}
We re-analysed the TTV measurements of the Kepler-29 system discovered in
\citep{Fabrycky2012} in our recent paper \citep{Migaszewski2016}. This
compact configuration of two massive super-Earth planets in $\sim 5$~Earth mass
range is separated at conjunctions by only $\simeq 0.01$~au. We found that the
planets are in 9:7~MMR.

For the analysis here we used osculating elements in Tab.~\ref{tab:tab1} for two
dynamical models of the system. The first $N$-body model accounts for the mutual
interactions of the planets. The Kepler-29 configuration has been also tested in
the framework of the RTBP with two different splitting schemes of the Hamiltonian.
We transformed the observational system to the RTBP model by fixing the inner
mass to zero and the outer planet eccentricity also to zero. (In fact, this
eccentricity may be very small, $\ec\simeq 0.001$ in the real configuration). This
example is used as a~transition model between low-dimensional dynamical
system and the full $N$-body formulation.
%
%
\section{Results and interpretation}
\label{sec3}
%
In this Section we describe the results of testing the chaotic indicators defined in
Sect.~\ref{sec1}, when applied to the systems defined in Sect.~\ref{sec2}, and
characterized in Tabs~\ref{tab:tab1} and~\ref{tab:tab2}.

Those configurations are non-integrable multi-dimensional conservative systems
exhibiting resonant structures. We aim to illustrate these structures using two-dimensional
dynamical maps (grids) composed of two canonical variables selected in a given
initial condition. Usually, we choose the semi-major axis -- eccentricity,
$(a,e)$-plane for a selected planet, since these elements are rescaled canonical
actions of the planetary Hamiltonian, Eq.~\ref{eq:18}-\ref{eq:19}. We
vary these parameters along the axes of the grid within certain ranges, and the
dynamical signatures of phase trajectories are then computed in each point of
the grid. The results are colour-coded and marked in two-dimensional maps. 

Fast indicators, like FLI and MEGNO, are designed to detect chaotic orbits
for typically $10^3$--$10^4$ characteristic periods \citep{Giordano2016},
associated with the
fundamental (proper) frequencies.  However, in multi-dimensional dynamical
systems, like planetary systems, the frequencies may span a range of
a few orders of magnitude, like
the mean motions (fast frequencies) and precessions of nodes
and pericentres (secular frequencies)
see, for instance \citep{Malhotra1998}. 
When these frequencies interact, various resonances emerge, like
the two-body and three-body mean-motion resonances, secular resonances
between precessional frequencies, and secondary resonances, which appear
inside the MMRs \citep{book:Morbidelli2002}.  Therefore the ``fast indicator''
feature, meaning a detection of chaotic behaviour for a relatively short
interval of time, must be related to {\em the local} instability time-scale. 
{\em The absolute} integration interval required to reveal chaotic
motions has always a particular dynamical context. In this paper we
usually refer to typical time-scale of two-body
MMRs expressed in units of the outermost planets' period. It is not necessarily the
same, as the time-scale of secular or secondary resonances, which
is usually much longer.

In our experiments, we aim to reliably characterise the 
MMRs structures that
may involve secondary resonances, as shown and justified below. Therefore we
considered time-scales covering as many as $10^5$--$10^6$ outermost orbits.
We also computed high-resolution scans, up to
$1024\times1024$ points, to avoid missing fine structures of the phase-space.
Such time-scales and map resolutions may be redundant for routine
computations. Yet they may cause a huge, non-realistic CPU overhead, depending
on the particular algorithms. 

For all numerical experiments, we used our multi-CPU, ``embarrassingly parallel''
farm code \code{$\mu$Farm} (Go\'zdziewski, in preparation) armed with a number of
different fast indicators, which makes use of the Message Passing Interface
(MPI) and GCC ver. 4.8. Intensive computations have been performed on Intel Xeon
CPU (E5-2697, 2.60GHz) of the \code{Eagle} cluster at the Pozna\'n Supercomputing
and Networking Center. We refer to this particular CPU quoting code execution
timings, and they should be used comparatively.

Finally, we do not intend to analyse the dynamical systems in detail. We focus
on the sensitivity of the fast indicators for fine structures in the
phase-space, associated with complex borders of chaotic and regular motions, the
presence of separatrices and secondary resonances. We stress that this paper has an experimental character,
regarding applications to the $N$-body dynamics.
We test the REM reliability and sensitivity through investigating various computing schemes, in
order to find the optimal one.

\subsection{System 1: FGL Hamiltonian system}
\label{subs:system1}
The Hamiltonian defined by Eq.~\ref{eq:9} and the corresponding symplectic map
version were studied for resonances and chaotic diffusion phenomena
\citep{Froeschle2000,Lega2003,Froeschle2005}, with the help of fast indicators FLI and
MEGNO \citep{Slonina2015}. The REM algorithm has been already tested for this
Hamiltonian system by \cite{Faranda2012} with the canonical map technique for
a relatively small time-span of $10^3$ iterations. 

To preserve a homogeneous computing environment, we computed the REM maps with
the symplectic SABA$_3$ scheme. For MEGNO, we used the symplectic tangent map
\citep{Mikkola1999}, in accord with Eq.~\ref{eq:7}. Also SABA$_3$ scheme has
been used. Dynamical maps are shown in the $(I_1,I_2)$-plane, and show a small
portion of the Arnold web for $\epsilon=0.01$. This value is significantly smaller from
$\epsilon=0.04$ which was found as the borderline value for the global overlap
of resonances, i.e., between Nekhoroshev and Chirikov regimes of the dynamics in
this system \citep{Froeschle2000}.

\begin{figure}
\centering
\vbox{
   \hbox{\includegraphics[width=0.47\textwidth]{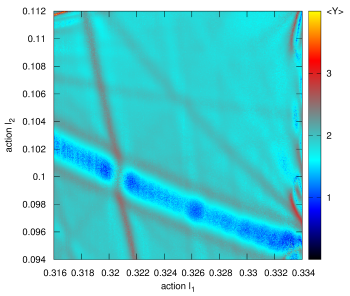}}
   \hbox{\includegraphics[width=0.47\textwidth]{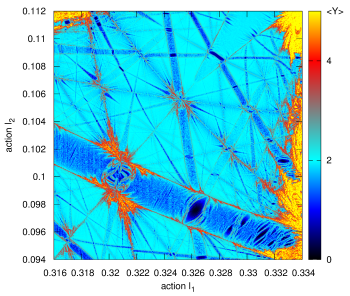}}
}
\caption{
MEGNO in a $1024 \times 1024$ grid of initial conditions in the
$(I_1,I_2)$-plane of actions for the FGL Hamiltonian. Perturbation parameter
$\epsilon=0.01$. The integrations were performed with the third-order
SABA$_3$-scheme, time-step of $h=0.29$, for $10^3$ ({\em upper plot}) and $10^4$
({\em bottom plot}) characteristic periods (time units), respectively. Integrations of MEGNO were interrupted if $\Y{}>10$. The
time-step provides the relative energy conservation to $\sim10^{-10}$. 
}
\label{fig:figure3}
\end{figure}

\begin{figure}
\centering
\vbox{
   \hbox{\includegraphics[width=0.47\textwidth]{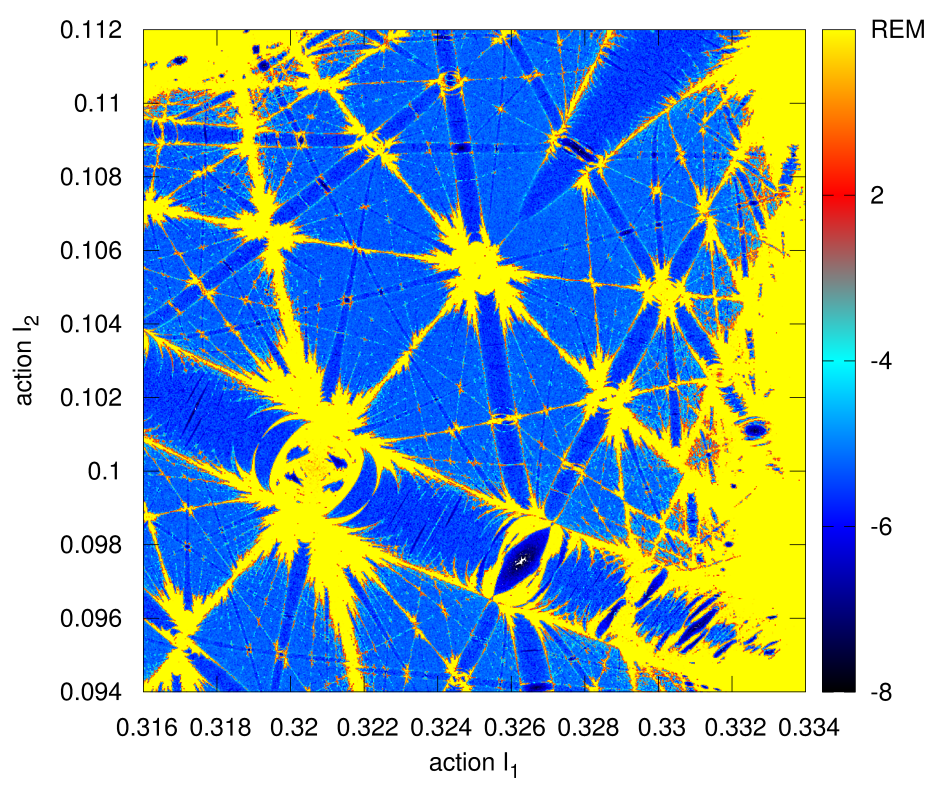}}
   \hbox{\includegraphics[width=0.47\textwidth]{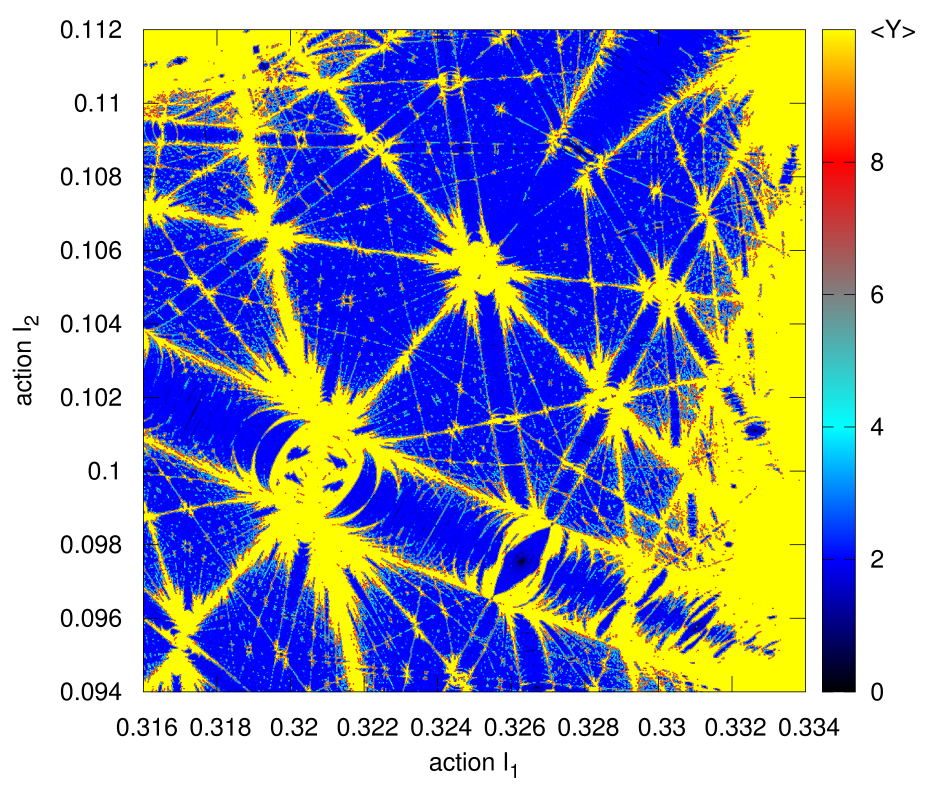}}
}
\caption{
A comparison of REM ({\em top panel}, note the logarithmic scale) and MEGNO
({\em bottom panel}, symplectic tangent map algorithm) for the FLG Hamiltonian.
The map is computed in a $1024 \times 1024$ grid of initial conditions in the
$(I_1,I_2)$-plane of actions. Perturbation parameter $\epsilon=0.01$. The
integrations were performed with the third-order SABA$_3$-scheme, time-step of
$h=0.29$ and for $10^6$ time units. Integrations of MEGNO were interrupted if
$\Y{}>10$. This time-step provides the relative energy conservation to
$\sim10^{-10}$. The CPU overhead for single initial condition is $\sim 1$~second
for REM, and between $0.1$ and $\sim 3$~seconds for MEGNO. 
}
\label{fig:figure4}
\end{figure}

We scanned a small fragment of the phase-space in the
$(I_1,I_2)$-plane with symplectic MEGNO for $T=10^3$
(upper panel of Fig.~\ref{fig:figure3}) and
$T=10^4$ (bottom panel of Fig.~\ref{fig:figure3}) time units, respectively. Given a small value of the perturbation parameter $\epsilon=0.01$, it
is clear that the $10^3$ periods integration interval is too short to reveal
chaotic motions that appear due to high-order resonances. Apparently, $10^4$ time
units is sufficient to detect main resonance structures. However, a complex
chaotic zone due to resonances overlap, which is seen at the right edge of the
MEGNO scans in Fig.~\ref{fig:figure3}, continuously develops for $10^5$ and
$10^6$ periods (Fig.~\ref{fig:figure4}). We also note that in order to
investigate the global diffusion, motion intervals as long as $10^8$ and $10^9$ 
characteristic periods must be considered, see \citep[][their Fig.~2]{Lega2003} or \citep{Slonina2015}.

Therefore, we extended the integration time to $T=10^5$, $10^6$ and $10^7$
characteristic periods, respectively. The results of the integrations for $10^6$ 
time units are illustrated in Fig.~\ref{fig:figure4} and they 
perfectly agree for both methods. Periodic (black), resonant (blue/dark blue or
grey) and chaotic (yellow/red or light grey) orbits are present in both maps
corresponding closely. We notice subtle resonant structures
between sharp (yellow/light grey) separatrices which are differentiated even
better from neighbouring trajectories in the REM map. 

For $T=10^7$ periods (not shown here), REM attains values as large as $10^3$
for chaotic orbits, and $10^{-4}$, for regular orbits. Nevertheless, only the
overall variability range is essential to detect all fine structures of the
phase-space, and we also found a perfect agreement of the derived REM scan with the
MEGNO map.
We note that some weak structures
e.g., around $(I_1=0.327,I_2=0.107)$, may be missing in the MEGNO map
for $T=10^6$ (Fig.~\ref{fig:figure4}) due to
non-optimal choice of the initial variations $\vec{\eta}$ required to
solve the deviation $\delta(t) \equiv \norm{\vec{\eta}}$. To avoid systematic
effects, we usually choose it randomly, following \cite{Cincotta2003}. However,
better strategies could be applied \citep{Barrio2009}, for instance, by
selecting the initial $\vec{\eta}$ as the unit vector parallel to $\nabla{\cal
H}$. 
On the other hand, the REM map for $T=10^4$ does not develop details seen in the
MEGNO scan for the same integration interval, which in this particular case may
be explained by longer saturation time-scale for REM than for MEGNO. This
effect is illustrated in two panels of Fig.~\ref{fig:figure3} for MEGNO. For stronger
perturbation $\epsilon=0.04$, or larger $(I_1,I_2)$-range, spanning lower-order
resonances, the equivalence of both algorithms is very close also for
$T=10^3$--$10^4$, see, for instance \citep{Faranda2012}.

The CPU overhead for one initial condition is very different for
both algorithms. For regular trajectories it is two times smaller for REM than
for MEGNO. For chaotic and strongly chaotic trajectories, the MEGNO CPU overhead
may be as small as $\sim 10\%$ of constant CPU overhead for REM, given the
chaotic signature of chaotic orbits may be examined ``on-line'', by tracking
whether the current value of $\Y<{\Y}_{\idm{lim}}$, where ${\Y}_{\idm{lim}} \gg
2$. The total integration time is similar, however the REM implementation could
be considered next to trivial.
%
%
\subsection{System 2: HD~37124, three sub-Jupiter system}
\label{subs:system2}
%
 Here we use the initial condition for HD~37124 system in
\citep{Gozdziewski2008}, which leads to dynamical structures in the semi-major
axes plane closely resembling the Arnold web in the model Hamiltonian,
Eq.~\ref{eq:9}.

\begin{figure}
    \centering
\vbox{
   \hbox{\includegraphics[width=0.47\textwidth]{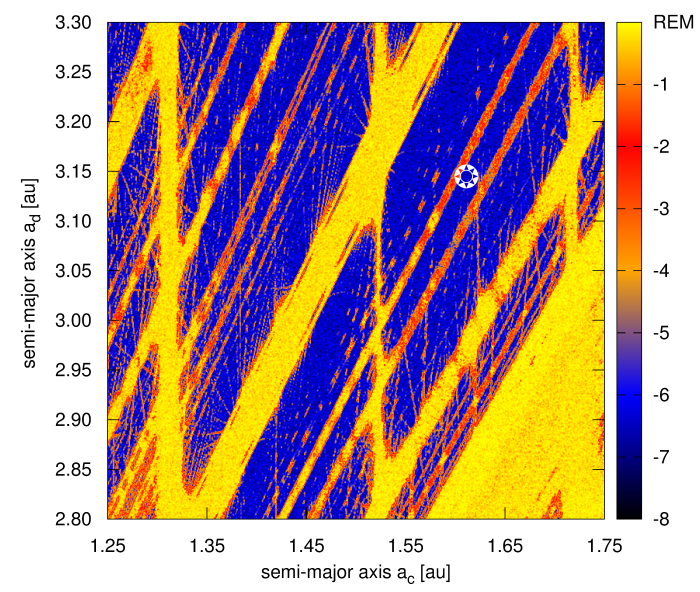}}
   \hbox{\includegraphics[width=0.47\textwidth]{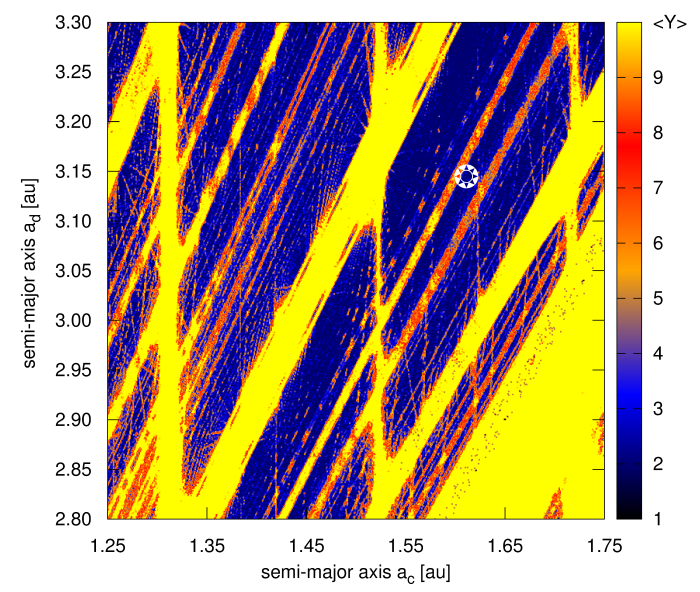}}
}
\caption{
REM ({\em top panel}, note the logarithmic scale) 
and MEGNO ({\em bottom panel}) maps for the HD~37124 system presented in $(\ac,\ad)$-plane.
SABA$_3$ REM algorithm with time-step of 5~days and forward integration time of
50~kyrs took $\sim 30$ seconds per initial condition. The CPU overhead for MEGNO
varied between $\sim 1$ to $\sim 22$ seconds, given limiting $\mean{Y}=10$.
The star symbol marks the nominal initial condition displayed in
Tab.~\ref{tab:tab1}. The resolution is $640\times640$ pixels.
}
\label{fig:figure5}
\end{figure}

Figure~\ref{fig:figure5} shows such a map in the $(\ac,\ad)$-plane. The grid
resolution is $640\times 640$ initial conditions, the integration time is
50~kyrs. The REM has been integrated with the SABA$_3$-scheme with the
time-step of 5~days, while for the Bulirsch-Stoer-Gragg \code{ODEX} integrator, the
relative and absolute accuracy has been set to $10^{-14}$. In this example, we
used this general purpose ODE solver as a reference, to obtain a reliable
dynamical map. Strong gravitational interaction between massive planets are
expected, and the tested configuration resides in collisional, very chaotic
zone.

Both dynamical maps agree very well, and all dynamical features may be found. We
note however, that this is rather a borderline case of REM application, due to
strongly chaotic regime. Also, due to fast linear growth of MEGNO for chaotic
orbits in this zone, unstable motions are quickly revealed. Hence the
integration time may be greatly reduced when some prescribed limit is reached.
This is not the case for REM, because, usually, the whole integration must be
performed before its value could be determined. However, the algorithm
provides reliable results even in such a difficult case.
%
%
\subsection{System 3: Kepler-26 planetary system near 7:5 MMR}
\label{subs:system3}
\begin{figure}
    \centering
\vbox{
   \hbox{\includegraphics[width=0.47\textwidth]{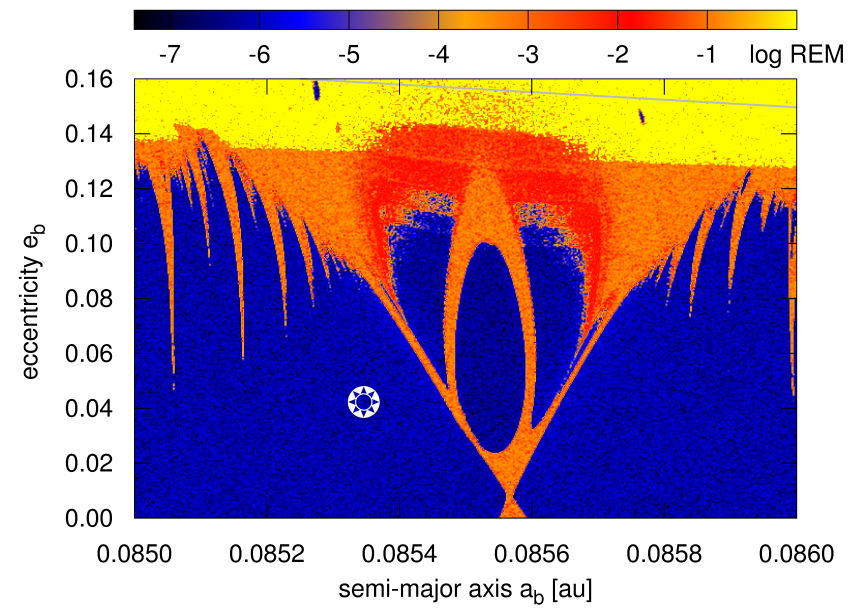}}
   \hbox{\includegraphics[width=0.47\textwidth]{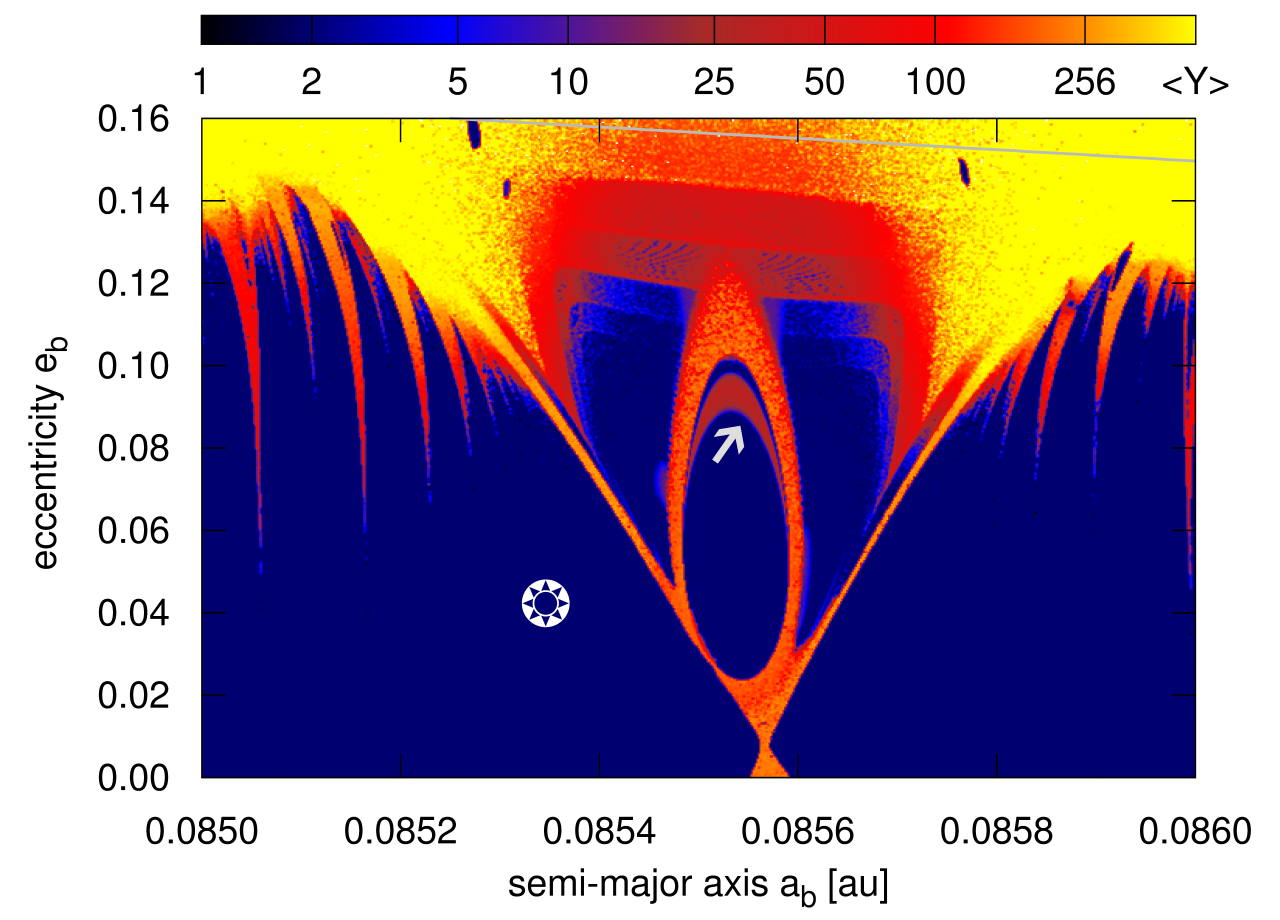}}
   \hbox{\includegraphics[width=0.47\textwidth]{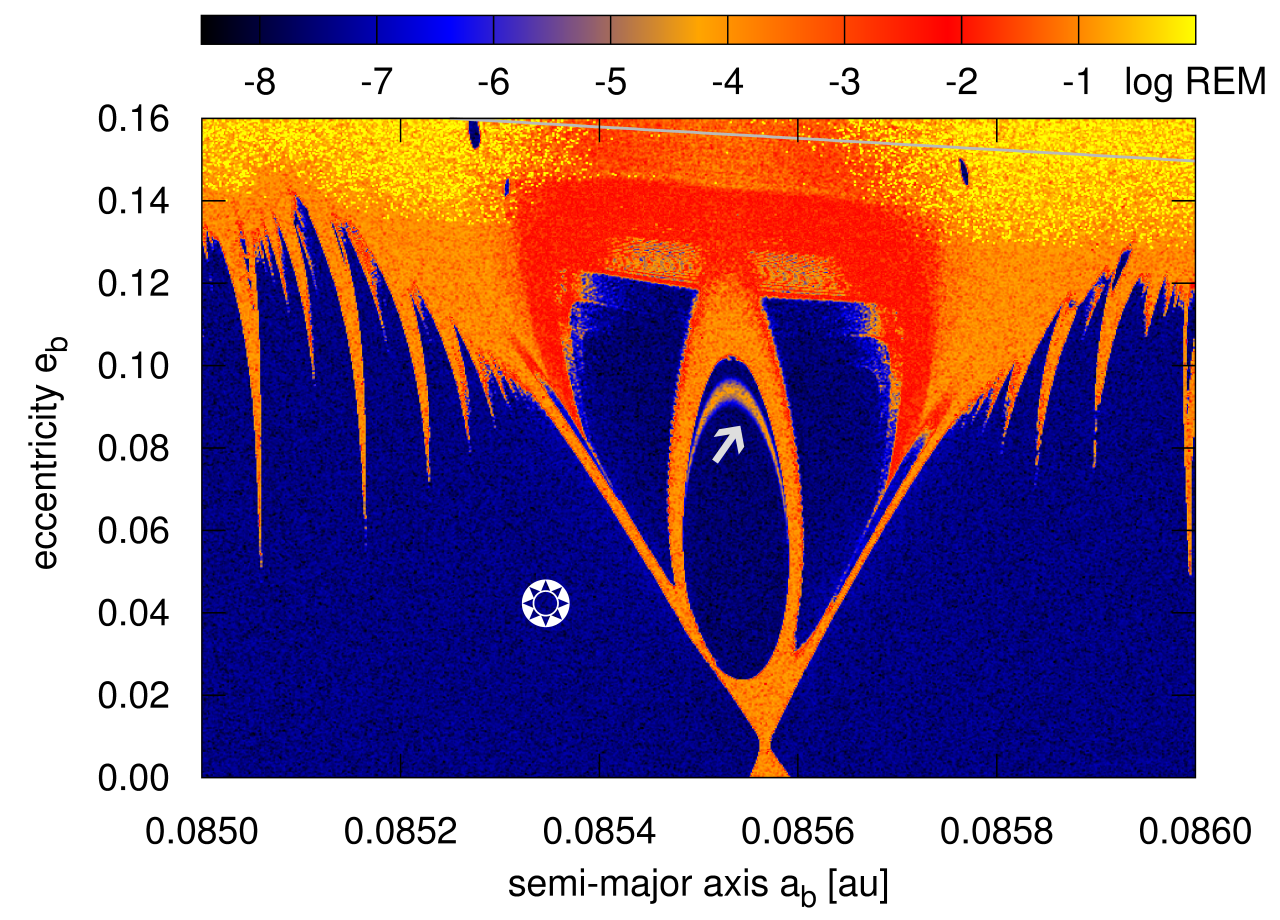}}   
}    
\caption{
MEGNO and REM dynamical maps for Kepler-26. {\em Top panel}: the REM map in
$(\ab,\eb)$-plane with the leapfrog-UVC(5) and time-step 0.25~days.
The forward integration interval $16$~kyrs. {\em Middle panel} is for symplectic
MEGNO map in the $(\ab,\eb)$-plane computed with SABA$_4$ scheme and time-step of
$0.5$~days integrated for $16$~kyrs ($\sim 5\times 10^5$ outermost orbits).
The maximum value of $\mean{Y}$ is equal to $256$. 
{\em Bottom panel:} the REM map computed with the leapfrog-UV$_\gamma$
algorithm, $\gamma=10^{-14}$, time-step of 0.25~days and the forward integration
interval of 5~kyrs ($\sim 1.5\times 10^5$ outermost orbits). 
White arrows show a~structure of weakly chaotic solutions
(it is absent in the top panel).
The resolution of
all maps is $800\times600$ points.
Thin grey curve in the top marks the mutual Hill radius separation of the
orbits. The perturbation parameter $\max\epsilon$ vary across the map between
$\sim 2.4 \times 10^{3}$ and $\sim 3 \times 10^{-3}$, see also
Tab.~\ref{tab:tab2}. The star symbol marks the nominal initial condition
displayed in Tab.~\ref{tab:tab1}. See the text for more details.}
\label{fig:figure6}
\end{figure}

%
The orbital period ratios of the inner pair of super-Earth in the Kepler-26
system are close to the second order 7:5 MMR. Dynamical maps in 
the $(\ab,\eb$)-plane shown in Fig.~\ref{fig:figure6} illustrate a complex shape of
the resonance. Both REM and MEGNO unveil its peculiar separatrix structure in
its interior part, which exhibits a few disconnected stable regions. 

We applied the most CPU efficient implementation of REM,
which is the second order leapfrog-UVC(5) algorithm
(Sec. \ref{sec4}). It is the mixed-variable scheme with Keplerian drift in
universal variables without Stumpff series \citep{Wisdom2016} and symplectic
correctors \citep{Wisdom2006} of the 5th order. For computing the MEGNO map, we
used the tangent map algorithm and the SABA$_4$ integrator. 

In the first experiment, the forward integration time
of 16~kyrs was the same for both algorithms. We recall that REM requires
effectively 32~kyrs integration, i.e., $5\times 10^5$ outermost orbits.
Then the overall structure of the 7:5 MMR and higher order MMRs are the same in both
maps. The algorithms reveal subtle stepping structure of chaotic configurations
(around $0.0855~\mbox{au}$ and eccentricity around $e_{\idm{c}}\sim0.12$) as well as tiny
islands of stable motion at the top of both maps. However, the elliptic shape of
strong chaos surrounding weaker chaotic motions present in the MEGNO map, 
marked with a white arrow, are
missing in the REM map. We attribute such fine structures to the presence of
secondary resonances \citep{book:Morbidelli2002} within the MMR zones. 

\begin{figure*}
\centerline{
\hbox{
   \hbox{\includegraphics[width=0.47\textwidth]{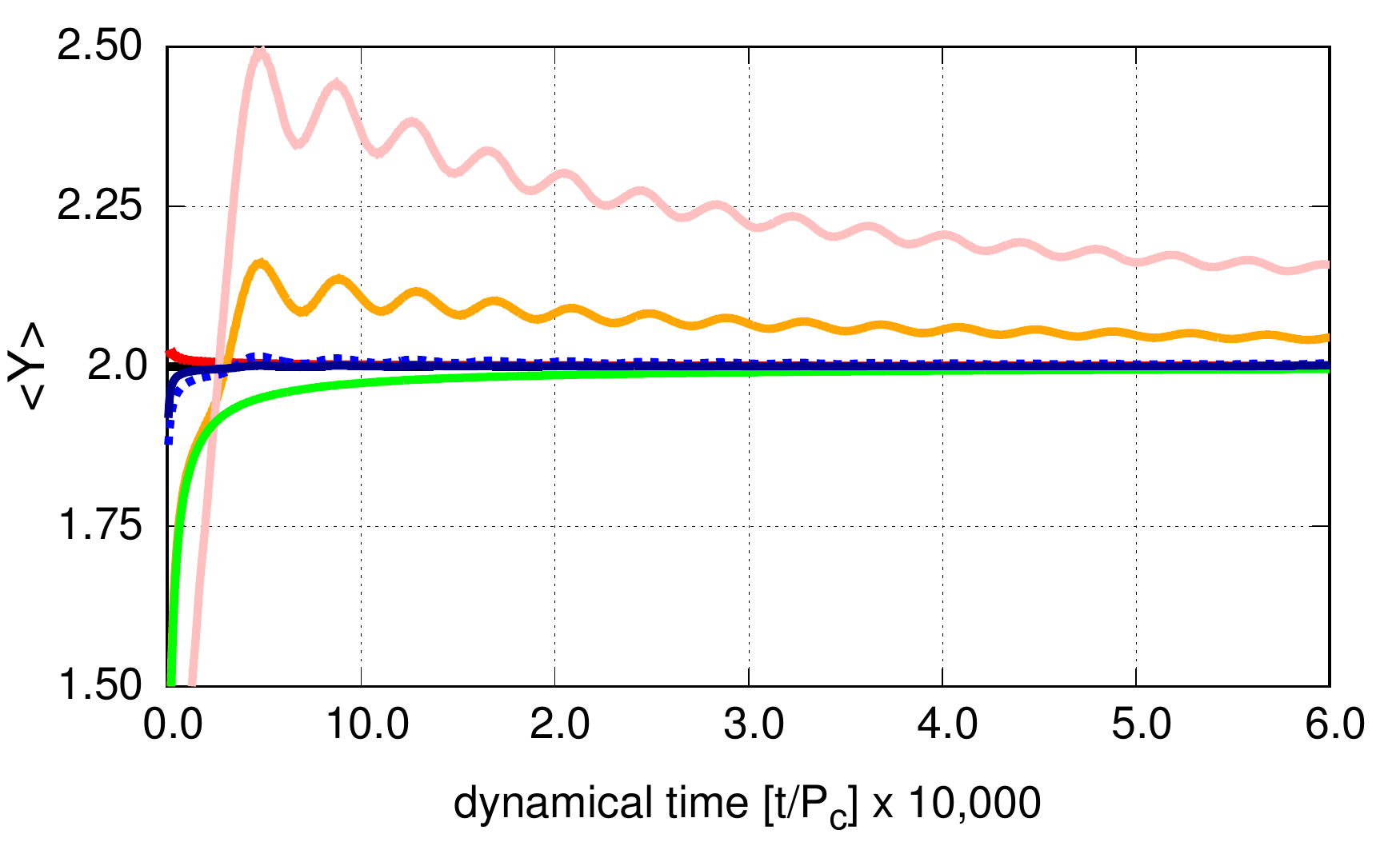}}
   \hbox{\includegraphics[width=0.47\textwidth]{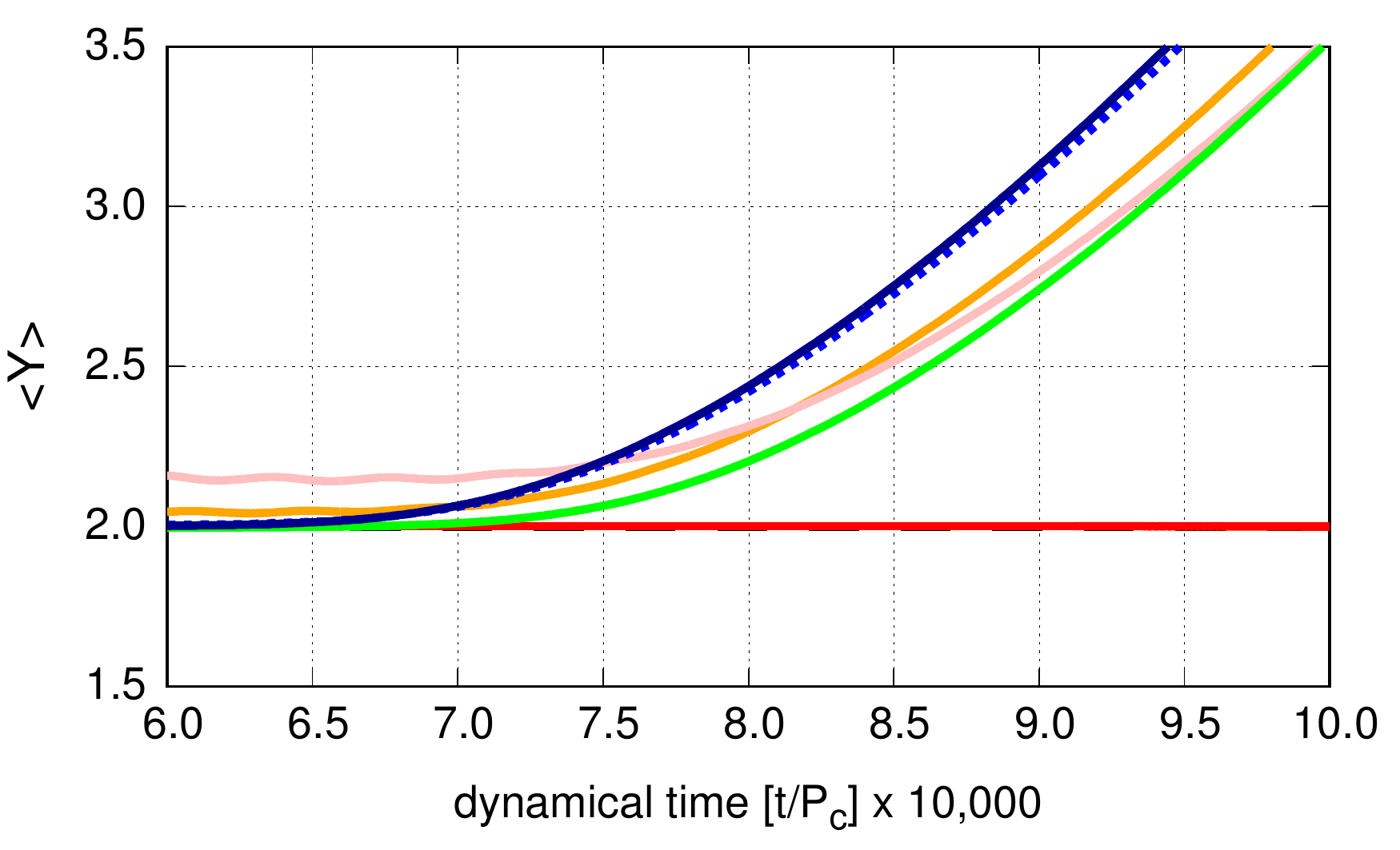}}
}    
}
\caption{
Temporal evolution of the MEGNO for a few initial conditions selected in the
arc-like structure of weakly chaotic Kepler-26 configurations inside the 7:5
MMR, marked with arrows in the dynamical maps in Fig.~\ref{fig:figure6}
(middle and bottom panels). Time is expressed in
units of the outermost period. {\em The left panel} illustrates apparently
regular solutions for $\sim 6\times 10^4$ outermost orbits. However, after 
additional $\sim
2\times 10^4$ outermost periods and more ({\em the right plot}), the MEGNO indicates chaotic solutions in agreement with
slow divergence. See the text for details.
}
\label{fig:figure7}
\end{figure*}

We selected a few initial conditions in the arc structure, and the MEGNO was
computed for these configurations 
to shed more light on their nature. The results are
illustrated in Fig.~\ref{fig:figure7}. The chaotic orbits in this region appear
as strictly regular up to $\sim 6 \times 10^4$ outermost periods, given the MEGNO
converged to~2 (the left panel in Fig.~\ref{fig:figure7}). However, for a longer integration interval the MEGNO
diverges slowly. This experiment shows that we would miss the chaotic arc
structure if the integration was restricted to the usual interval of $10^4$
outermost orbital periods, and extending the integration time to 
$\sim 10^5$ outermost orbits is unavoidable. We extended the integration time
even more, as the safety factor. 

In the arc region, the chaos may be called
as {\em slow} in contrast to of the other parts of the map, in which the MEGNO
indicates chaotic orbits for $\sim 10$--$100$~times shorter interval
({\em hard chaos}). 
In such a case, the ``purely'' numerical error growth does not make it possible
to detect weakly chaotic orbits by the REM algorithm. 
Therefore we used the
leapfrog-UV$_{\gamma}$ variant (see Sect.~5.1) that relies in perturbing the
initial condition vector, $\vec{x}_0 = \vec{x}_T + \gamma
\vec{\eta}$,
($\gamma=10^{-14}$) at the end of the first interval of integration $(t=T)$.
This simple modification brings a
dramatic improvement of the REM sensitivity for chaotic motions. The results    illustrated in the bottom
panel of Fig.~\ref{fig:figure6} are fully consistent with the MEGNO map in the
middle panel. We also note that {\em the total} integration interval for REM of
$2T = 5$~kyrs is similar
to the minimal integration time required to reveal the weakly chaotic orbits
with MEGNO, see Fig.~\ref{fig:figure7}.
In that case the CPU
overhead of $\sim 8$~s is constant for REM, and varies between $\sim
1$--16~seconds for MEGNO integrated for 5~kyrs 
(strongly chaotic and regular orbits,
respectively).

Furthermore, the REM map involves a signature of the collision zone of orbits
defined geometrically as the solution of $a_{\idm{b}} (1+a_{\idm{b}}) =
a_{\idm{c}} (1-a_{\idm{c}})$. {\em A dynamical border} of this zone is marked as
a change of shades across the REM map, around $\eb \simeq 0.14$. This zone
appears below the collision curve determined by the semi-major axis
$(a_{\idm{c}}-R_{\idm{H}})$, where $R_{\idm{H}}$ is the mutual Hill radius for
circular orbits
\[
 R_{\idm{H}} = \sqrt[3]{\frac{m_{\idm{b}}+m_{\idm{c}}}{3M_{\star}}}\,
 \frac{a_{\idm{b}}+a_{\idm{c}}}{2},
\]
and $m_{\idm{b,c}}$, $a_{\idm{b,c}}$ are the masses and semi-major axes of the
planets, $M_{\star}$ is the stellar mass. The borderline is marked with thin, grey
curve in the dynamical maps. This feature illustrates that the leapfrog implementations
used in our experiments are robust for such near-collisional configurations, in spite of the
step size that was kept constant across the whole grid.

We conclude that the REM detected all MMR's structures and the overall shape of
chaotic zones with relatively very small CPU overhead. This experiment brings a
universal warning that if we are interested in a comprehensive characterisation of
the fine structures of the MMRs, the time-scales of possible
resonances must be examined with great care.

%
\subsection{System 4: the Laplace resonance in Kepler-60}
\label{subs:system4}
\begin{figure}
    \centering
\vbox{
   \hbox{\includegraphics[width=0.47\textwidth]{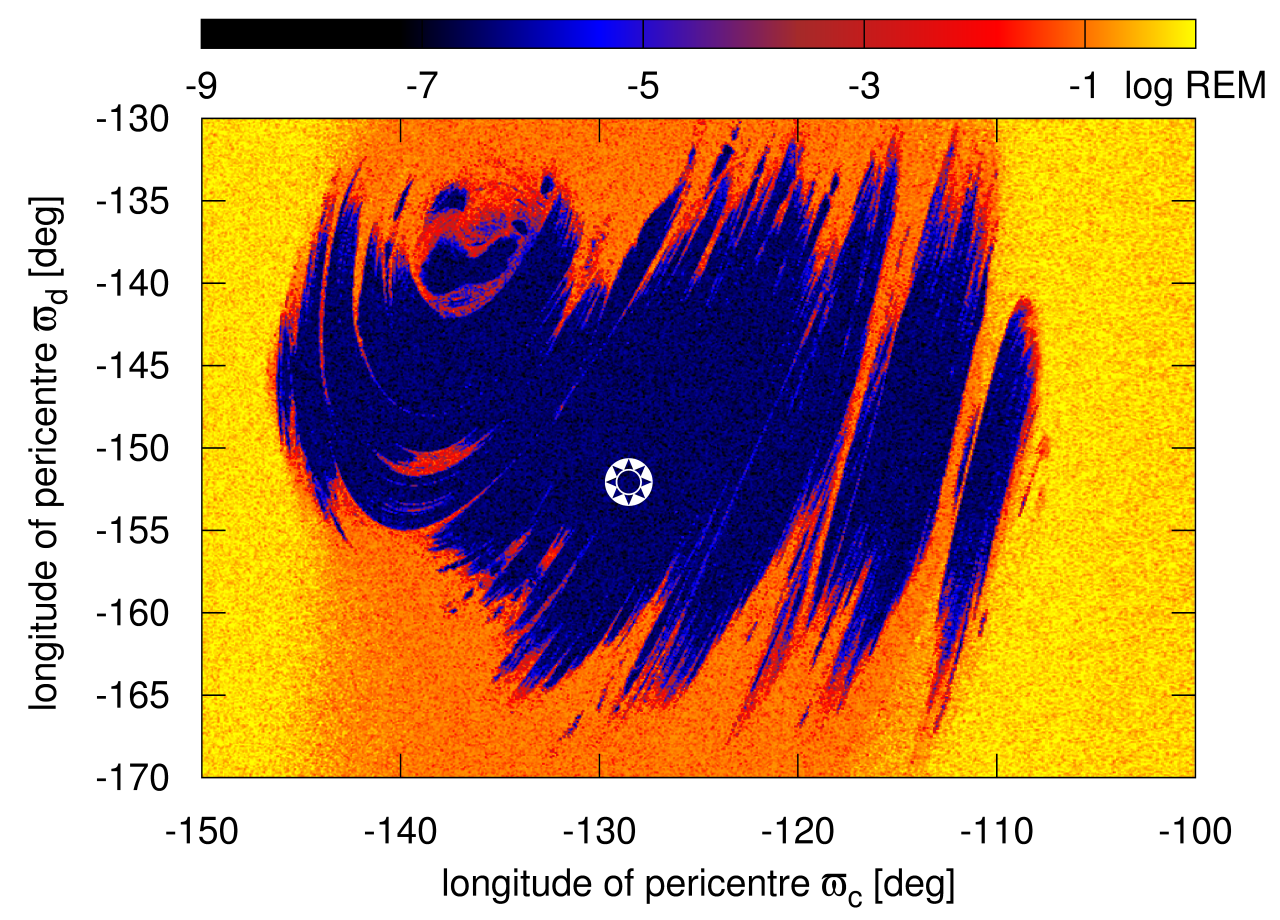}} 
   \hbox{\includegraphics[width=0.47\textwidth]{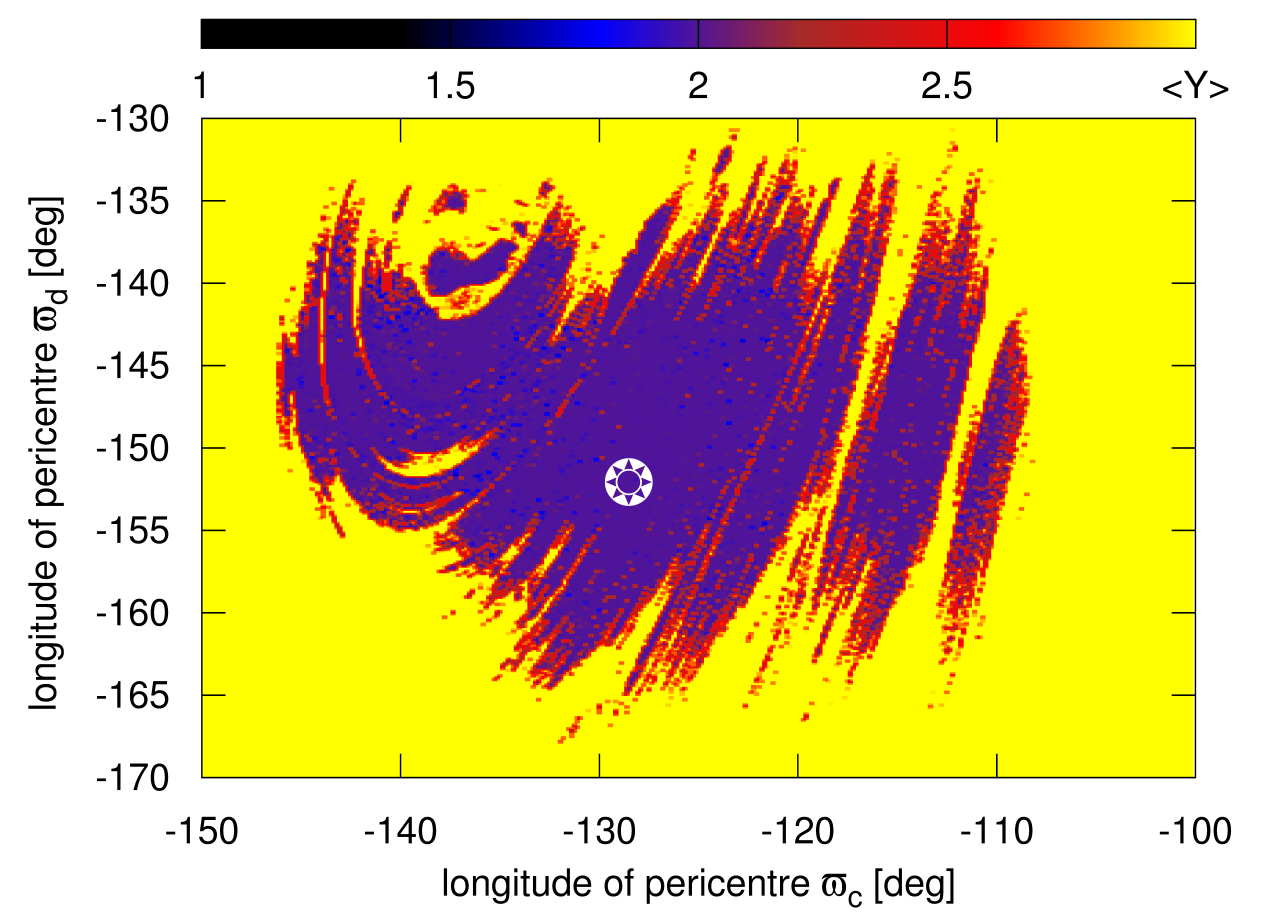}} 
}
\caption{
The REM {(\em top panel)} and MEGNO {(\em bottom panel)}
dynamical maps for the Kepler-60 system in the
$(\varpi_{\idm{c}},\varpi{\idm{d}})$-plane. The initial condition is displayed
in Tab.~\ref{tab:tab1} and marked here with the star symbol. Note that grid
resolutions are different, $800\times600$ for REM, and $720\times720$ for MEGNO.
Integration time is $16$~kyrs for MEGNO and forward integration interval
of $16$~kyrs for REM.
}
\label{fig:figure8}
\end{figure}

%
The Kepler-60 system has been comprehensively analysed in
\citep{Gozdziewski2016}, also regarding its dynamical structure. In
Figure~\ref{fig:figure8} we illustrate non-published MEGNO map (bottom panel) in
the ($\varpi_{\idm{c}},\varpi{\idm{d}})$-plane that reveals a complex structure
of the Laplace resonance around one of the best-fitting solutions (marked with a
star symbol) to the TTV measurements in \citep{Rowe2015}, see Table~\ref{tab:tab1}.
The top panel shows a high resolution REM map derived with the leapfrog-UVC(5)
integrator for $18$~kyrs, with the time-step of $0.125$~d. With this time-step,
the CPU overhead is huge, $\sim 80$~s per stable initial condition, i.e., still
about two times smaller than the mean CPU time for MEGNO with the SABA$_4$ and
the same time-step and forward integration interval. A significant fraction of
the grid is spanned by strongly chaotic configurations, which are detected by
MEGNO within a~few seconds.
This CPU time may be reduced with larger time-step, since our setup of
this experiment is very conservative. We note that the long integration interval of
$5\times 10^5$ outermost orbital periods has been selected in order to reveal
potentially slow chaotic diffusion, as in the FGL example (see Figs.~\ref{fig:figure3} and \ref{fig:figure4}). The initial condition describing the Kepler-60
system in the zero-th order three-body Laplace resonance unveils
qualitatively the same Arnold-web structures in the semi-major axes planes.
%
%
\subsection{System 5: Kepler-36 planetary system in 7:6 MMR}
\label{subs:system5}
\begin{figure}
    \centering
\vbox{
   \hbox{\includegraphics[width=0.47\textwidth]{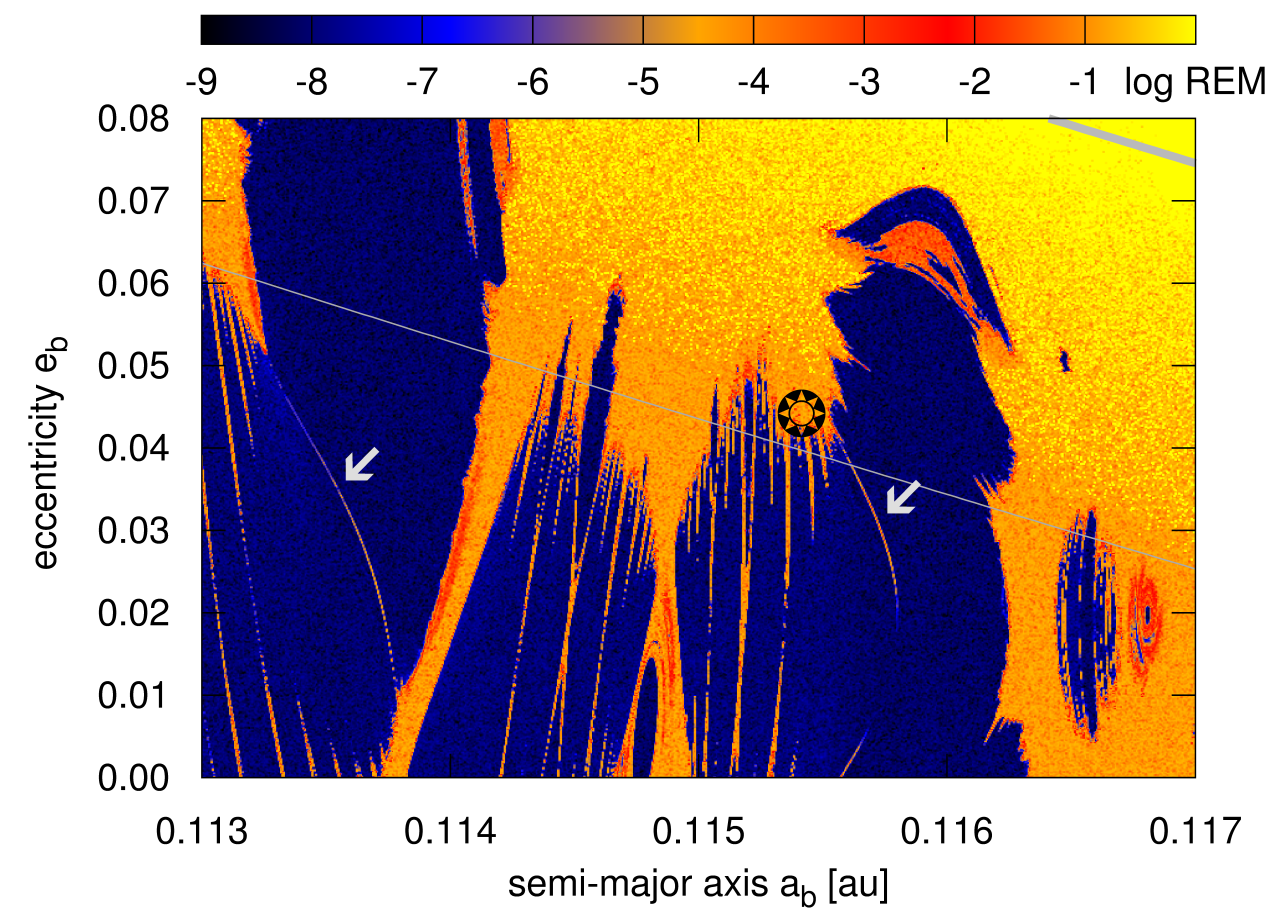}}
\hbox{\includegraphics[width=0.47\textwidth]{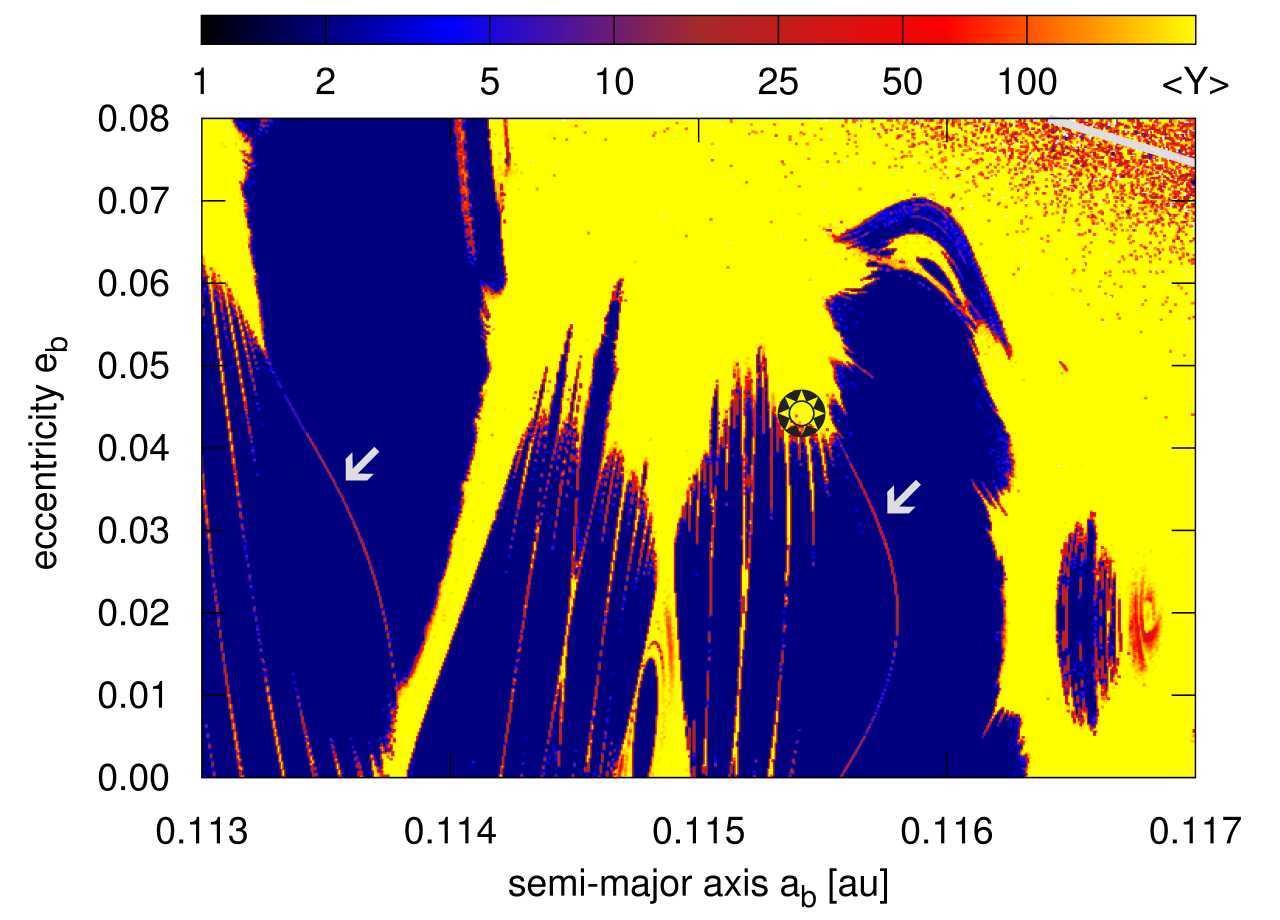}}   
   \hbox{\includegraphics[width=0.47\textwidth]{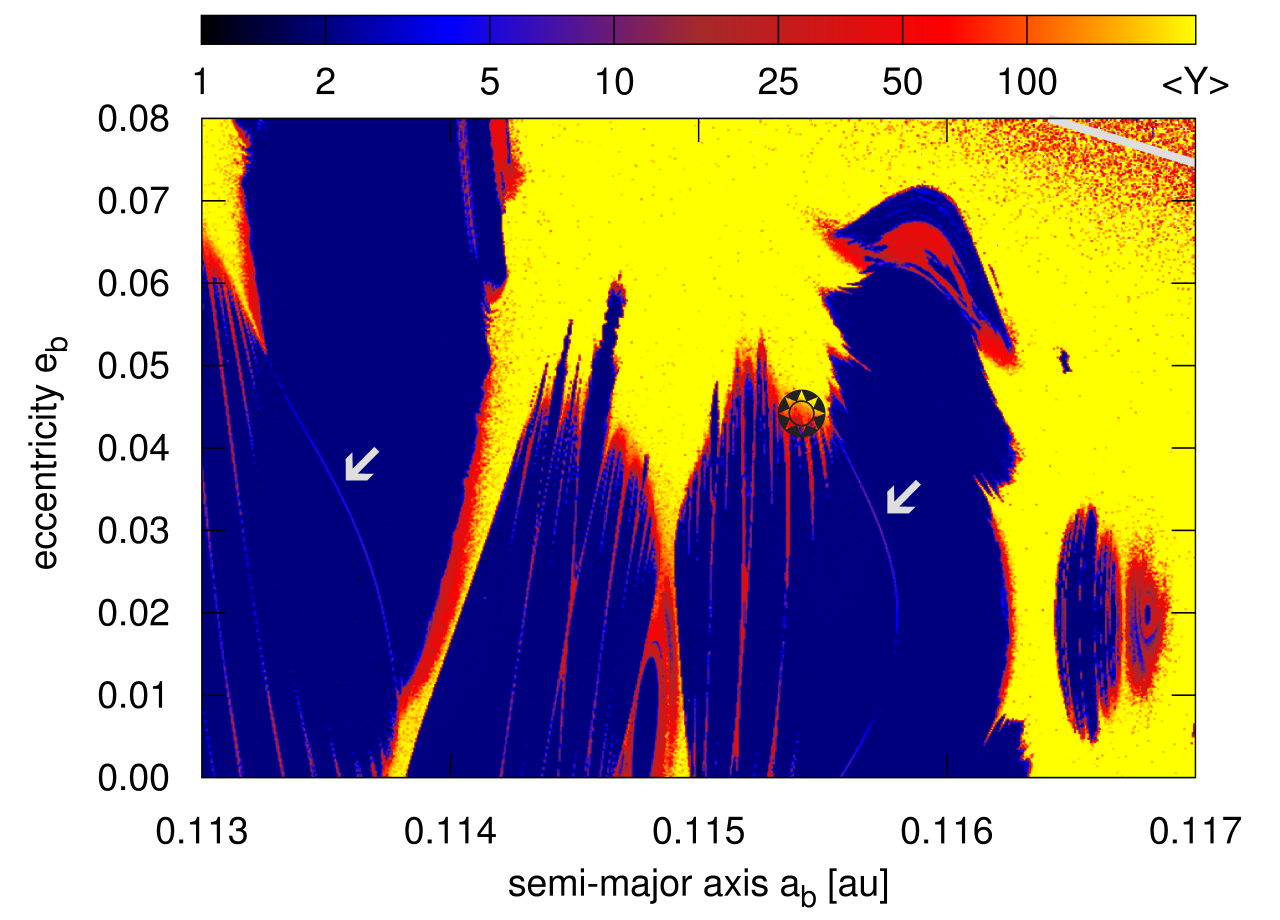}}
}
\caption{
MEGNO and REM comparison for the Kepler-36 planetary system. {\em Top panel} is for the second order
leapfrog-UV$_{\gamma}$ REM map in $(\ab,\eb)$-plane, forward integration
interval is 2~kyrs with CPU overhead of 3~s per initial condition and the
magnitude of random perturbation is $\gamma=10^{-14}$. The CPU overhead is about
of $4$~s.
{\em Middle and bottom panels} are for 
the symplectic MEGNO with 4th order SABA$_4$ scheme, time-step
$0.25$~days and the integration interval is 36 and 3~kyrs, respectively. 
For the bottom map, the CPU overhead is about $16$~s
per stable orbit. The resolution is $800\times600$. The star symbol marks the
nominal initial condition displayed in Tab.~\ref{tab:tab1}. 
Thick light-grey curve in the upper-right corner marks the collision 
line of orbits. Thin light curve in the top panel is for the mutual
Hill radius separation of the orbits.}
\label{fig:figure9}
\end{figure}

%
Dynamical maps in the $(a_{\idm{b}},e_{\idm{b}})$-plane for Kepler-36
\citep{Deck2012}, near to the first order 7:6~MMR are presented in
Fig.~\ref{fig:figure9}. 
We integrated the MEGNO map (middle panel in Fig.~\ref{fig:figure9}) for
36~kyrs ($\sim 10^6$ outermost orbits) with the 4th order SABA$_4$ scheme
and the tangent map algorithm \citep{Gozdziewski2008} with the time-step
0.25~days.  It looks like essentially the same as the map for 3~kyrs (bottom panel of
Fig.~\ref{fig:figure9}) spanning $\sim 8\times 10^4$ outermost orbits.
However, we note two fine unstable arcs marked with white arrows, which
are not well ``developed'' for the shorter integration interval.

The leapfrog-UV(5) REM computed for the integration interval of 36~kyrs with
time-step of $0.25$~days conserves the energy to $10^{-9}$ in relative scale.
While the dynamical map (not shown here) reveals globally the same chaotic and
regular solutions, two arcs marked with arrows in the MEGNO-panels in
Fig.~\ref{fig:figure9} are missing in the REM map. These features appear due to
weakly chaotic solutions with longer instability time-scale than in the main
part of the dynamical map, similar to the Kepler-26 model.

However, when the REM integration is done with the leapfrog-UV$_{\gamma}$ scheme
with time-step $0.25$~days and $\gamma=10^{-14}$, the weakly chaotic structures
are present already for the forward integration time of 2~kyrs ({\em only} $\sim
5\times 10^4$ outermost orbits). Then the CPU overhead per initial condition is
$\sim 3$~s, and between $1$ and $16$ seconds for MEGNO integrated for 3~kyrs.
(We note that the weak, arrow-marked structures in Fig.~\ref{fig:figure9} do not
appear clearly for 2~kyr MEGNO integration). In the later case, the CPU
overhead depends on the local value of mLCE, since we have set-up rather large
limit of $\Y_{\idm{lim}}=256$, which was used to classify initial condition as
strongly chaotic. 
Figure~\ref{fig:figure9} shows a very good agreement between the maps of both
indicators. The maps reveal a complex structure spanned by two MMRs, 6:5~MMR
centred around $a_{\idm{b}}\simeq 0.1135~\mbox{au}$, and 7:6~MMR centred around
$a_{\idm{b}}\simeq 0.1155~\mbox{au}$. From these two first order resonances an
extended overlap zone emerges. We note a large range of REM values spanning 7~orders of
magnitude. The border of the dynamical collision zone of the orbits may be
clearly seen as a change of shades across the map, which is very close to a
thick, grey curve determined by the mutual Hill radius separation from the
geometrical collision curve (thick grey curve, Fig.~\ref{fig:figure9}). All
major structures are fully recovered, in spite of the proximity to
the collisional region.

This example shows that the REM algorithm modified with small
perturbation of the initial conditions after the forward integration actually outperforms the MEGNO
symplectic fourth-order SABA$_4$ scheme, providing the same sensitivity for
chaotic orbits, with even smaller CPU cost for the REM dynamical maps.
%
\subsection{System 6: stable chaos in 9:7 MMR of Kepler-29?}
\label{subs:system6}
%
%

The Kepler-29 system has been found to be the most challenging example in our
sample, and a demanding testbed for the fast indicator algorithms investigated in
this paper.

\begin{figure}
\centerline{
\vbox{   
   \hbox{\includegraphics[width=0.47\textwidth]{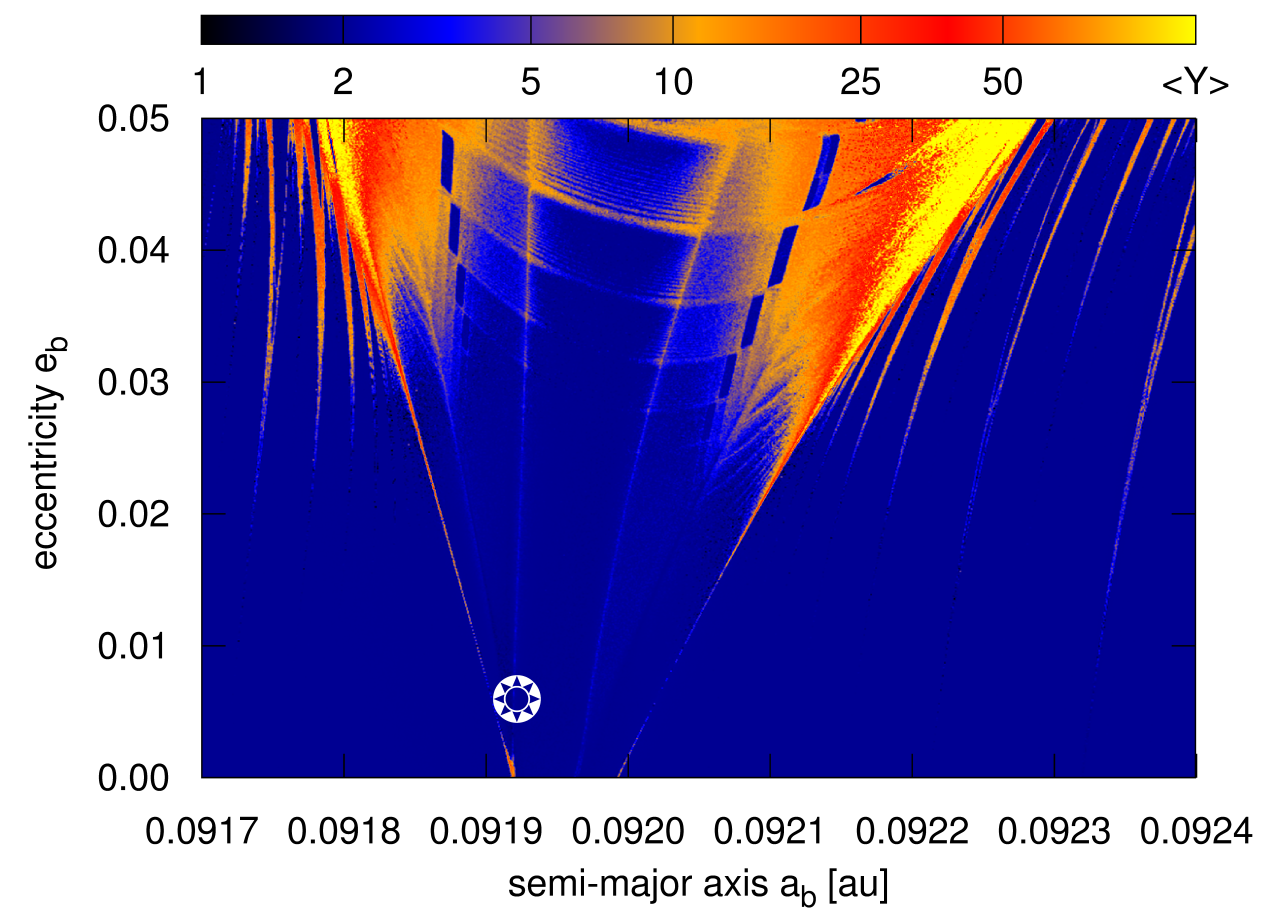}}
   \hbox{\includegraphics[width=0.47\textwidth]{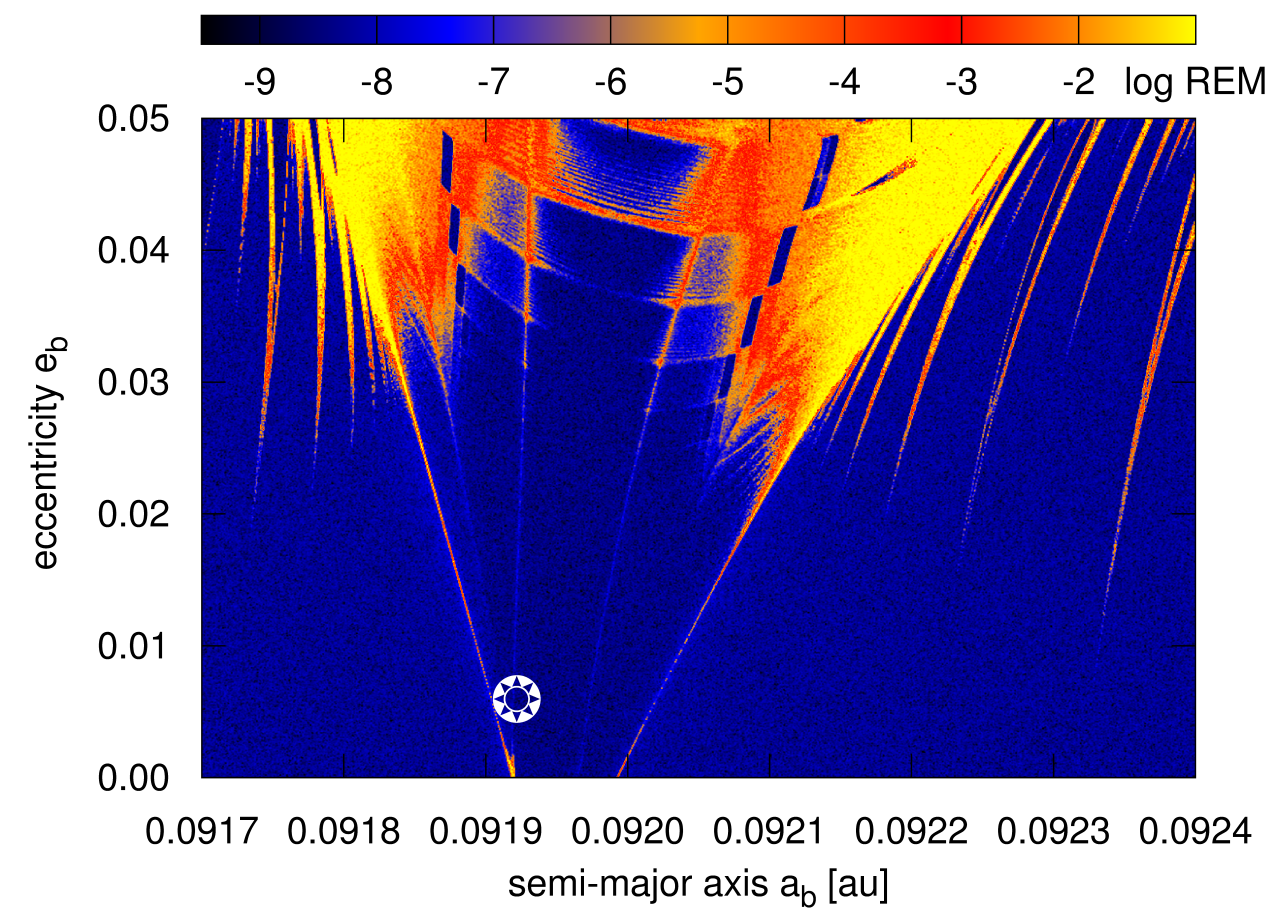}}   
} 
}   
\caption{
Dynamical maps for Kepler-29 in the
$(a_{\idm{b}},e_{\idm{b}})$-plane. {\em The upper panel} is for symplectic MEGNO map
with SABA$_4$ integrator, time-step of $0.5$~days, integrated for 1.2~kyrs
($3.3 \times 10^4$ outermost periods). {\em The bottom panel} is for the REM map
with the leapfrog-UVC(5) integrator, time-step of $0.25$~days, and the integration
interval is $2\times 1.2$~kyrs. 
The resolution is $1024\times768$ points.
The star symbol marks the nominal initial condition displayed in
Tab.~\ref{tab:tab1}.
}
\label{fig:figure10}
\end{figure}

\begin{figure*}
\centerline{ 
\vbox{
   \hbox{\includegraphics[width=0.47\textwidth]{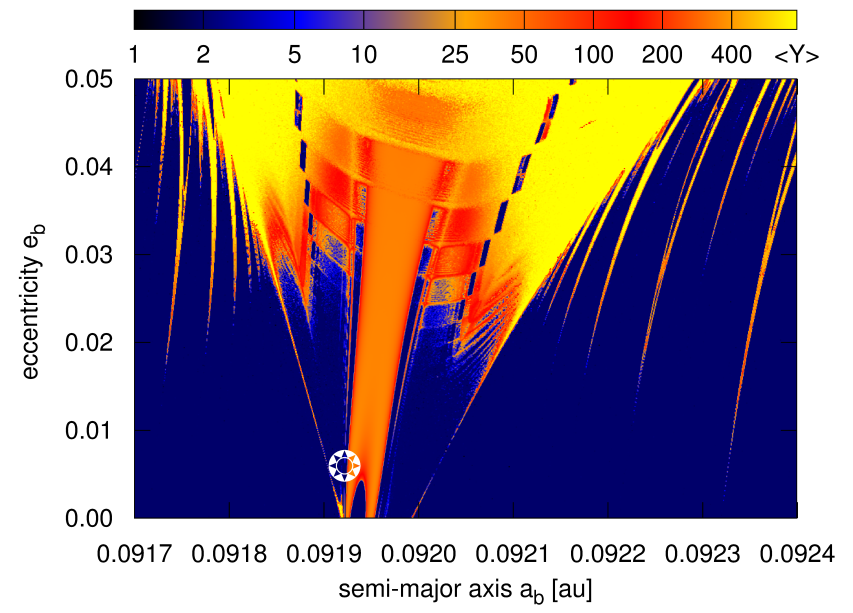}}
   \hbox{\includegraphics[width=0.47\textwidth]{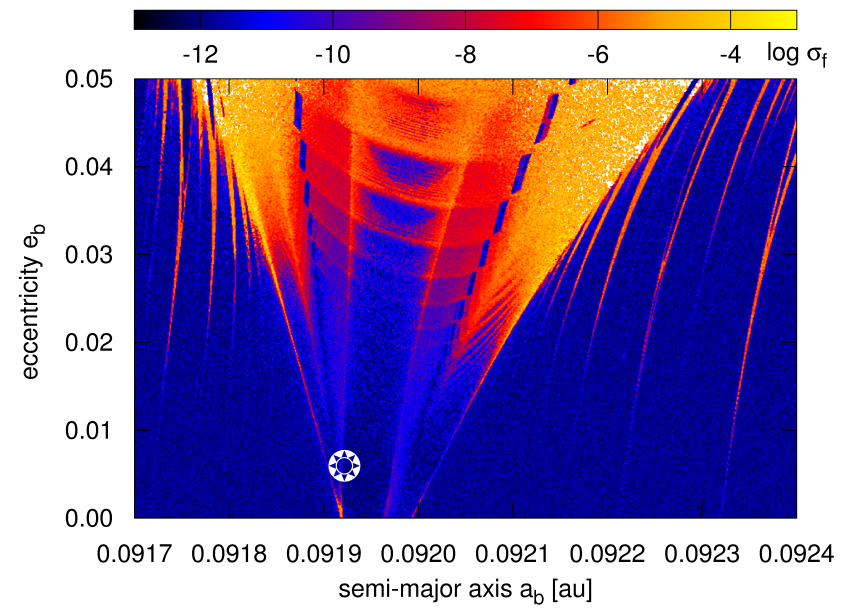}}
}
\vbox{   
   \hbox{\includegraphics[width=0.47\textwidth]{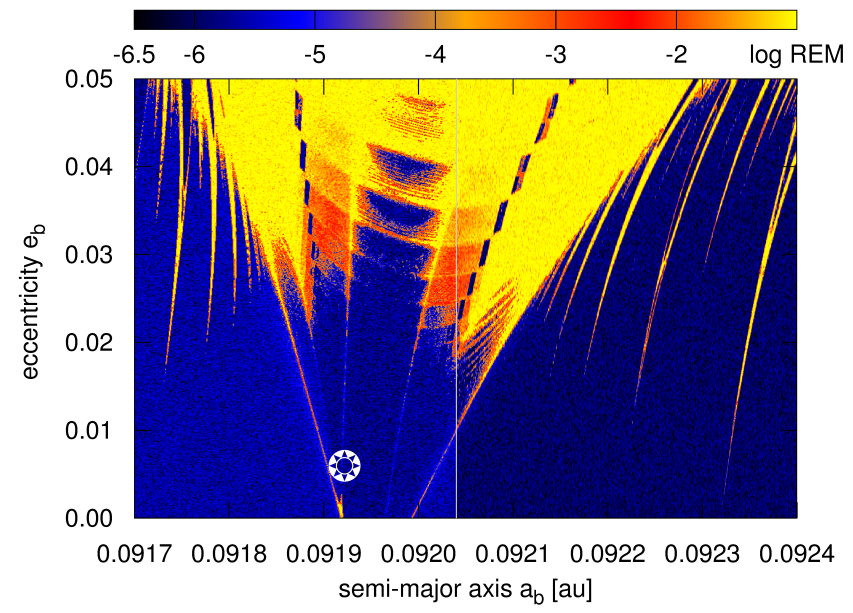}}
   \hbox{\includegraphics[width=0.47\textwidth]{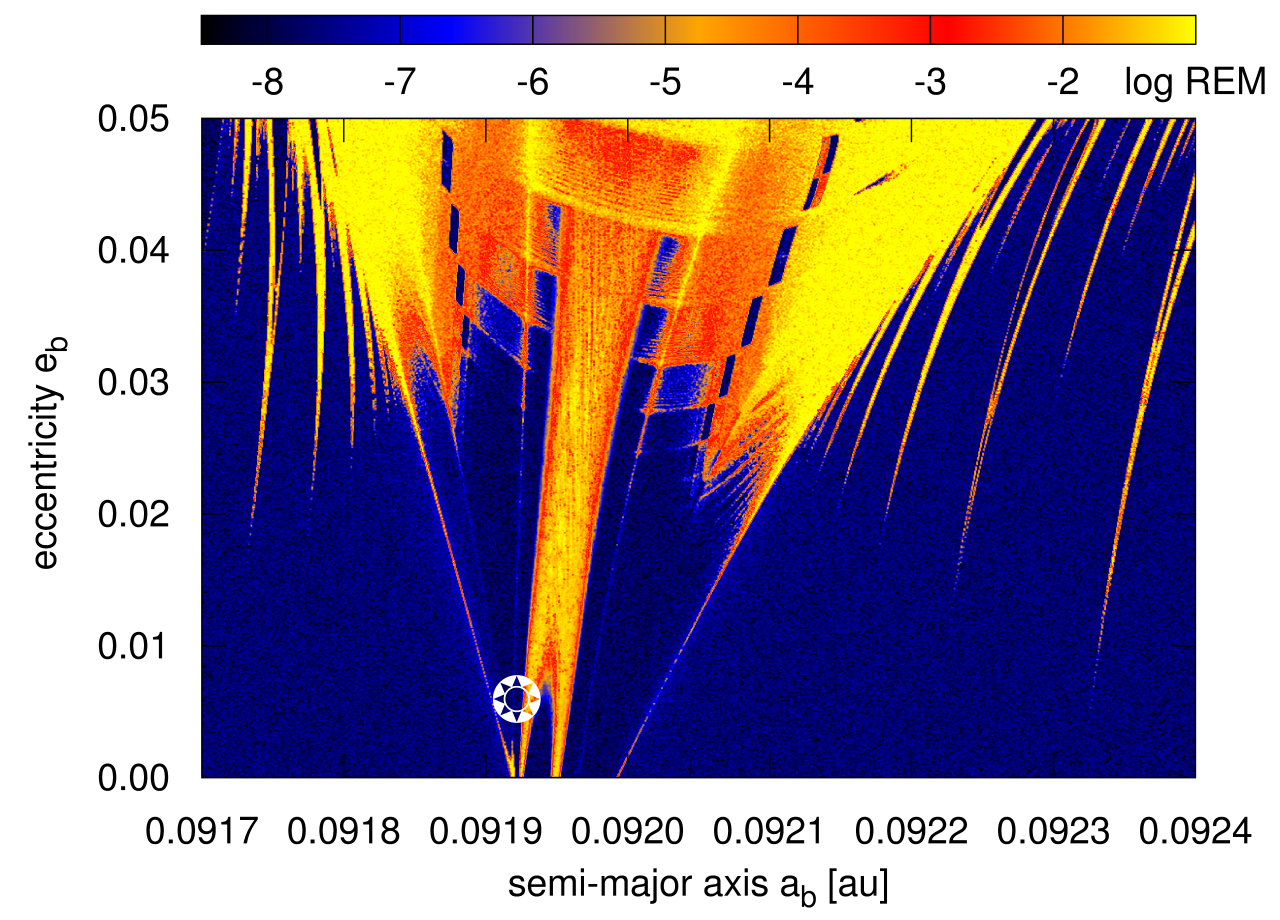}}   
} 
}   
\caption{
Dynamical maps for Kepler-29 in the $(a_{\idm{b}},e_{\idm{b}})$-plane. 
{\em Top-left panel} 
for symplectic
MEGNO map with SABA$_4$ integrator, time-step $0.5$~days, integrated for 72~kyrs
($2\times10^6$ outermost periods).
{\em Top-right panel} is for the $N$-body REM map, divided in two parts: the
left is for SABA$_4$ with time-step $h=0.25$~d, and the right one is for
SABA$_3$ with time-step $h=0.5$~days, forward integration interval is 36~kyrs
({$10^6$ outermost periods}). 
{\em Bottom-left panel} is for the diffusion frequency 
of the mean motion of the inner planet, the
total integration spans 6~kyrs, or $2\times2^{22}$ time-steps of
0.5~day ($\sim 2\times 10^5$ outermost periods).
{\em Bottom-right panel} is for the REM map with the
leapfrog-UV$_\gamma$ integrator, time-step 0.25~days, 
and the forward integration interval is 3~kyrs 
$\simeq 10^5$ outermost periods. 
The resolution of all grids is equal to $1024\times768$ points. The
star symbol marks the nominal initial condition displayed in
Tab.~\ref{tab:tab1}.
}
\label{fig:figure11}
\end{figure*}

In Fig.~\ref{fig:figure10} we present the REM and MEGNO maps computed for
$3\times 10^4$ outermost orbits, equivalent to $\sim1.2$~kyrs interval which 
should be typically sufficient to
reveal chaotic motions associated with the two-body mean motion resonances. The
map in the upper panel of Fig.~\ref{fig:figure10} has been obtained with
the symplectic MEGNO algorithm with SABA$_4$ and a step size of $0.25$ days,
respectively. The bottom left panel shows the REM dynamical map obtained
with the leapfrog-UVC(5) scheme and for the same forward integration interval of
1.2~kyrs. Apparently, both maps agree perfectly. The overall shape of the 9:7
MMR is clearly recovered in both maps, and major structures are the same in
the~region of moderate eccentricities.
However, keeping in mind that the MEGNO integration interval may be too short,
as in the Kepler-26 example, we extended the integration interval up to $2\times
10^6$ orbits (72~kyrs). This experiment reveals a wide chaotic strip in the
centre of the V-shaped MMR (top-left panel in Fig.~\ref{fig:figure11}). We note
that mLCE in the central strip is as large as $\sim 0.02/\mbox{yr}^{-1}$, given that
the maximal value of ${\Y}=768$ has been reached for 72~kyrs, and we
approximate mLCE $\equiv \lambda = 2{\Y}$, in accord with Eq.~\ref{eq:7}.
Actually, we know {\em a posteriori} that
the integration time to detect this structure with the help of
MEGNO is $\simeq 3$~kyrs and it corresponds to $6\times 10^4$ outermost
periods. Yet we show again that the usual ``rule of thumb'' choice of $10^4$ 
outermost periods for integrating MEGNO would be not sufficient, as
we demonstrated in Fig.~\ref{fig:figure7} for the Kepler-26 system.

Surprisingly, for the same long, total integration time of 72~kyrs, the REM with
SABA$_{4}$ and leapfrog-UVC(5) integrators do not ``see'' the wide chaotic strip in
the middle of the 9:7~MMR. Indeed, the top-right panel of Fig.~\ref{fig:figure11}
shows the REM map computed with the symplectic SABA$_{4}$ scheme. A thin,
vertical grey line across this map marks the change of the time-step from
$0.25$~days to $0.5$~days. The longer time-step has no impact on the results besides
smaller REM (darker shade).

We confirmed the discrepancy with the third fast indicator, the FMFT. We choose
the sampling time-step of $0.5$~days and $N=2^{22}$ for the same grid of initial
conditions as for MEGNO and REM (Fig.\ref{fig:figure11}). This is equal to $T \sim
2\times 10^5$ outermost periods, hence one order of magnitude longer
interval than usually required by MEGNO to reveal low-order two body MMRs. No
signs of geometric instability have been found in the problematic zone, in the
sense of a variation of the osculating elements and the proper mean motions
(bottom-left panel in Fig.~\ref{fig:figure11}). Moreover, we found a very
close agreement of the REM and FMFT signatures. These maps could be hardly
distinguished one from the other.  

The FMFT experiment reveals a very slow chaotic diffusion of the orbital elements, similar to the
Kepler-29 and Kepler-36 cases, yet in much more extended zone. Therefore we
applied the REM algorithm with the middle-interval perturbation. In this experiment
we choose the middle-interval perturbation of the state vector as
$\gamma=10^{-14}$, and we integrated the system with the leapfrog-UV$_\gamma$
scheme (Sect. 5.1). The time-step of $0.25$~days and the forward integration
interval is only 3~kyrs, i.e., the minimal integration time for MEGNO to
reveal the instability. For this time interval, the dynamical REM map in
the bottom-right panel of Fig.~\ref{fig:figure11} fully corresponds to the MEGNO
map integrate for 72~kyrs, and it reveals both all major and tiny structures
of the 9:7~MMR. The CPU overhead is in this
case only $\simeq 5$ seconds, which is roughly two time less than for
SABA$_4$--MEGNO integrated for the same interval.

The FMFT experiment helps us to explain the different signatures of the indicators
by the so called ``stable chaos'' phenomenon. This phenomenon was discovered by
\cite{Milani1992,Milani1997} for asteroid motions. It is found to be due to high
order MMRs with Jupiter in combination with secular perturbations on the
perihelia of the asteroids. The amazingly complex structure of the 9:7~MMR in
the Kepler-29 system is likely related to the secondary resonances which are characteristic
for low-eccentricity systems and appear due to a commensurability of the
resonant frequency with the apsidal libration frequency
\citep[e.g.,][]{book:Morbidelli2002}. While a detailed analysis of the Kepler-29
system is beyond the scope of this paper, it may be a clear evidence of the
stable chaos for the Kepler-29 planets in low-order 9:7 MMRs. This is unusual since large mLCE appear due to secular
interactions of relatively low-dimensional, two planets system only. We found
a similar effect, though much subtle, in the Kepler-26 system.

The results for Kepler-29 are the most clear indication of a possibility of a
non-unique classification of particular unstable (chaotic) orbits by different
fast indicators  due to locally varied time-scales of instability. In the
\kepler{}-systems, the slow chaotic diffusion of orbital elements clearly appears in the regions spanned
by MMRs. Regarding the canonical REM algorithm, for these weakly chaotic
solutions the numerical errors are too small to provide sufficient Lyapunov error
and sufficiently distant shadow orbit. By enforcing this perturbation
by adding an appropriate $\gamma\vec{\eta}$ term {\em only once} after the forward
integration interval, we enhance the sensitivity of the algorithm for chaotic
orbits. Given that the perturbation is very small (at the $10^{-14}$ level), both 
the REM signature for regular orbits and the energy conservation
 are not affected (see Sect. 5 for more details). 

%
%
\subsection{System 7: Kepler-29 as the restricted three body problem}
\label{subs:system7}
%
In the last experiment, we test a modified configuration of the Kepler-29 system
(Tab.~\ref{tab:tab1}) as the RTBP configuration, which is close to the 9:7~MMR
in the $N$-body model. We made this experiment to illustrate some differences
that may appear when REM is computed with different splittings of the same
Hamiltonian. 

\begin{figure}
    \centering   
\vbox{
   \hbox{\includegraphics[width=0.47\textwidth]{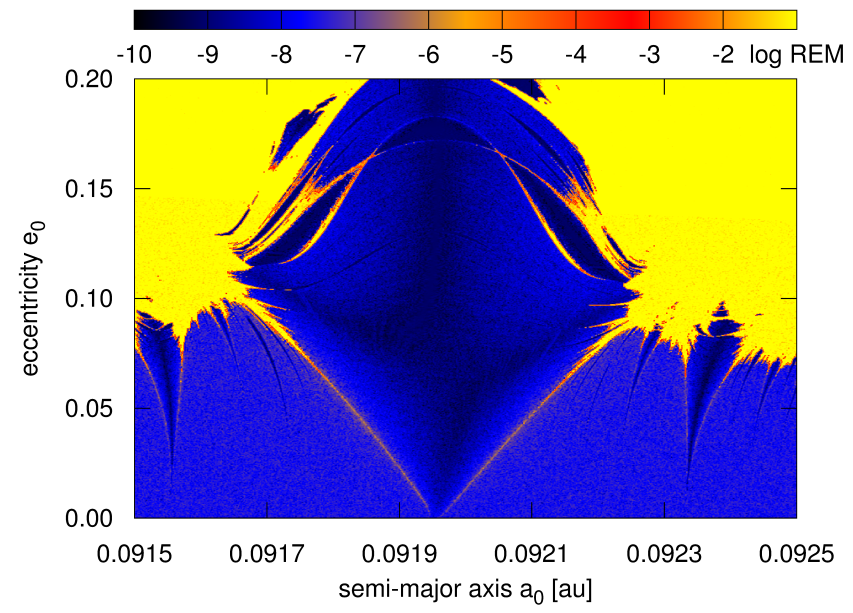}}
   \hbox{\includegraphics[width=0.47\textwidth]{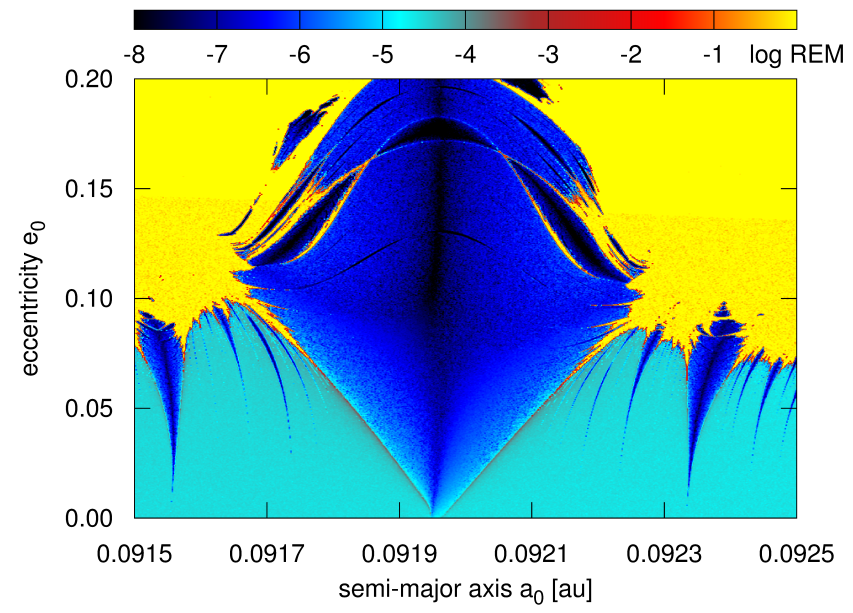}}
   \hbox{\includegraphics[width=0.47\textwidth]{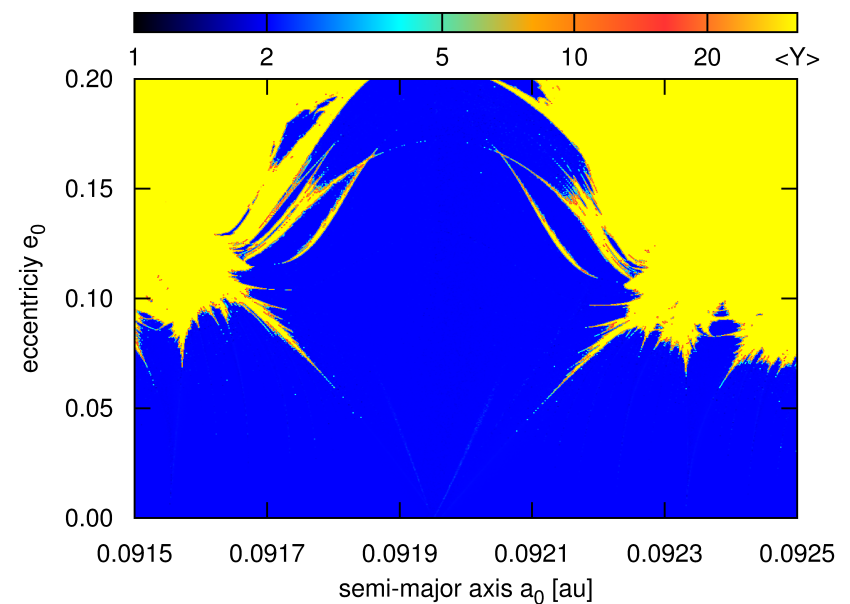}}
} 
\caption{
Dynamical REM maps for the $N$-body and RTBP models (Sect.~\ref{subs:systems}
and \ref{subs:r3bp}) for the Kepler-29-like system in the $(a_0,e_0)$ plane of the
mass-less planet (Tab.~\ref{tab:tab2}). The top panel is for the $N$-body REM map
with leapfrog-UVC(8) scheme, step size 0.25~d,
the middle panel for the REM map derived for the RTBP-Hamiltonian
integrated with the 4th order Yoshida scheme and time-step of 0.06125~d, and the bottom
panel is for the MEGNO map computed with the ODEX integrator,
the relative and absolute accuracy set to $10^{-15}$. For the REM maps
the forward integration interval is 3.6~kyrs ($\sim10^5$ periods of the binary), 
which is the same as for the MEGNO
map. The grid resolution is equal to $900\times768$ points.
}
\label{fig:figure12}
\end{figure}

The results are illustrated in Fig.~\ref{fig:figure12}. A map in the top panel
has been obtained in the framework of the $N$-body problem (Sect.~\ref{sec2})
with the leapfrog-UVC(8) algorithm with step size of $0.25$~days and integration
time of 3.6~kyrs. The middle panel shows the REM dynamical map obtained with the
4th order Yoshida integrator, and the forward integration interval of 3.6~kyrs
($10^5$ revolutions of the binary). However, due to the particular Hamiltonian
splitting (Sect.~\ref{subs:r3bp}), which is ``blind'' for the planetary
character of the model investigated, the step size has to be as small as
$0.0625$~d to conserve the energy at $\sim 10^{-8}$~level.

The overall shape of the 9:7 MMR is clearly recovered in both maps, and the
major structures are the same. However, significant differences of the absolute
REM values appear in the regions of the central, V-shaped MMR, as well as in
higher-order MMRs shown as smaller ``drops'' out of the central structure. The
background level of REM for stable orbits of $10^{-7}$--$10^{-6}$ can be the
basis to identify regular orbits. 

The RTBP map derived with the Yoshida scheme exhibits more clear differentiation
of regular orbits. We attribute it to a combination of two numerical effects.
One is the different sensitivity for stable-resonant and stable-quasiperiodic
orbits (we recall the FGL Hamiltonian example). For the Yoshida integrator,
there is also a numerical instability of the ``drift'' (Eq.~\ref{eq:14}), which
effectively means the rotation by angle $2h$. It results in the energy drift
\citep{Petit1998}. Indeed, we found that the Yoshida scheme exhibits such a
strong, linear energy drift re-inforced by smaller step sizes. This numerical
instability has likely a different impact on  the REM index in stable resonant
regions and in stable quasi-periodic zones. They are strongly discriminated as
dark-blue (dark grey) and light-cyan (light-grey) regions in the bottom REM map
in Fig.~\ref{fig:figure12}.

Yet the $N$-body variant of REM outperforms the RTBP model in the CPU overhead.
A single initial condition was integrated with the leapfrog-UVC(5) scheme for
$4.4$~seconds, while the 4th order Yoshida integrator required $\sim
7.7$~seconds, though the energy error is worse by 1-2 orders of magnitude.

For reference, we also computed the MEGNO map (the bottom panel in
Fig.~\ref{fig:figure12}), with the ODEX integrator, for the same interval
of 3.6~kyrs. For this integration time the separatrices
of the 9:7 MMR, its fine structure as well as lower-order MMRs appear as 
much less clear than in the REM maps. We note that
this result does not change when we use the SABA$_4$ integrator.

We conclude that the leapfrog-UVC(5) REM algorithm may be used for investigating the
dynamical structure of 2-planet \kepler{} systems, if they could be described
in the framework of RTBP. We also note that the RTBP could be easily generalized
with perturbations like primaries oblateness, radiation, and other conservative
effects. As long as such perturbed problems could be solved with symplectic and
reversible algorithms, REM may be the method of choice, given its straightforward
implementation and a great sensitivity for chaotic orbits. 
%
\section{Numerical setup and CPU efficiency} 
\label{sec4}
%
The most important feature of integrators used to compute the dynamical maps in
Sect.~\ref{sec3} is the time-reversibility, closely related to conservation of
the first integrals \citep{book:Hairer2006}. Usually, as much as $10^5$--$10^6$
outermost orbital periods must be considered when we want to investigate
large volumes or fine structures of the phase-space of the \kepler{}
planetary systems. Therefore the CPU overhead is the next critical factor for
choosing integration schemes. We focus on low-eccentric planetary systems, when
constant time-step is permitted due to relatively small mutual perturbations. We
aim to analyse the most relevant integrators features, like the maximal reliable
time-step, total integration time and preservation of the first integrals of
motion, when used to compute the dynamical maps in Sect.~\ref{sec3}. We use the
Kepler-26 and Kepler-36 systems as testbed configurations.
\subsection{Keplerian solvers and the leapfrog implementations} 
\label{subs:LFUV}
The classic ``planetary'' leapfrog scheme \citep{book:Hairer2006}, and its
derivatives, as the SABA$_n$/SBAB$_n$ schemes \citep{Laskar2001} or Yoshida
integrators \citep{Yoshida1990}, are composed of the Keplerian ``drift'', which
propagates the system along Keplerian orbits, and ``a~kick'', which corresponds to
the linear advance of the momenta. This is the genuine \cite{Wisdom1991}
scheme, known as the mixed-variable symplectic leapfrog. A crucial factor for
implementing this algorithm is an accurate and fast solver for propagating the
initial conditions at Keplerian orbit. In our implementation, we used the
Keplerian drift code of \cite{Levison1994} in their \code{SWIFT} package, which
become a de-facto numerical standard. A version of the leapfrog and higher order
schemes with the DL drift are postfixed with ``-DL'' throughout the text. We
also used a new, improved Keplerian solver by \cite{Wisdom2016}, kindly provided
by the authors (Jack Wisdom, private communication). This solver is based on the
universal variables \citep{book:Stumpff1959}, but without Stumpff series. The
REM variants with this solver are postfixed by ``-UV''. 

Furthermore, to improve the accuracy of the classical leapfrog integrations, we
used symplectic correctors introduced by \cite{Wisdom2006}. Our most
``sophisticated'' leapfrog REM implementation is then the leapfrog-UVC($n$)
algorithm with Wisdom correctors of order $n=1,3,5,7,8$.

Finally, we made extensive numerical experiments to improve the REM 
sensitivity to slow chaotic diffusion inside the MMRs in the Kepler-26, Kepler-36 and Kepler-29
systems. The sensitivity may be greatly enhanced by introducing
a random and very small perturbation of the state vector
(initial condition) $\vec{x}_T$ at
the end of the forward integration interval ($t=T$). It 
becomes the initial condition $\vec{x}_0$ for the backward integration:
\[
  \vec{x}_0 \equiv \vec{x}_T + \gamma \vec{\eta},
\]
where, in accord with Eq.~\ref{eq:A11}, $\gamma$ is the magnitude of the
perturbation, and $\vec{\eta}$ is the unit vector with random components. Here,
we choose $\gamma \sim 10^{-14}$, which provides the energy conserved well
bellow the limit introduced by the integrator scheme itself. This step may be
considered as simulating the error growth after much longer integration
interval, or by selecting a shadow orbit nearby the tested solution. 
We call this variant of the REM as the leapfrog-UV$_{\gamma}$
algorithm (UV$_{\gamma}$, i.e., the leapfrog with the Keplerian drift in
universal variables and the $\gamma$-perturbation added at the end of the first
interval of integration).
%
%
\subsection{Time-reversibility and CPU overhead of SABA$_n$ schemes}
\label{subs:CPUsaba}
%
Without the round-off errors, a symmetric integrator would be time-reversible
independently of the constant step size \citep{book:Hairer2006}. When the
round-off errors are present, the algorithm introduces certain systematic errors
depending on the number of steps. Therefore the REM final values may subtly depend on
the time-step, Hamiltonian splitting, and  total integration time. 

\begin{figure}
\centering
\includegraphics[width=0.47\textwidth]{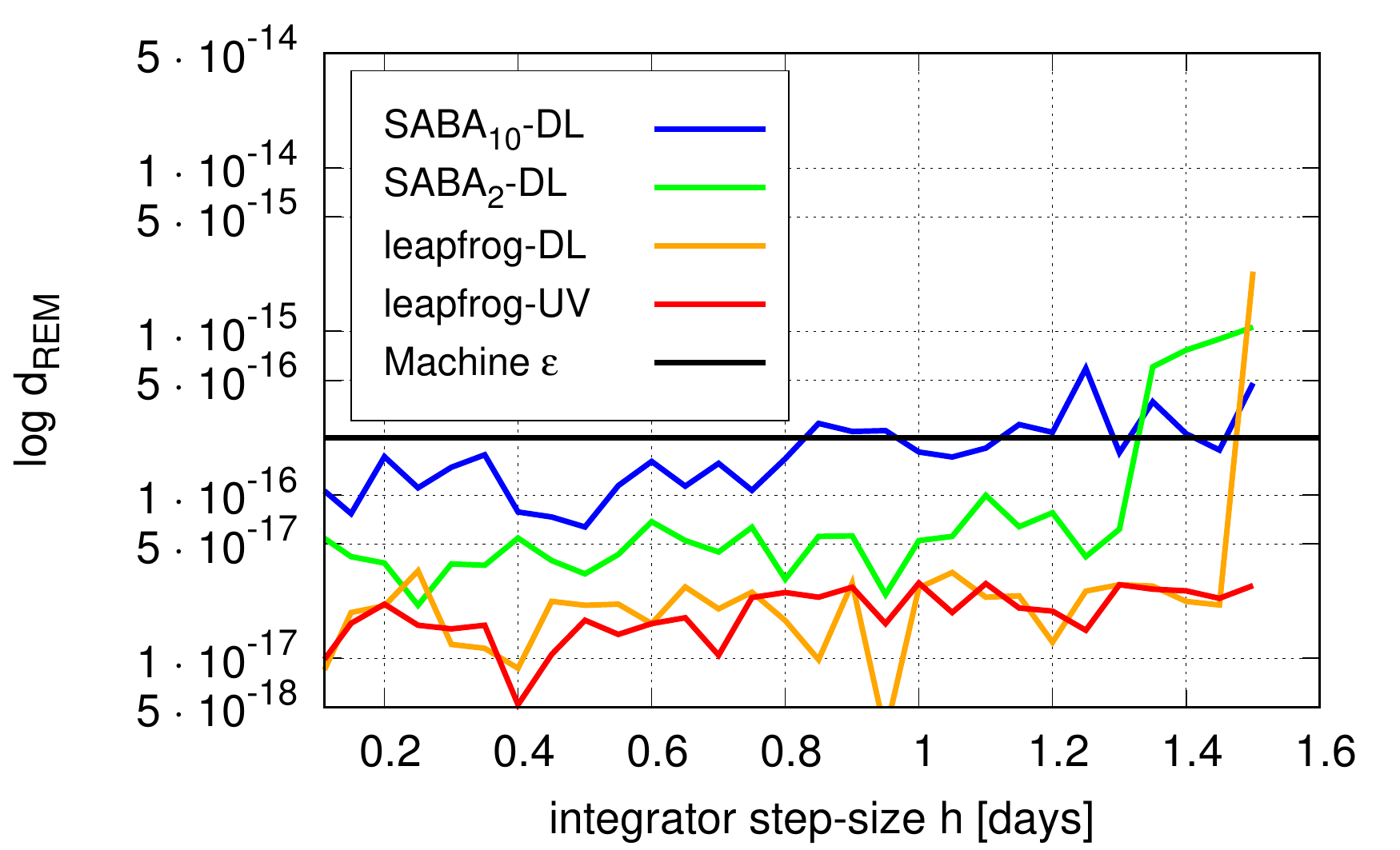}
\caption{
Time-reversibility test of SABA$_n$ and the leapfrog schemes, postfixed with
DL and UV, which stand for the Keplerian drift implemented in the
\citep{Levison1994} and \citep{Wisdom2016} Keplerian solvers,
respectively.  The time-reversibility breaks when the time-step is too
large.
}
\label{fig:figure13}
\end{figure}

Fig.~\ref{fig:figure13} illustrates numerical single-step reversibility for the
second order and the 10th-order SABA$_n$ schemes as well as the leapfrogs with
DL and UV solvers. In this test, we perform one forward integration time-step
$h$ and then the backward one for $-h$. Clearly all schemes are time-reversible up to
machine-precision (IEEE floating-point arithmetic, MACH$~\sim 2.2 \times
10^{-16}$), as expected, for a wide range of time-steps. In fact, the
reversibility is even better than the MACH value, since the calculations were
performed on \code{INTEL}-architecture CPU with registers of 80~bits.  

\begin{figure}
\centering
\includegraphics[width=0.47\textwidth]{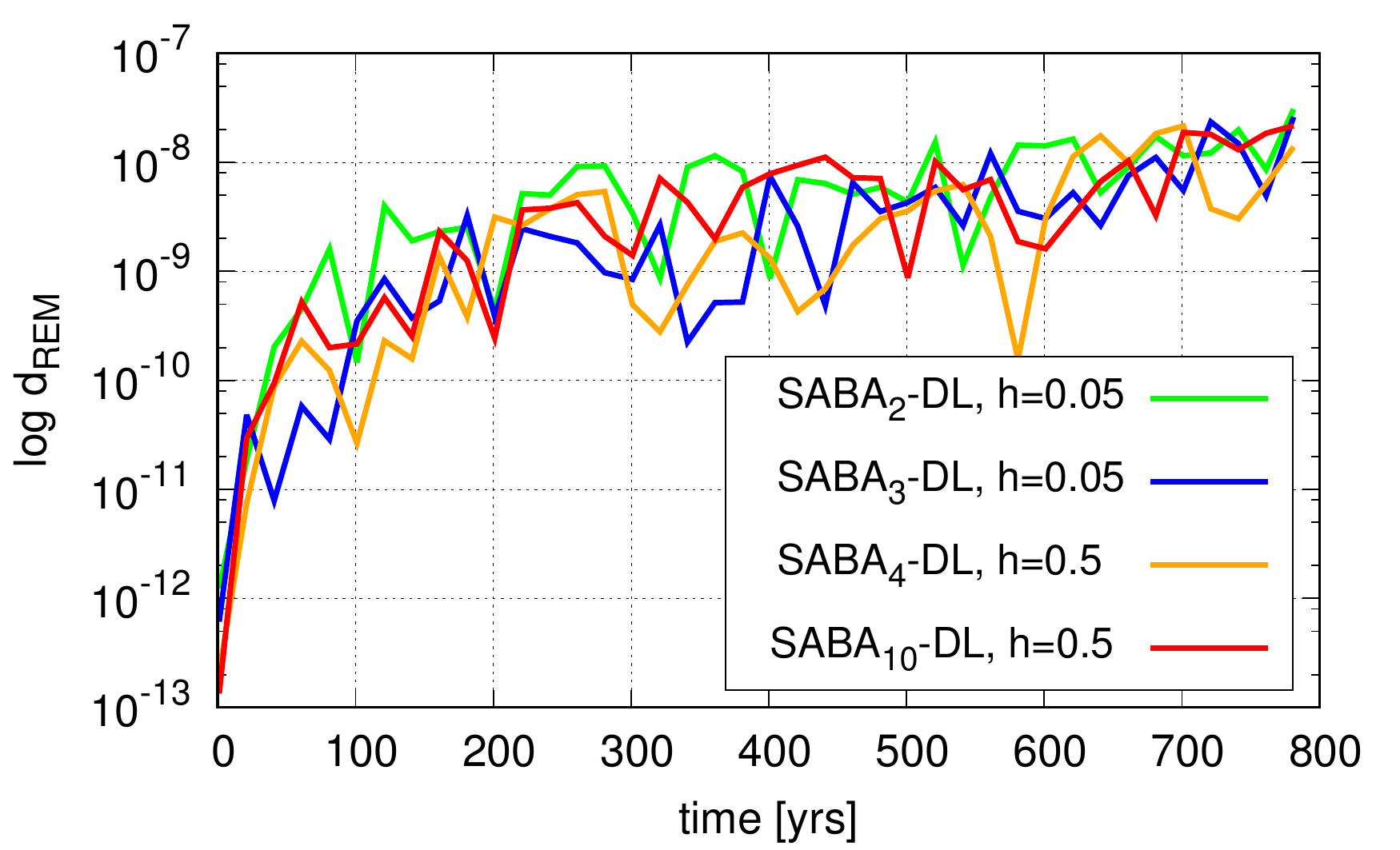}
\caption{
Time-reversibility test of SABA$_n$ schemes for $800$~yrs.  We choose a
stable HD~37124 configuration to test.  SABA$_2$ (green/black line), SABA$_3$ (blue/dark-grey line) final REM values
are illustrated for the time-step $h=0.05$ days, and SABA$_4$ (orange/grey line) and
SABA$_{10}$ (red/light-grey line) for $h=0.5$ days.  Depending on selected scheme, the energy is
preserved with a different precision but for all integrator schemes the
relative error does not exceed $10^{-9}$ in the relative scale.
}
\label{fig:figure14}
\end{figure}

For a longer forward time interval, equal to $800$~yrs and large number of
steps, the final REM value for a stable orbit slowly increases with total number
of time-steps (Fig.~\ref{fig:figure14}), essentially uniformly
for different order methods and step sizes.  For this relatively
short integration time, REM is preserved to $~10^{-7}$.

\begin{figure}
\centering
\includegraphics[width=0.47\textwidth]{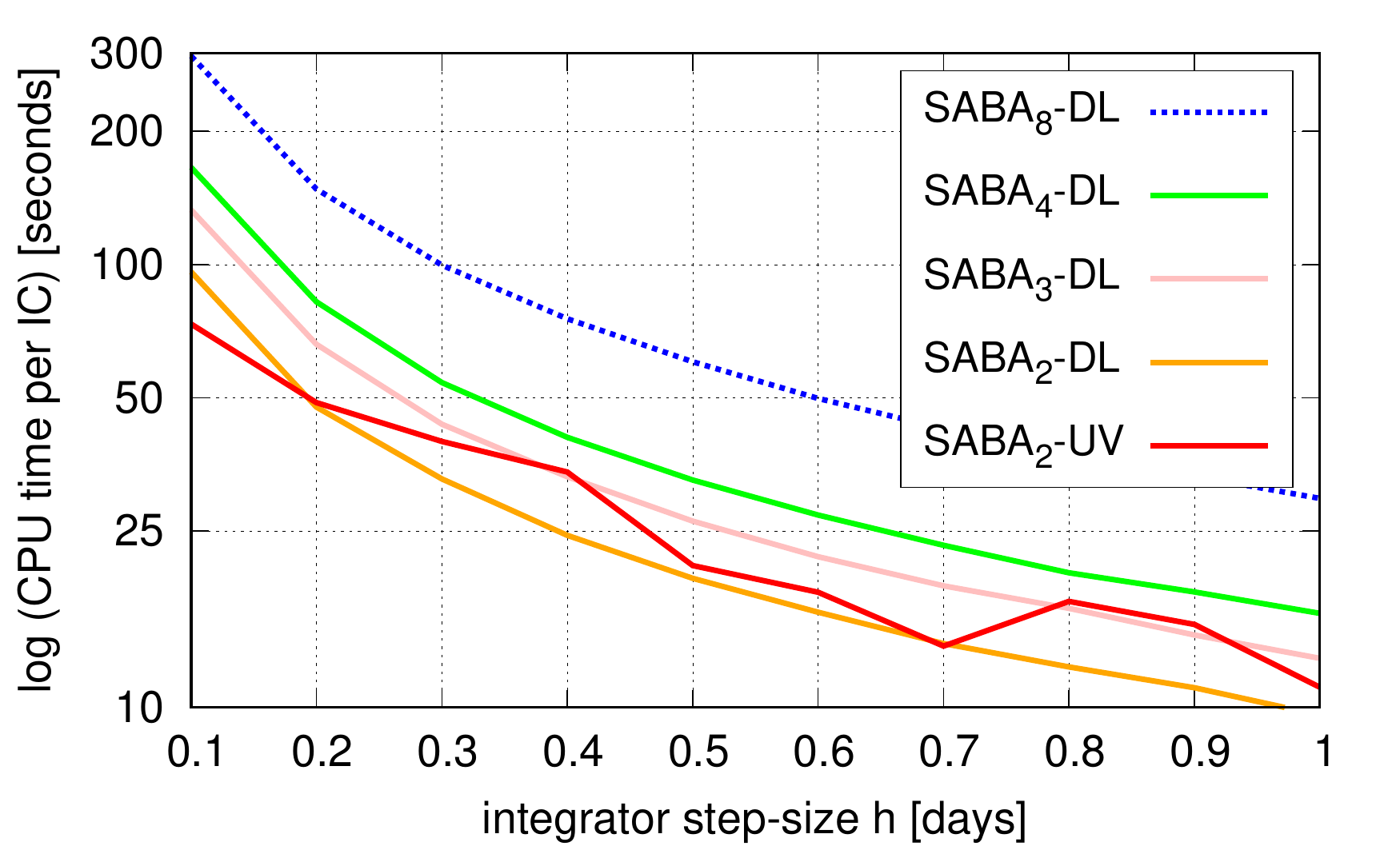}
\caption{
A relative CPU overhead for REM with different SABA$_n$ schemes
postfixed 
with -DL and -UV, which stand for 
the Keplerian drift implemented in the \citep{Levison1994} and \citep{Wisdom2016}
Keplerian solvers, respectively.
The CPU time is
expressed in seconds per single initial condition and total integration
time of $2\times 10$~kyrs.
}
\label{fig:figure15}
\end{figure}

\begin{figure}
\centering
\includegraphics[width=0.47\textwidth]{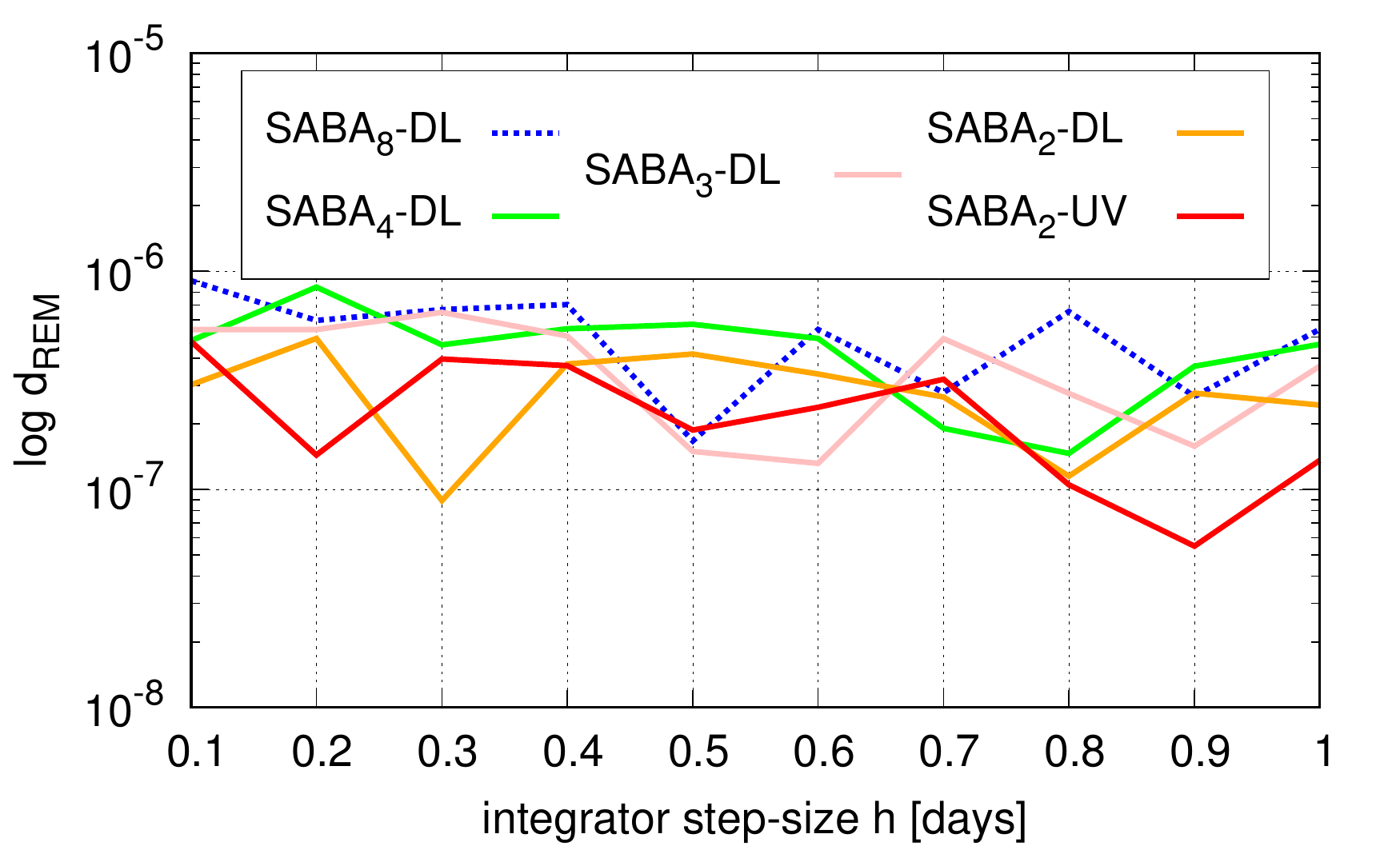}
\caption{
REM values for a range of time-steps and total integration time of $2\times
10$~kyrs.  A stable configuration of the Kepler-26 planetary system
(Tab.\ref{tab:tab1}) is tested.  SABA$_{2,4,8}$ integrators are
postfixed with -DL and -UV, which stand for the Keplerian drift implemented
in the \citep{Levison1994} and \citep{Wisdom2016} solvers, respectively.
}
\label{fig:figure16}
\end{figure}

Fig.~\ref{fig:figure15} presents the relative CPU overheads for SABA$_n$
schemes for the REM integrations of a stable orbit in the Kepler-26 system. The
time-step was changed between $0.1$ and $1$ days. The forward integration time
is fixed to $10$~kyrs. For short time-steps $\sim 0.1$~days, which correspond to
$1/170$ of the outermost orbital period ($\sim 17.25$~days), the CPU time would
be essentially non-realistic and unacceptable for massive integrations with
high-order methods, like SABA$_{6}$ or SABA$_{10}$. For lower-order SABA$_n$
integrators, the CPU overhead is still significant, and depends weakly on the
Keplerian solvers. We observed some gain of accuracy and performance when using
the UV-drift code. At the same time, the reversibility test in
Fig.~\ref{fig:figure16} suggests that the REM value depends a little on the
integrator scheme used for a wide range of time-steps. This could mean that
low-order SABA$_n$ algorithms should be preferred for REM calculations to the
higher order integrators, provided that
a reasonable relative energy conservation
of $10^{-7}$--$10^{-8}$ is guaranteed for regular orbits.
%
\subsection{SABA$_n$ vs. the second order leapfrog}
\label{subs:SABAvsLF}
%
The results illustrated in Fig.~\ref{fig:figure14} and close to uniform behaviour
of REM inspired us to test the second order, classic leapfrog algorithm. Its CPU
overheads may be greatly reduced by concatenating subsequent half-steps. For
instance, the sequence drift-kick-drift, once initialized with half-step drift,
may be continued by full time-steps drift-kick sequence, reducing the number of
the force calls. The integration sequence is finalized with
half-step drift, when the end-interval result of the integration is required.
This is the REM case.
\begin{figure}
\centering
\includegraphics[width=0.47\textwidth]{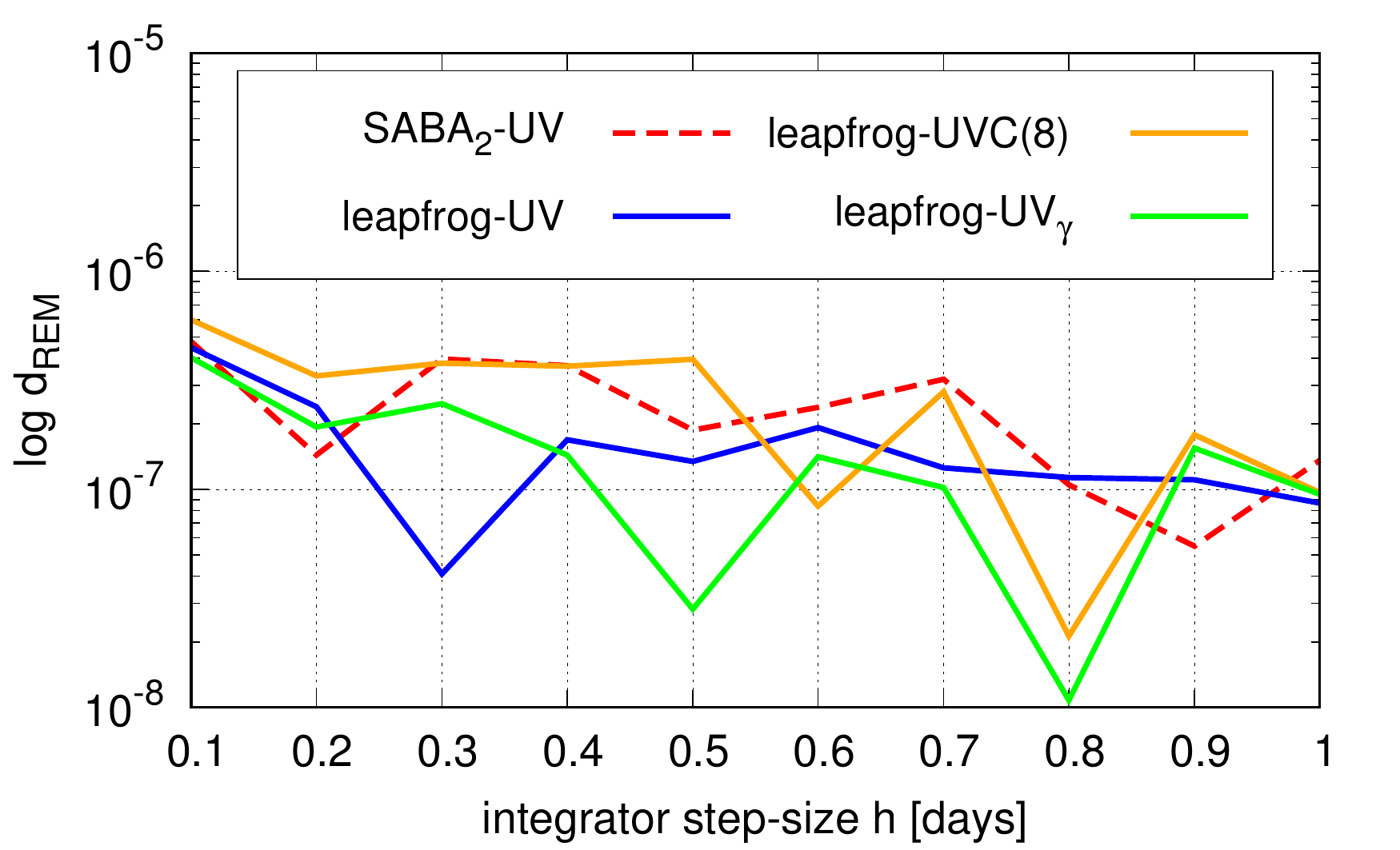} 
\caption{
Reversibility test for different leapfrog schemes: leapfrog-UV with the UV drift
(blue/dark-grey thin line), leapfrog-UV(8) with the 
UV solver and \citep{Wisdom2006} correctors of the eight 
order (orange/grey thick line), and with 
the UV solver and  $\gamma$ perturbation (leapfrog-UV$_\gamma$ 
with $\gamma=10^{-14}$, green/light-grey curve). 
For a reference, SABA$_2$ scheme with the UV drift 
is illustrated (SABA$_2$-UV, dashed curve).
}
\label{fig:figure17}
\end{figure}
\begin{figure}
\centering
\includegraphics[width=0.47\textwidth]{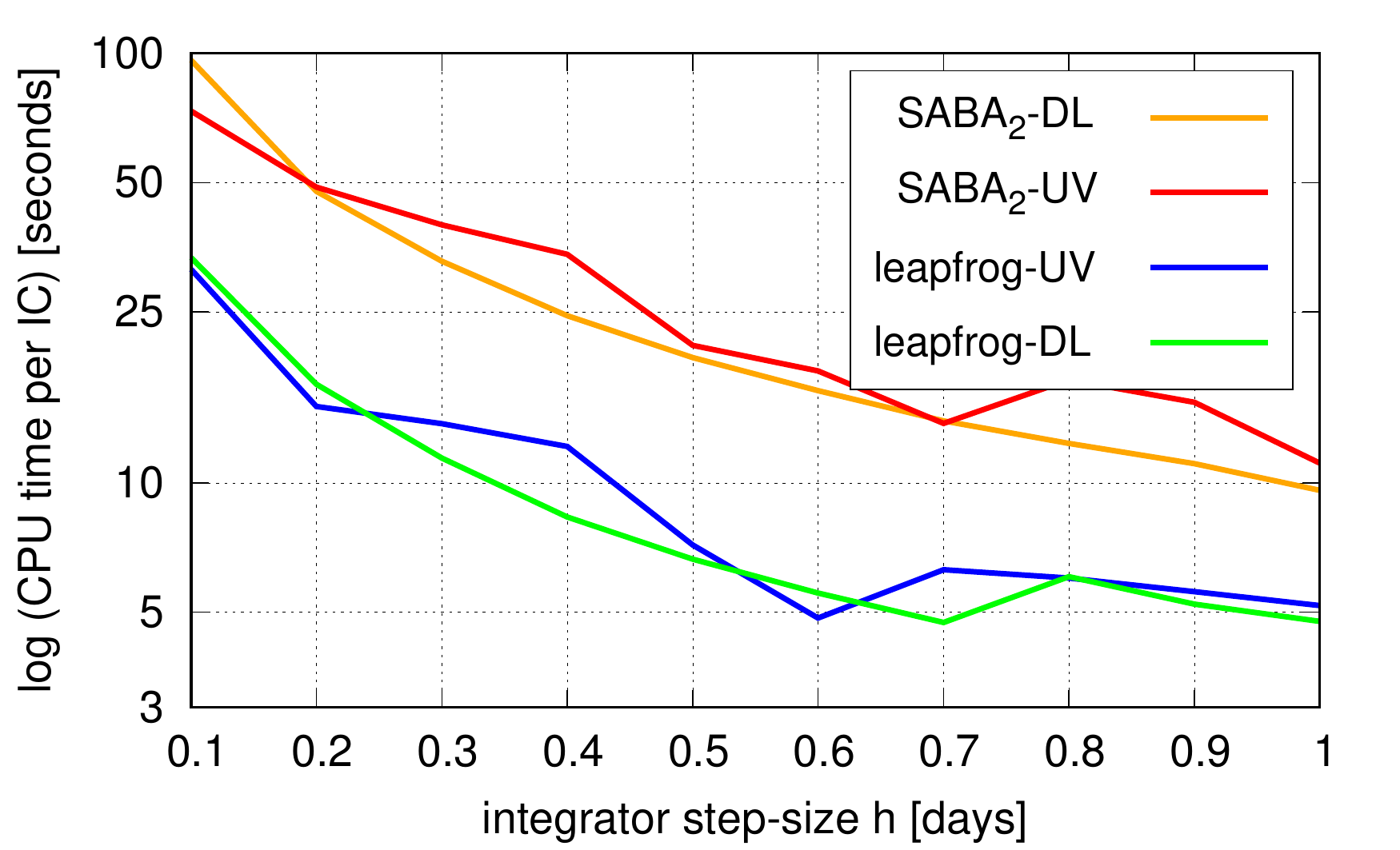}
\caption{
A comparison of REM CPU overhead for variants of the leapfrog: SABA$_2$ with
the DL and UV drifts  (orange/light grey and red/dashed light grey lines), and
leapfrog with the DL and UV drifts (blue and green lines/thick grey lines).
Total integration time is $2\times 10$ kyrs.
}
\label{fig:figure18}
\end{figure}
Figure~\ref{fig:figure17} illustrates the REM outputs for a stable configuration
in the Kepler-26 system, when integrated for the forward interval of $10$~kyrs
and different variants of the leapfrog algorithm. The step size is varied
between $0.1$ and $1$~days, though we warn the reader that $h>0.5$~day may introduce
numerical instability for chaotic orbits. This test shows that all tested schemes,
including the $\gamma$-perturbed variant of the leapfrog-UV$_\gamma$
with $\gamma=10^{-14}$,
provide similar REM outputs.  We note that REM fluctuations spanning roughly $1$~order of
magnitude do not have likely any systematic meaning, given a very small
statistics of measurements.

However, quite surprising results are provided by the CPU time test
illustrated in~Fig.~\ref{fig:figure18}.
Given the classic leapfrog variants optimized by the concatenation of sub-steps, these schemes systematically
outperform SABA$_2$ almost by two-times, independent of the step size in a
range of $[0.1,1]$~days. We found that uncorrected leapfrog fails the
REM test for shorter time-steps than its corrected variant. 
\begin{figure}
\centering
\includegraphics[width=0.47\textwidth]{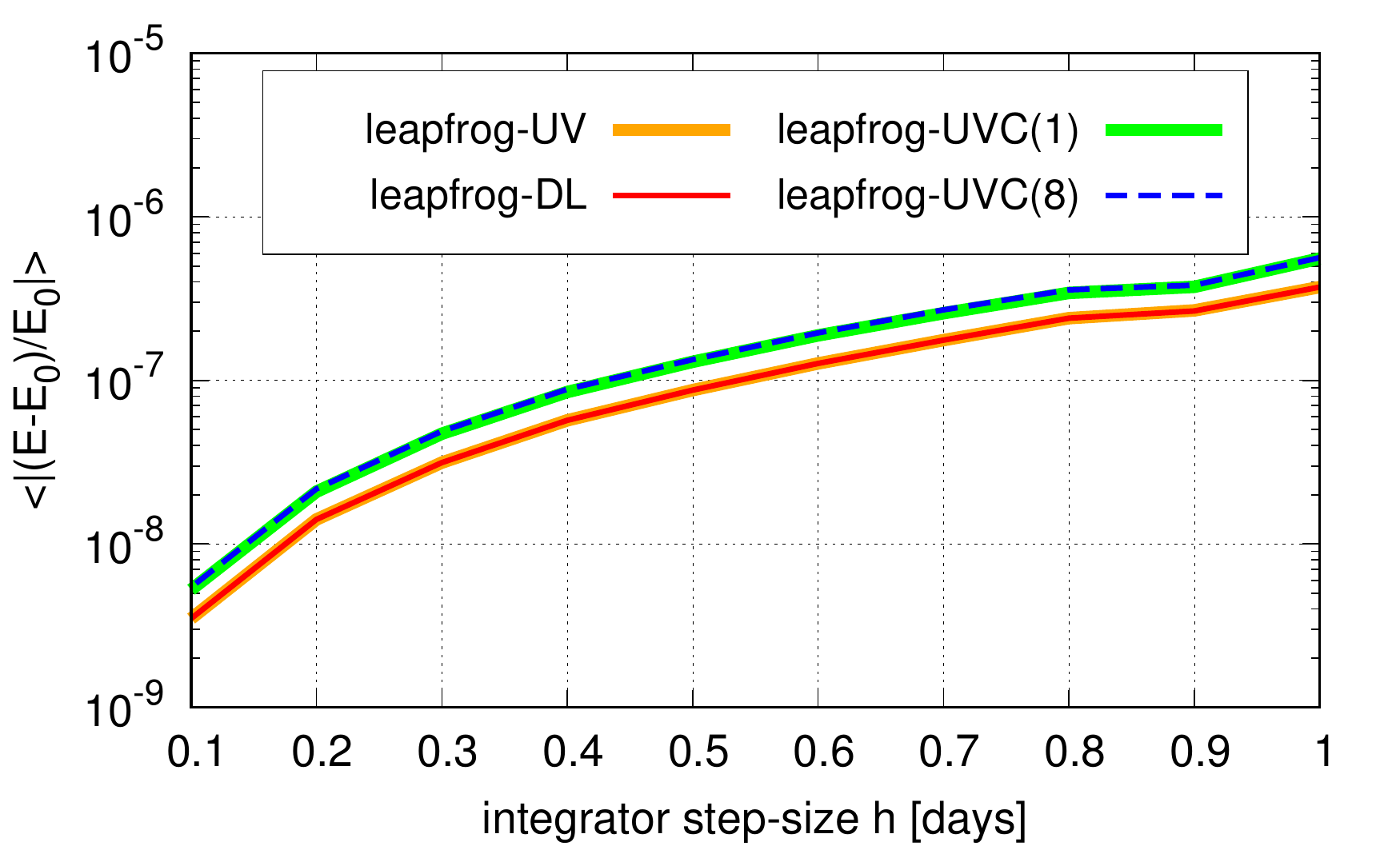}
\caption{
Mean error of the energy for the leapfrog variants tested in this paper. The
Kepler-26 initial condition was examined (Tab.~\ref{tab:tab1}). Here,
leapfrog-DF means the second order leapfrog with \citep{Levison1994} Keplerian
drift, leapfrog-UV means the leapfrog with Keplerian drift code by
\citep{Wisdom2016}, leapfrog-UVC($n$) is for this algorithm and
\citep{Wisdom2006} correctors of order 1 and 5, respectively.
}
\label{fig:figure19}
\end{figure}

\begin{figure}
\centering
  \vbox{
    \hbox{\includegraphics[width=0.47\textwidth]{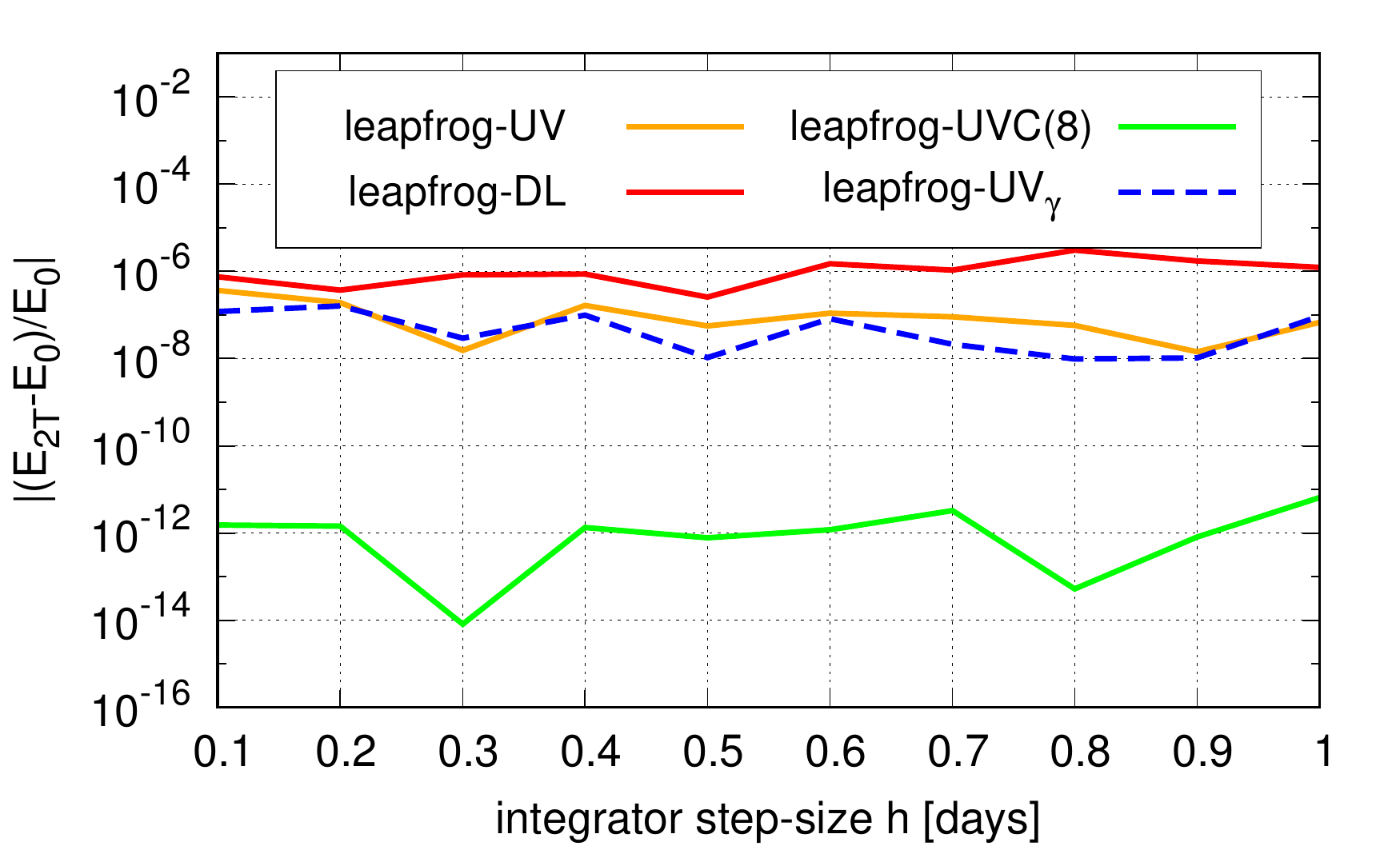}}
    \hbox{\includegraphics[width=0.47\textwidth]{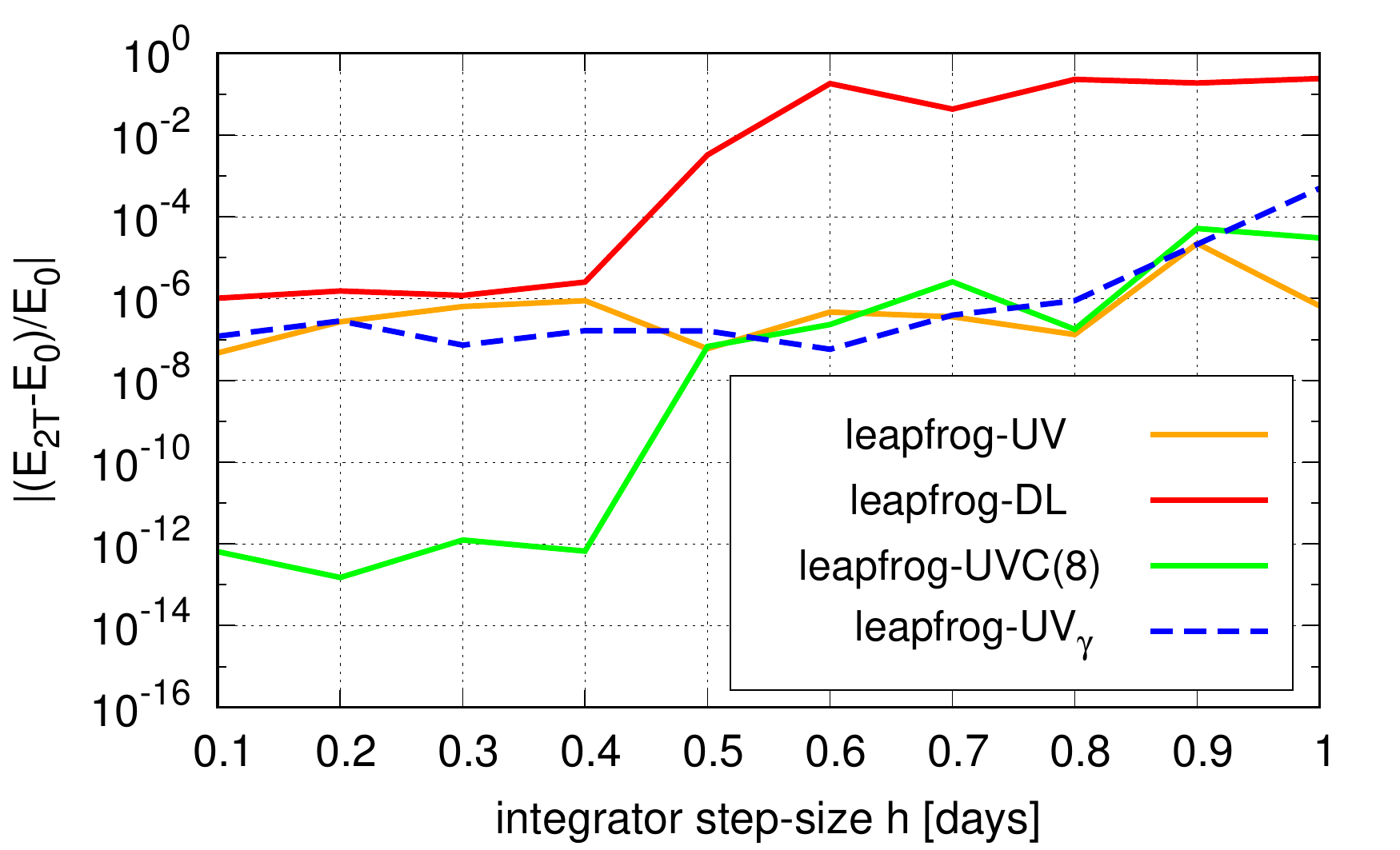}}
  }
\caption{
Energy error after integrating the REM value for the leapfrog variants tested in
this paper (see captions of the previous figures), for Kepler-26 ({\em top
panel}), and Kepler-29 ({\em bottom panel}) initial conditions, see
Tab.~\ref{tab:tab1}. 
The magnitude of perturbation 
of the initial condition after the forward integration, $\gamma=10^{-14}$.
See the text and captions of the previous figures for the meaning of labels.
}
\label{fig:figure20}
\end{figure}
For sufficiently small step sizes, the corrected leapfrog with Keplerian drift
by \cite{Wisdom2016} may be the less CPU demanding REM algorithm, still
providing reliable results, as compared to MEGNO computed with high-order SABA
integrators, or the non-symplectic Bulirsch-Stoer-Gragg scheme. To illustrate
that, in Fig.~\ref{fig:figure19} we computed the mean error of the energy for
$10$~kyrs of the Kepler-26 system (Tab.~\ref{tab:tab1}). We used
four variants of the second order leapfrogs. Even for step sizes as large as
$1$~day, the mean error of energy is $\sim 10^{-6}$, and with some gain with the
symplectic correctors.

Next Fig.~\ref{fig:figure20} is for the energy error computed with the REM
estimation, i.e., after the interval $t=2T \equiv T+\|-T\|$, relative to the
initial value at $t=0$, where $T$ is the forward integration time. We tested two
systems, Kepler-26 (top panel), and Kepler-29 (bottom panel). For this
particular numerical setup, the Wisdom correctors improve the energy
conservation by a~few orders of magnitude, essentially for zero CPU cost. This
certainly improves the REM estimate for regular orbits, by reducing the
deviation introduced by the surrogate Hamiltonian solved by the leapfrog, from
the true one. A small, middle-interval change of the initial 
condition in the $\gamma$-perturbed variant of the REM, based
on the leapfrog-UV$_\gamma$ scheme ($\gamma=10^{-14}$) does not introduce any
impact on the energy conservation w.r.t. the unperturbed version. Moreover, the results for Kepler-29 bring a~clear warning: too
large step size may cause numerical instability of the Keplerian solvers, as
well as diminish the great gain of accuracy provided by the correctors. In fact,
our large-scale numerical tests in the previous section 
for Kepler-29 failed with step sizes
longer than $0.5$~days.

Our experiments with \kepler{} systems in Tab.~\ref{tab:tab1} show that
step sizes of $\sim 1/40$ of the innermost orbital period provide the optimal
conservation of the energy $\sim 10^{-8}$--$10^{-9}$ in the relative scale.
However, a fine tuning of the step size may be required for systems of interest,
given their proximity to collision and strongly chaotic regions of motion.
%
\section{Conclusions}
\label{conclusion}
%
In this paper we propose an application of the fast indicator called REM, based
on the time-reversibility of Hamiltonian ODEs, to a~particular class of planetary systems.
They are characterized by quasi-circular orbits and relatively small mutual
perturbations. The REM algorithm has been introduced elsewhere. Our numerical
application of REM for planetary systems presented in this paper can be regarded
as an extension of the analytic theory for quasi-integrable non-linear
symplectic maps.

Besides presenting the theoretical aspects, we show that REM is equivalent to
variational algorithms, like mLCE, FLI and MEGNO, provided that dynamical
systems of interest may be investigated with symplectic and symmetric numerical
algorithms. Such systems span the FGL Hamiltonian exhibiting the Arnold web, the
restricted three body problem and a few multiple systems discovered by the 
\kepler{} mission. The \kepler{} planetary systems are the main target of our analysis,
since their eccentricities are damped by the planetary migration, and a low
range of eccentricities is typical. Moreover, the \kepler{} systems are very
compact, and are found in 2-body and 3-body MMRs, forming resonant chains.
This leads to rich dynamical behaviours. 

Revealing the phase-space structures of these dynamically complex systems is
possible thanks to CPU efficient fast indicators. We found that REM may be such a
useful numerical technique, particularly for investigating the short-term,
resonant dynamics of the \kepler{} systems. Given its simple implementation, it
provides essentially the same results, as much more complex algorithms based on
variational equations or the frequency analysis.

We show that a value of REM $\sim 10^{-6}$ is reached for stable orbits,
weakly depending on orbital and
physical parameters of Kepler-26, Kepler-29, Kepler-36 and Kepler-60 systems,
respectively, for the integration intervals as much as $\sim 10^6$ orbital
periods of the outermost planet, and maximal eccentricities reaching collisional
values. MMR's structures, and stability zones are found similarly as
with the MEGNO algorithm. However, we also found systematic discrepancies in
detecting chaotic orbits within the MMRs if the REM algorithm
relies only on the numerical errors behaviour. In such a direct variant, it is sensitive 
to chaotic motions similar to FMA or MEGNO, but it may ignore some subtle chaotic
structures with a small diffusion of the fundamental frequencies. Such structures are likely associated
with the ``stable chaos'' phenomenon. 

We found however, 
that a very small, random perturbation of 
the initial conditions after the forward integration step
greatly enhances the REM sensitivity even for such  slow chaotic diffusion. This
$\gamma$-perturbed REM variant is fully consistent with the analytical assumptions and a 
derivation of the Lyapunov error. It may be understood as a form of
the shadow orbit approach used to compute the mLCE, or a simulation of the
numerical error attained after a very long integration interval.
 
We may distinguish between different time-scales of chaotic diffusion
comparing outputs of the unmodified and $\gamma$-perturbed versions of the
REM.  The perturbed variant may be efficiently implemented as an additional backward
integration with the modified (perturbed) initial condition.  Another approach may rely
in comparing the outputs of the unmodified REM and from the MEGNO run.

One of the crucial aspects of investigating large volumes of the phase-space is
the CPU overhead. Though the REM could use any symplectic and time-reversible
integration scheme, we found that its most CPU efficient and still reliable
implementation may be provided by the classic leapfrog scheme. Its variant with
the Keplerian solver based on the universal variable and symplectic correctors
exhibits at least 2-times less CPU overhead, as compared to all other symplectic
integration algorithms tested in this paper. For weakly perturbed systems, REM
may be equally or more CPU efficient than MEGNO and other algorithms of the
variational class. This means that high-resolution dynamical maps for
time-scales of $10^4$--$10^5$ outermost orbital periods, as found in our extensive experiments, which are
sufficient to visualise major and minor structures of the 
two-body and three-body MMRs, may be computed with a single workstation.

The REM may be a particularly useful and easy to implement
numerical tool for low-dimensional conservative dynamical systems,
like the FLG Hamiltonian, variants of 
the restricted three body problem with different
perturbations, the Hill problem, models with 
galactic potentials, the rigid-body and attitude
dynamics. It is CPU efficient and accurate 
fast indicator if the right-hand
sides of the equations of motion imply complex variational equations.
The algorithm is also very attractive from the didactic point of view.
Given the leapfrog CPU efficiency and reliability, the implementation of
REM for planetary dynamics requires essentially the knowledge of the
Keplerian motion.

We believe that the REM method could be also implemented with the time-reversibility
requirement only, following \cite{Faranda2012}. This could make it possible to
apply the algorithm for a wider class of systems, like the regularized three
body problem \citep[see, for instance,][]{Dulin2013}, and its variants. Besides symplectic symmetric
integrators, there are also known symmetric schemes like symmetric Runge-Kutta
and collocation methods (e.g., Gauss, LobattoIIIA--IIIB), as well as high-order
symmetric composition methods \citep{book:Hairer2006}. We intend to investigate
these integrators for REM analysis in future papers, as well as to provide more
arguments for applications of this interesting and appealing algorithm.
%
%
\section{Acknowledgements}
%
We thank Claudia Giordano for a thorough reading of the manuscript and for
critical and informative comments that greatly improved this work.  We are
very grateful to Chris Moorcroft for the language corrections.  K.G.  thanks
the staff of the Pozna\'n Supercomputer and Network Centre (PCSS, Poland)
for their generous, professional and continuous support, and for providing
powerful computing resources (grants No.  195 and No.  313).  This work has
been supported by Polish National Science Centre MAESTRO grant
DEC-2012/06/A/ST9/00276.
%
\appendix
\appendix
\section{REM, forward and Lyapunov errors analysis}
\label{appendix}
We briefly introduce here the definition of Lyapunov error (LE), forward error
(FE) and reversibility error (RE) for symplectic maps. We refer to symplectic
maps since they are invertible and in the linear case the eigenvalues of the
matrix and its inverse are the same allowing analytical results to be obtained
on the asymptotic equivalence of FE and RE for random perturbations. We first
consider a linear map in $\Real^{2d}$ 
\begin{equation}\label{eq:A1}
 \vec{x}_n = \m{A} \vec{x}_{n-1} = \m{A}^n \vec{x}_0 ~,
\end{equation}
where $\m{A}^n$ is the $n-th$ iteration of $\m{A}$. 
The linear map is symplectic if $\m{A}$  satisfies the condition
\begin{equation}\label{eq:A2}
\begin{split}
& \m{A}~\m{J}~\m{A}^T = \m{A}^T~\m{J}~\m{A}= \m{J} ~, \\
&\m{J}=\begin{pmatrix}
0& 1\\
-1& 0
\end{pmatrix} ~.
\end{split}
\end{equation}
A non linear map 
\begin{equation}\label{eq:A3}
 \vec{x}_n= \m{M}(\vec{x}_{n-1}) ~,
\end{equation}
is defined to be symplectic if its Jacobian matrix  $D\m{M}(\vec{x})$ defined by
\begin{equation}\label{eq:A4}
D\m{M}_{jk}= \partial \m{M}_j/\partial \vec{x}_k ~,
 \end{equation}
is symplectic. Above $\m{M}_j$ is the $j$-th element of the symplectic map
$\m{M}$, and $x_k$ is the $k$-th component of the vector $\vec{x}$. For
simplicity from now on we shall refer to symplectic maps of $\Real^2$ namely to
area preserving maps. We shall analyze in detail the case of integrable maps in
normal form. Using action angle variables $\vec{x}=(\theta,\iota)$ the map reads 
\begin{equation}\label{eq:A5}
\begin{split}
& \theta_n = \theta_{n-1} + \Omega (\iota_{n-1}) ~,\\
& \iota_n = \iota_{n-1} ~.
\end{split}\end{equation}
The tangent map is constant in this case and reads
\begin{equation}\label{eq:A6}
D\m{M}= 
\begin{pmatrix}
 1& \alpha \\
 0 & 1                    
\end{pmatrix}
 ~,
\end{equation}
where $\alpha=\Omega'(\iota_n)=\Omega'(\iota_0)$.
We consider also the representation of $\m{M}$
in Cartesian coordinates $\vec{x}=(x,y)$
\begin{equation}\label{eq:A7}
\begin{split}
& \vec{x}_n = \m{R} \left( \Omega \right) \vec{x}_{n-1} ~, \qquad \quad
 \Omega = \Omega \left( \frac{\left \| \vec{x}_{n-1} \right \|^2}{2}\right) ~.   \\
 & \m{R}(\Omega)=\begin{pmatrix}
\cos \Omega & \sin \Omega \\
-\sin \Omega   &  \cos \Omega
\end{pmatrix} ~.
\end{split}\end{equation}
related to the action angle coordinates by
\begin{equation}\label{eq:A8}
\begin{split}
& y = \sqrt{2\iota} \cos{\theta} ~, \\
& z = - \sqrt{2\iota} \sin{\theta} ~.
\end{split}\end{equation}
In this second case it is important to stress the fact that the tangent map is not
constant. The dependence of the rotation frequency on the distance gives a
peculiar structure to the tangent map which reads
\begin{equation}\label{eq:A9}  
  (DM)_{ij}= R_{ij}(\Omega) + R'_{ik}(\Omega)\,\Omega'\,x_j x_k
\end{equation}  
or using a  compact  notation 
\begin{equation}\label{eq:A10}  
  D\m{M}(\vec{x})= \m{R}(\Omega)+ \Omega' \m{R}'(\Omega) \,\vec{x} \vec{x}^T ~.
\end{equation}  
As a consequence the explicit general calculation of the errors is not trivial.
The results we obtain suggest what may be expected from symplectic numerical
integration schemes when applied to integrable Hamiltonian systems expressed in
Cartesian coordinates.

\subsection{Lyapunov error}

First we define the Lyapunov error showing
its relation with the maximal Lyapunov Characteristic Exponent
(mLCE).
Taking a vector $\vec{x}_0$  and its displacement
in the phase space  $\vec{x}_{\gamma,\,0}$ defined as
\begin{equation}\label{eq:A11}
 \vec{x}_{\gamma,\,0}=\vec{x}_0+\gamma\vec{\eta}_0 ~,
\end{equation}
where $\eta_0$  is an arbitrary  versor (unit vector), and $\gamma$  a
small parameter, then the perturbed and unperturbed maps read
\begin{equation}\label{eq:A12}
\begin{split}
& \vec{x}_n = \m{M}(\vec{x}_{n-1})=\m{M}^n(\vec{x}_0) ~,\\
& \vec{x}_{\gamma,\,n}=\m{M}(\vec{x}_{\gamma,\,n-1})=\m{M}^n(\vec{x}_{\gamma,\,0}) ~.
\end{split}
\end{equation}
Now, when the parameter $\gamma$  is very small we can expand the
tangent orbit up to first order in $\gamma$ , at step $n$, as
\begin{equation}\label{eq:A13}
\vec{x}_{\gamma,\,n} = \vec{x}_n+ \gamma\,\vec{\eta}_n+ O(\gamma^2) ~.
\end{equation}
From  eq. \eqref{eq:A12} and \eqref{eq:A13} we obtain the recurrence for $ \vec{\eta}_n$ 
\begin{equation}\label{eq:A14}
\vec{\eta}_n= D\m{M}(\vec{x}_{n-1}) \,\vec{\eta}_{n-1} ~.
\end{equation}
The Lyapunov error $d_n^{(L)}$  defined as 
the norm of the displacement  in the phase space is given by 
\begin{equation}\label{eq:A15}
d_n^{(L)} = \left\| \vec{x}_{\gamma,\,n}- \vec{x}_n \right\|    
 = \gamma \left\| \vec{\eta}_n \right\| + O(\gamma^2 ) ~.
\end{equation}
Now, the definition of the maximal Lyapunov Characteristic
Exponent (mLCE) $\lambda$ reads as
\begin{equation}\label{eq:A16}
 \lambda = \lim_{n\rightarrow \infty }\,\frac{1}{n} \,\,
 \log\left \| \vec{\eta}_n \right \| =
 \lim_{n\rightarrow \infty} \frac{1}{n}\lim_{\gamma\rightarrow 0}
 \left[  \log\left( \frac{d_n^{(L)}}{\gamma} \right) \right] ~.
\end{equation}
we then use this general result for different cases.

\subsection{Lyapunov error for linear canonical maps}

The evaluation of LE  when the map is linear $M(\vec{x})=\m{A}\vec{x}$
and $\m{A}$ is in canonical form, is a simple exercise and we quote the
results for comparison with the FE and RE errors considered in P16.
We notice that the Lyapunov distance $d_n^{(L)}$ is related to the norm
of the displacement vector $\vec{\eta}_n$ by  \eqref{eq:A15}.
\space
\subsubsection{Parabolic case}
\space
The canonical form of the matrix $\m{A}$ is 
\begin{equation}\label{eq:A17}
\m{A}=
\begin{pmatrix}
1 & \alpha \\
0 & 1
\end{pmatrix} ~,
\end{equation}
with $ \alpha=\Omega'\left(\beta\right)>0$.
So that setting $\vec{\eta}_0=(\eta_x,\eta_y)$ we have
\begin{equation}\label{eq:A18}
\left\| \vec{\eta}_n \right\| =
 \left( 1 +2\eta_x\eta_y n\alpha+ n^2\alpha^2 \,n_y^2 \right)^{1/2} ~.
\end{equation}
The growth is linear unless when $\eta_y=0$ in that case
$\left\| \vec{\eta}_n \right\| = 1$ just as  when  $\alpha=0$. 
The integrable map in action-angle coordinates is amenable
to this case: indeed the tangent map of equation \eqref{eq:A10}
is given by \eqref{eq:A17} where
$\alpha= \Omega'(\iota_n)=\Omega'(\iota_0)$.
\space
\subsubsection{Elliptical case}
\space
The canonical matrix is the rotation of a fixed angle $\m{A}=\m{R}(\omega)$. 
Thus the Euclidean norm is invariant 
\begin{equation}\label{eq:A19}
 \left\| \vec{\eta}_n \right\| = \left\| \vec{\eta}_0 \right\| = 1 ~.
\end{equation}
\space
\subsubsection{Hyperbolic case}
\space
For the hyperbolic canonical case the matrix $\m{A}$  reads
\begin{equation}\label{eq:A20}
\m{A}=
\begin{pmatrix}
 e^{\lambda} & 0 \\
0 & e^{-\lambda}
\end{pmatrix} ~,
\end{equation}
and we have
\begin{equation}\label{eq:A21}
\left\| \vec{\eta}_n \right\| = (\eta_x^2 e^{2\lambda n} +  \eta_y^2 e^{-2\lambda n} ) ~.
\end{equation}
This case is of interest
because hyperbolic systems have  orbits which diverge exponentially
with  $n$. The orbits are  fully chaotic 
if the phase space is compact. An example is given by
the automorphisms of the torus $\Toro^2$ (linear maps with integer coefficients
and unit determinant) such as the Arnold cat map.

A generic linear map $\m{M}(\vec{x})= \m{B}(\vec{x})$  can always be set in canonical form
with a similarity transformation $\m{B}=\m{U}\m{A}\m{U}^{-1}$.
Since the trace is invariant the elliptic case corresponds to $|\Tr(B)|<2$,  the parabolic case
to $\Tr(B)=2$ and the hyperbolic case to $\Tr(B)>2$. Denoting with  $\m{V}=\m{U}^T\m{U}$ 
a symmetric positive matrix with unit determinant and $\vec{\chi}_0=\m{U}^{-1} \vec{\eta}_0$   we have
$\left\| \vec{\eta}_n \right\|^2= \vec{\chi}_0\cdot (\m{A}^n)^T\,\m{V} \m{A}^n \,\vec{\chi}_0$ therefore
the result depends on the coefficients $a,b,c$ of the matrix  $\m{V}$. 
In the elliptic  case $\left\| \vec{\eta}_n \right\|^2$  has oscillating terms in $n$,
however the asymptotic behavior in $n$ is the same as in the canonical case.

\subsection{Lyapunov error integrable canonical maps}

This section is an extension of the results obtained in P16. We consider 
here just the canonical maps in the elliptic case which corresponds to the usual
integrable case. 
The tangent map is no  longer constant and is given by equation \eqref{eq:A10}.
In order to compute $\vec{\eta}_n$  by iterating \eqref{eq:A14} and using the chain rule we can write
\begin{equation}\label{eq:A22}
 D\m{M}^n(\vec{x}_0)= \m{R}(n\Omega) + n\Omega'\,\m{R}'(n\Omega) \,\vec{x}_0\vec{x}_0^T ~.
\end{equation}
where the index $'$ stays for the derivative over the coordinate. 
Taking into account that $\left\| \vec{x}_n \right\| =  \left\| \vec{x}_0 \right\| $  we set
$\Omega=\Omega(\left\| \vec{x}_0 \right\|^2/2)$ and the same for $\Omega'$, thus we obtain 
\begin{equation}\label{eq:A23}
\begin{split}
 \left\| \vec{\eta}_n \right\| & = 
 \bigg( \vec{\eta}_0 \cdot \left( \m{R}^T(n\Omega) 
 + n\Omega'\,\vec{x}_0\,\vec{x}_0^T\,\m{R}'^T(n\Omega) \right) \cdot \\
 &  \cdot \left(  \m{R}(n \Omega )+n\Omega' \m{R}'(n \Omega )
 \vec{x}_0\,\vec{x}_0^T  \right)\vec{\eta}_0 \bigg)^{1/2} = \\
& = \left( 1 + 2 \,n \Omega' \, \vec{\eta}_0 \cdot \vec{x}_0 \, \vec{\eta}_0 \cdot \m{J}\vec{x}_0 +
n^2(\Omega')^2\,\vec{x}_0 \cdot \vec{x}_0 \, (\vec{\eta}_0\cdot \vec{x}_0)^2 \right)^{1/2} ~,
\end{split}
\end{equation}
where we have taken into account $\m{R}^T\m{R}'=\m{J}$.
Comparing this equation with equation \eqref{eq:A18} it is possible to observe
how, in the integrable non linear case, a linear and a quadratic
term in $n$ appear. This is precisely what happens in the parabolic case 
(see eq. \eqref{eq:A17}) which corresponds to the integrable non linear map written
in action angle coordinates, whose tangent map is constant. 
In general the error depends on  $\vec{\eta}_0$  and when it is perpendicular
to $\vec{x}_0$  then  $\left\| \vec{\eta}_n \right\| =1 $ as for a constant rotation.
The same happens in action angle coordinates when the displacement
along the action vanishes ($\eta_y=0$ in \eqref{eq:A15}).
This is a characteristic property of Lyapunov methods: the dependence on
the initial deviation vector, namely, the choice of initial condition
for the tangent map may change the  value of mLCE (\cite{Barrio2009}).

\subsection{Forward error}

In this section we introduce the forward error (FE) defined as the
displacement of the perturbed orbit $\vec{x}_{\gamma,\,n}$ with
respect to the exact one, both with the same
initial point $\vec{x}_0$.
If the perturbation is due to the round-off  the exact map $\m{M}(\vec{x})$
generating the orbit $\vec{x}_n$ cannot be numerically
computed unless we use higher precision.  For this reason  we propose 
to use the reversibility error (RE) since for symplectic maps  
asymptotic equivalence results can be proved for random perturbations, see next section.
We start with the definition of the random error vector $\gamma~\vec{\xi}$  with
linear independent components and with the properties 
\begin{equation}\label{eq:A24}
\begin{split}
 & \mean{\xi_i} = 0 ~,\\
 & \mean{\xi_i\xi_j} = \delta_{ij} ~.
\end{split}\end{equation}
This means that the random vectors have zero mean and unit
variance. The amplitude of the noise is $\gamma$ and  for each realization
of the random process we have
\begin{equation}\label{eq:A25}
\vec{x}_{\gamma,\,n}=\m{M}_\gamma( \vec{x}_{\gamma,\,n-1})=  \m{M}( \vec{x}_{\gamma,\,n-1})
   + \gamma \,\vec{\xi}_{n} \qquad \quad n\geq 1  ~,
\end{equation}
with $\vec{x}_{\gamma,\,0}=\vec{x}_0$  meaning that we start from the same point in
the phase space.  The random vectors chosen at any iteration have independent components
\begin{equation}\label{eq:A26}
\mean{(\vec{\xi}_n)_i\,(\vec{\xi}_m)_j}= \delta_{m,\,n} \, \delta_{i,\,j} ~. 
\end{equation}
We introduce the stochastic process defined by  
\begin{equation}\label{eq:A27}
 \vec{\Xi}_n = \lim_{\gamma\rightarrow 0 } 
 \,\, \frac{\vec{x}_{\gamma,n} - \vec{x}_n }{\gamma} 
 = \lim_{\gamma\rightarrow 0 } \,\,\frac{\m{M}^n_{\gamma}(\vec{x}_0) 
 - \m{M}^n(\vec{x}_0) }{\gamma} ~, 
\end{equation}
To eliminate fluctuations affecting the  FE   we consider
the following definition of the forward distance
\begin{equation}\label{eq:A28}
 d_n^{(F)}= \mean{ \left\| \vec{x}_{n,\,\gamma}-\vec{x}_n \right\|^2 }^{1/2} ~.
\end{equation}
The limit of  $d_n^{(F)}/\gamma$ is just the mean square deviation of
the process $\vec{\Xi}_n$ whose average is zero. As a consequence 
from \eqref{eq:A28} we obtain 
$ 
d_n^{(F)}  =\gamma \mean{ \left\| \vec{\Xi}_n \right\|^2 }^{1/2}  +O(\gamma^2) ~.
$
A  recurrence for $\Xi_n$ is easily found observing that from \eqref{eq:A27}
\begin{equation}\label{eq:A29}
\vec{\Xi}_n  = \lim_{\gamma \rightarrow 0} \frac{\vec{x}_{\gamma,\,n}-\vec{x}_n}{\gamma}
= D\m{M}(\vec{x}_{n-1})\vec{\Xi}_{n-1} +\vec{\xi}_n  ~,
\end{equation}
valid for $n\ge 1$ with initial condition $\vec{\Xi}_0=0$. 
The solution  is 
\begin{equation}\label{eq:A30}
\vec{\Xi}_n= \sum_{k=1}^n \, D\m{M}^{n-k}(\vec{x}_k) \,\vec{\xi}_k  ~.
\end{equation}
If we perturb the initial condition $\vec{x}_{\gamma,\,0}=\vec{x}_0+\gamma~\vec{\xi}_0$
the recurrence starts with  $\vec{\Xi}_0=\vec{\xi}_0$ and \eqref{eq:A30} holds with the sum 
starting from  $k=0$  rather than $k=1$.
In P16 we have shown that
\begin{equation}\label{eq:A31}
\mean{\vec{\Xi}_n\cdot\vec{\Xi}_n}= \sum_{k=1}^n \,\Tr \left( \Tr\,(D\m{M}^{n-k}(\vec{x}_k) )^T  
D\m{M}^{n-k}(\vec{x}_k)  \right) ~.
\end{equation}
\subsubsection{Forward error for linear canonical maps}

Let the  linear map be  $M(\vec{x})=\m{A} \vec{x}$ where $\m{A}$ is the
canonical form previously described. 
Taking \eqref{eq:A31} into account with $D\m{M}^k=\m{A}^k$  the global error is obtained from
\begin{equation}\label{eq:A32}
 \mean{\vec{\Xi}_n\cdot \vec{\Xi}_n }= \sum_{k=0}^{n-1}\,\Tr \bigl(\, \m{A}^k)^T\,\m{A}^k \,\bigr) ~.
\end{equation}
\space
\subsubsection{Parabolic case}
\space
The matrix $\m{A}$ is given by \eqref{eq:A17} so that from \eqref{eq:A32} we have
\begin{equation}\label{eq:A33}
\mean{\vec{\Xi}_n \cdot \vec{\Xi}_n}^{1/2}= \left[ \sum_{k=0}^{n-1} \, (2+\alpha^2k^2) \right]^{1/2}
= {\alpha \over \sqrt{3}}\,n^{3/2} \,O(n^{1/2}) ~.  
\end{equation}
\space
\subsubsection{Elliptical case}
\space
The matrix $\m{A}$ is the rotation matrix (see, \eqref{eq:A19}) so that
\begin{equation}\label{eq:A34}
 \mean{\vec{\Xi}_n\cdot\vec{\Xi}_n}^{1/2}= \left[\sum_{k=0}^{n-1} \, 2\right]^{1/2}
= (2n)^{1/2} ~.
\end{equation}
\space
\subsubsection{Hyperbolic case}
\space
The matrix $\m{A}$ is given by \eqref{eq:A20} so that
\begin{equation}\label{eq:A35}
 \mean{\vec{\Xi}_n \cdot \vec{\Xi}_n}^{1/2}= \left[\sum_{k=0}^{n-1} \,(e^{-2k\lambda}+e^{2k\lambda}\right]^{1/2}
= e^{\lambda n}  + O(e^{-\lambda n}) ~.
\end{equation}
A generic map $\m{B}$ is conjugated to its canonical form $\m{A}$ by  a similarity transformation
$\m{B}=\m{U}\m{A}\m{U}^{-1}$. In this case the variance of $\vec{\Xi}_n$ are still given by
\eqref{eq:A35} where $\m{A}$ is replaced by $\m{B}$  and  $ \Tr (\,(\m{B}^n)^T\,\m{B}^n)=  \Tr (\,\m{V}^{-1}\,(\m{A}^n)^T\,\m{V}\,\m{A}^n)$
where $\m{V}=\m{U}^T\m{U}$ is a symmetric positive matrix with unit determinant.
Explicit results can be found in P16. Asymptotically in $n$ the behavior of the variance of $\vec{\Xi}_n$
and consequently $d_n^{(F)}$  is the same as for the corresponding canonical maps.
\subsection{Forward error for integrable canonical maps}

We recall that the canonical form of an integrable map with an elliptic fixed point
is given by a rotation matrix $\m{R}(\Omega)$ an that according to \eqref{eq:A31}  
\begin{equation}\label{eq:A36}
D\m{M}^{n-k}(\vec{x}_k)=\m{R}((n-k)\Omega)+(n-k) \Omega'\,\m{R}'((n-k)\Omega)\vec{x}_k\,\vec{x}_k^T ~.
\end{equation}
Now proceeding step by step we compute the value $\mean{\vec{\Xi}_n\cdot\vec{\Xi}_n}$.
We first consider the matrix product
\begin{equation}\label{eq:A37}
\begin{split}
  & D\m{M}^{n-k}  (\vec{x}_k)^T   \,D\m{M}^{n-k}(\vec{x}_k)  = 
 \left(  \m{R}^T + (n-k)\Omega'\,\vec{x}_k\,\vec{x}_k^T \,(\m{R}')^T  \right) \times \\
& \times \left(  \m{R} + (n-k)\Omega' \,\m{R}'\, \vec{x}_k\,\vec{x}_k^T \,\right) 
= \m{I}+(n-k)^2\,\Omega'^2\,\vec{x}_k\vec{x}_k^T\,\m{R}'^T\m{R}'\,\vec{x}_k\vec{x}_k^T\,+  \\
& + (n-k)\Omega'\left( \vec{x}_k\,\vec{x}_k^T \,\m{R}'^T \m{R} + \m{R}^T \m{R}' \vec{x}_k\,\vec{x}_k^T \right) ~.
\end{split}
\end{equation}
Taking into account that  $ (\m{R}')^T \m{R}'=\m{I}$  and  $\m{R}^T \m{R}' =\m{J}$
plus the additional identities  $\Tr(\m{J}\,\vec{x}_k^T\vec{x}_k) $  and 
$\Tr (\vec{x}_k^T\vec{x}_k )=\vec{x}_k\cdot\vec{x}_k$  we obtain 
\begin{equation}\label{eq:A38}
 \Tr(\,D\m{M}^{n-k} (\vec{x}_k)^T \,D\m{M}^{n-k} (\vec{x}_k)\,) = 2+{\Omega'}^2 \,\left\| \vec{x}_0 \right\|^4 \,(n-k)^2 ~.
\end{equation}
Observing that  $\vec{x}_k\cdot\vec{x}_k=\vec{x}_0\cdot\vec{x}_0 $ the final result reads 
\begin{equation}\label{eq:A39}
\begin{split}
 \mean{\vec{\Xi}_n\cdot \vec{\Xi}_n}= & \sum_{k=1}^n \, \Tr( (D\m{M}^{n-k} (\vec{x}_k)^T \, D\m{M}^{n-k}(\vec{x}_k) ) = \\
&  = 2n + {\Omega'}^2 \,\left\| \vec{x}_0 \right\|^4\,\sum_{k=1}^n (n-k)^2  ~.
\end{split}
\end{equation}
The previous result gives the following asymptotic behavior of FE 
\begin{equation}\label{eq:A40}
d_n^{(F)} \sim \,\frac{\gamma}{\sqrt{3}}\,\Omega'\,\left\| \vec{x}_0\right\|^2 \, n^{3/2} ~.
\end{equation}

\subsection{Reversibility error}

We consider the reversibility error (RE)  for random perturbations presenting
cases in which it is asymptotically equivalent to the FE.
Here we extend the proof to integrable maps in
canonical form.  The inverse map at step $n+1$ is affected by a random error
$\gamma~\vec{\xi}_{-n-1}$ according to
\begin{equation}\label{eq:A41}
 \vec{x}_{\gamma,\,-n}= \m{M}^{-1}_\gamma~(\vec{x}_{\gamma,\,-n+1})=  \m{M}^{-1}(\vec{x}_{\gamma,\,-n+1})+ \gamma~\vec{\xi}_{-n} ~, 
\end{equation}
just as we have considered the direct map, see \eqref{eq:A25}.
The perturbed inverse map is not the inverse of the perturbed map,
indeed
\begin{equation}\label{eq:A42}
\begin{split}
\m{M}^{-1}_\gamma(\m{M}_\gamma(\vec{x}_0)) & = \m{M}^{-1}_\gamma(\m{M}(\vec{x}_0)+\gamma~\vec{\xi}_1 ) = \\
& = \vec{x}_0 +\gamma D\m{M}^{-1}(\vec{x}_1) \vec{\xi}_{1}+ \gamma~\vec{\xi}_{-1} + O(\gamma^2) ~,
\end{split}
\end{equation}
where both  $\vec{\xi}_1$ e $\vec{\xi}_{-1}$  are independent stochastic vectors.
We introduce  the random vector  $\vec{\Xi}_{-m,n}$ such that $\gamma~\vec{\Xi}_{-m,n}$
defines  the global error after $n$ iterations with $\m{M}_\gamma$ and $m$ iteration with
$\m{M}^{-1}_\gamma$ namely
\begin{equation}\label{eq:A43}
\vec{\Xi}_{-m,\,n}= 
  \lim_{\gamma\rightarrow 0 }\,\frac{ \m{M}^{-m}_\gamma\,(\vec{x}_{\gamma,n})-\vec{x}_{n-m}}{\gamma} ~.
\end{equation}
Using equation \eqref{eq:A26} we define for $m=n$ the displacement between the initial condition in
the phase space after $n$  iterations with the
perturbed map $\m{M}_\gamma$ and with  the perturbed inverse map  $\m{M}_\gamma^{-1}$ 
\begin{equation}\label{eq:A44}
\vec{\Xi}_n^{(R)} \equiv \vec{\Xi}_{-n,n}=  
\lim_{\gamma\rightarrow 0 }\,\frac{ \m{M}^{-n}_\gamma\,(\m{M}_\gamma^n(\vec{x}_0))-\vec{x}_0}{\gamma} ~.
\end{equation}
In order to  compute  $ \vec{\Xi}_n^{(R)}$ we may use 
for $\vec{\Xi}_{m,n}$ the recurrence relation \eqref{eq:A29} with respect to $m$ 
replacing  the map $\m{M}$ with $\m{M}^{-1}$ and taking into account
that the initial displacement  $\vec{\Xi}_{0,n}$ is not zero.
We obtain the  recurrence directly observing that 
\begin{equation}\label{eq:A45}
\begin{split}
 \vec{\Xi}_{-m,\,n} & =\lim_{\gamma\rightarrow 0 }\, 
 \frac{\m{M}_\gamma^{-1}(\vec{x}_{\gamma,n-m+1})- \m{M}^{-1}(x_{n-m+1})
+\gamma~\vec{\xi}_{-m}} {\gamma}= \\
& = D\m{M}^{-1}(x_{n-m+1})\,\vec{\Xi}_{-m+1,n} +\vec{\xi}_{-m} \qquad \quad m \geq 1 ~.
\end{split}
\end{equation}
The initial condition $\vec{\Xi}_{0,n}$ in this case, according to \eqref{eq:A45}, is 
\begin{equation}\label{eq:A46}
\vec{\Xi}_{0,\,n} = \lim_{\gamma\rightarrow 0 }\,\frac{ \vec{x}_{\gamma,n} -\vec{x}_{n}} {\gamma} = \vec{\Xi}_n ~.
\end{equation}
The solution is the same as for the forward error with a non vanishing initial condition namely
\begin{equation}\label{eq:A47}
\begin{split}
 \vec{\Xi}_{-m,n}= D\m{M}^{-m} (\vec{x}_n) \vec{\Xi}_n +\sum_{k=1}^m\,D\m{M}^{-(m-k)} (\vec{x}_{n-k}) \vec{\xi}_{-k} ~.
\end{split}
\end{equation}
The stochastic process related to the reversibility error is
\begin{equation}\label{eq:A48}
\vec{\Xi}_n^{(R)}= \vec{\Xi}_{-n,n}= D\m{M}^{-n} (\vec{x}_n) \vec{\Xi}_n +\sum_{k=1}^n\,D\m{M}^{-(n-k)} (\vec{x}_{n-k} )
\vec{\xi}_{-k} ~.
\end{equation}
This  brings to the follow definition of the reversibility distance  
\begin{equation}\label{eq:A49}
d_n^{(R)} = \mean{\left\| \m{M}_\gamma^{-n}(\m{M}_\gamma^n(\vec{x}_0))-\vec{x}_0\right\|^2}^{1/2} ~,
\end{equation}
which is related to the mean square deviation of the reversibility error  
$\vec{\Xi}^{(R)}_n$  by  
$d_n^{(R)}= \gamma \,\mean{ \left\|  \vec{\Xi}_n^{(R)} \right\|^2 }^{1/2}  +O(\gamma^2)$  where 
\begin{equation}\label{eq:A50}
\begin{split}
& \mean { \vec{\Xi}_n^{(R)} \cdot \vec{\Xi}_n^{(R)} } = \sum_{k=1}^n \,\Tr \left( \,(D\m{M}^{-(n-k)} (\vec{x}_{n-k}) )^T
D\m{M}^{-(n-k)} (\vec{x}_{n-k})\,\right) + \\    
& + \sum_{k=1}^n \,\Tr\left( \,(D\m{M}^{-n}(\vec{x}_n) D\m{M}^{n-k}(\vec{x}_k)\,)^T
D\m{M}^{-n}(\vec{x}_n) D\m{M}^{n-k}(\vec{x}_k)\,\right) ~.
\end{split}
\end{equation}
\subsection{Reversibility error for linear canonical maps}
Letting the map be $\m{M}(\vec{x})=\m{A} \vec{x}$   where $\m{A}$ is a real matrix in canonical form,
the process   $ \vec{\Xi}_n^{(R)}$ becomes 
\begin{equation}\label{eq:A51}
 \vec{\Xi}_n^{(R)}= 
\sum_{k=1}^n\,\m{A}^{-k} \vec{\xi}_k + \sum_{k=1}^n\,\m{A}^{-(n-k)}  \vec{\xi}_{-k} ~.
\end{equation}
and its  variance is 
\begin{equation}\label{eq:A52}
\begin{split}
\mean{ \Vert  \vec{\Xi}_n^{(R)} \Vert^2 }  & = \sum_{k=1}^n \Tr(\,(\m{A}^{-k})^T\m{A}^{-k}\,)+ 
\sum_{k=0}^{n-1} \Tr(\,(\m{A}^{-k})^T\m{A}^{-k}\,)= \\
& = 2 \,\,\sum_{k=0}^{n-1} \Tr(\,(\m{A}^{-k})^T\m{A}^{-k}\,)+
\Tr(\,(\m{A}^{-n})^T\m{A}^{-n} -\m{I} \,) ~.
\end{split}
\end{equation}
\space
\subsubsection{Parabolic case}
\space
\begin{equation}\label{eq:A53}
\mean{ \Vert  \vec{\Xi}_n^{(R)} \Vert^2}^{1/2} = \Bigl(  2 \mean{ \Vert  \vec{\Xi}_n \Vert^2} + n^2\alpha^2 \Bigr) ^{1/2} ~.
\end{equation}
\space
\subsubsection{Elliptic case}
\space
\begin{equation}\label{eq:A54}
 \mean{ \Vert  \vec{\Xi}_n^{(R)} \Vert^2}^{1/2} =    \Bigl(   2 \mean{ \Vert  \vec{\Xi}_n \Vert^2} \Bigr) ^{1/2} ~. 
\end{equation}
\space
\subsubsection{Hyperbolic case}
\space
\begin{equation}\label{eq:A55}
\mean{ \Vert  \vec{\Xi}_n^{(R)} \Vert^2}^{1/2} =  \Bigl(   2 \mean{ \Vert  \vec{\Xi}_n \Vert^2} +
e^{2\lambda n}+e^{-2\lambda n}-2  \Bigr)^{1/2} ~.
\end{equation}
The forward and reversibility errors are 
asymptotically proportional one with the other, and at  the leading order in $n$ and  first order in $\gamma$.

\subsection{Reversibility error for canonical integrable maps}

In order to evaluate the mean square deviation of $\vec{\Xi}_n^{(R)}$ for an 
integrable map in canonical (normal) form \eqref{eq:A9} we use \eqref{eq:A52}
where $D\m{M}^k(\vec{x})$ is given by  \eqref{eq:A37}.
If we take into account that $\m{R}^{-k}(\Omega)=\m{R}(-k\Omega)$ 
then the first sum in the r.h.s. of  \eqref{eq:A52} is the same as for the FE, namely 
\begin{equation}\label{eq:A56}
\begin{split}
 \sum_{k=1}^n \,& \Tr\left( \,(D\m{M}^{-(n-k)} (\vec{x}_{n-k}) ) ^T
D\m{M}^{-(n-k)} (\vec{x}_{n-k})\,\right) = \\
&= 2n+{\Omega'}^2 \,\left\| \vec{x}_0\right\|^2 \,\sum_{k=1}^n\,(n-k)^2 ~.
\end{split}
\end{equation}
To evaluate the second sum in the r.h.s. of \eqref{eq:A52}
we first consider a single term contributing to it
\begin{equation}\label{eq:A57}
\begin{split}
& D\m{M}^{-n}(\vec{x}_n) D\m{M}^{n-k}(\vec{x}_k)  = 
\left( \m{R}(-n\Omega) - n \Omega' \,\m{R}'(-n\Omega)\,\vec{x}_n\,\vec{x}_n^T \,\right) \cdot \\
& \cdot \left(  \m{R}((n-k)\Omega) +  (n-k)\Omega'\m{R}'((n-k)\Omega) \,\vec{x}_k\,\vec{x}_k^T \,  \right) = \\
& = \m{R}(-k\Omega) + (n-k)\Omega'\, \m{R}(-n\Omega)\, \m{R}'((n-k)\Omega) \, \vec{x}_k\,\vec{x}_k^T - \\ 
& - n \Omega' \,\m{R}'(-n\Omega) \,\vec{x}_n\,\vec{x}_n^T \, \m{R}((n-k)\Omega) - \\   
& - n(n-k){\Omega'}^2\,\m{R}'(-n\Omega)\,\vec{x}_n\,\vec{x}_n^T\,\m{R}'((n-k)\Omega)\, \vec{x}_k\,\vec{x}_k^T ~.
\end{split}
\end{equation}
To evaluate equation \eqref{eq:A57} and the trace of the matrix times its transpose, we use the 
the following relations
\begin{equation}\label{eq:A58}
\begin{split}
& \m{R}^T(\alpha) \m{R}'(\alpha) =  \m{R}(-\alpha) \m{R}'(\alpha)= \m{J} \qquad
 {\m{R}'}^T(\alpha)\m{R}(\alpha) = \m{J}^T=-\m{J} \\  
& \m{R}'(\alpha)\m{R}^T(\alpha) = \m{R}'(\alpha) \m{R}(-\alpha) = \m{J}  \qquad 
  \m{R}(\alpha){\m{R}'}^T(\alpha)=\m{J}^T=-\m{J} \\ 
& \m{R}(-\alpha) \m{J}\m{R}(\alpha) = \m{J} ~. \qquad \qquad
\end{split}
\end{equation}
where  $\m{J}$  is the matrix defined by \eqref{eq:A2} with $\m{I}=1$.
We  show first  the last term in the r.h.s. of \eqref{eq:A57}  vanishes
\begin{equation}\label{eq:A59}
\begin{split}
& \vec{x}_n^T\,\m{R}'((n-k)\Omega) \,\vec{x}_k  = \vec{x}_0^T  \m{R}(-n\Omega)\m{R}'((n-k)\Omega) \m{R}(k\Omega)\vec{x}_0= \\
& = \vec{x}_0^T \m{R}(-k\Omega)\,\m{R}(-(n-k)\Omega)\m{R}'((n-k)\Omega) \m{R}(k\Omega)\vec{x}_0= \\
& = \vec{x}_0^T \m{R}(-k\Omega)\,\m{J}\,\m{R}(k\Omega)\vec{x}_0= \vec{x}_0^T\,\m{J} \vec{x}_0=0 ~,
\end{split}
\end{equation}
since the matrix $\m{J}$ is antisymmetric.

The next step is to evaluate the following product where we introduce the following 
notation $\m{R}_k=\m{R}(k\Omega)$ and $\m{R}'_k=\m{R}'(k\Omega)$
\begin{equation}\label{eq:A60}
\begin{split}
& (D\m{M}^{-n}(\vec{x}_n)  D\m{M}^{n-k}(\vec{x}_k))^T \,D\m{M}^{-n}(\vec{x}_n)  D\m{M}^{n-k}(\vec{x}_k)  = \\
 & = \Bigl( \,\m{R}_{k} +  (n-k)\Omega'\,\vec{x}_k\,\vec{x}_k^T \,{\m{R}'}^T_{n-k}  \m{R}_{n}\,  
 - n \Omega' \,\m{R}_{-(n-k)} \,\vec{x}_n\,\vec{x}_n^T \,{\m{R}'}^T_{-n}  \,\Bigr)  \times \\
& \times \Bigl( \,\m{R}_{-k}    +   (n-k)\Omega'\, \m{R}_{-n}\, \m{R}'_{n-k}  \, \vec{x}_k\,\vec{x}_k^T   
 - n \Omega' \,\m{R}'_{-n} \,\vec{x}_n\,\vec{x}_n^T \, \m{R}_{n-k} \,\Bigr) ~. 
\end{split}
\end{equation}
Developing the product in \eqref{eq:A60} we have $9$ terms: the identity, four terms linear in $\Omega'$
whose trace is zero and four terms quadratic in $\Omega'$ which are all equal.
Indeed the trace of terms linear in $\Omega'$ is 
\begin{equation}\label{eq:A61}
\begin{split}
&\Tr\Bigl( \m{R}_{k} \m{R}_{-n}\, \m{R}'_{n-k} \,\vec{x}_k\,\vec{x}_k^T   \Bigr)   = \Tr\Bigl( \m{J}\, \vec{x}_k\,\vec{x}_k^T \Bigr) =0 ~,  \\
& \Tr\Bigl( \m{R}_{k}  \,\m{R}'_{-n} \,\vec{x}_n\,\vec{x}_n^T \, \m{R}_{n-k}  \Bigr)  =  \Tr\Bigl( \m{J}\,\vec{x}_n\,\vec{x}_n^T   \Bigr)  =0 ~, \\
& \Tr\Bigl( \m{R}_{-n}\, \m{R}'_{n-k}  \, \vec{x}_k\,\vec{x}_k^T \m{R}_{-k} \Bigr)  = \Tr\Bigl( \m{J} \, \vec{x}_k\,\vec{x}_k^T  \Bigr)=0 ~,  \\
& \Tr\Bigl( \m{R}_{-(n-k)} \,\vec{x}_n\,\vec{x}_n^T \,{\m{R}'}^T_{-n} \m{R}_{-k} \Bigr)    = \Tr\Bigl( \vec{x}_n\,\vec{x}_n^T \m{J} \, \Bigr)=0 ~,
\end{split}
\end{equation}
where we have systematically used the property  $\Tr(\m{A}\m{B})=\Tr(\m{B}\m{A})$.
The trace of the first term  quadratic in $\Omega'$  is given by $(n-k)^2{\Omega'}^2$ times
\begin{equation}\label{eq:A62}
\begin{split}
& \Tr \Bigl( \vec{x}_k\,\vec{x}_k^T \,{\m{R}'}^T_{n-k}  \m{R}_{n}\,\m{R}_{-n}\, \m{R}'_{n-k}  \, \vec{x}_k\,\vec{x}_k^T \Bigr)  = \\
& =  \Tr\Bigl(\vec{x}_k\,\vec{x}_k^T \, \vec{x}_k\,\vec{x}_k^T \Bigr ) = (\vec{x}_k\cdot \vec{x}_k)^2= \left\| \vec{x}_0\right\|^4 ~,
\end{split}
\end{equation}
where we have used  ${\m{R}'}^T(\alpha) \m{R}'(\alpha)=\m{I}$.
The trace of the  second quadratic in $\Omega'$  is given by  $-(n-k)n\,{\Omega'}^2$   times
\begin{equation}\label{eq:A63}
\begin{split}
& \Tr\left( \vec{x}_k\,\vec{x}_k^T \,{\m{R}'}^T_{n-k} \m{R}_{n}\,\m{R}'_{-n} \,\vec{x}_n\,\vec{x}_n^T \, \m{R}_{n-k}  \right)  = \\
& = \Tr\left(  \vec{x}_k\,\vec{x}_k \, {\m{R}'}^T_{n-k}  \m{R}_{n-k} \,\m{R}_k\,\m{R}'_{-n}\,\m{R}_n\,\m{R}_{-k} \,\vec{x}_k\,\vec{x}_k \right) = \\
& = \Tr\left(\vec{x}_k\,\vec{x}_k \,(-\m{J})\, \m{R}_k\,\m{J},\m{R}_{-k} \,\vec{x}_k\,\vec{x}_k   \right)=      -\left\| \vec{x}_0 \right\|^4 ~.
\end{split}
\end{equation}
The trace of the  third  quadratic in $\Omega'$  is   $-(n-k)n\,{\Omega'}^2$  times
\begin{equation}\label{eq:64}
\begin{split}
& \Tr\left(  \m{R}_{-(n-k)} \,\vec{x}_n\,\vec{x}_n^T \,{\m{R}'}^T_{-n} \m{R}_{-n}\, \m{R}'_{n-k}  \, \vec{x}_k\,\vec{x}_k^T  \right) = \\
& = \Tr\left(  \,\vec{x}_k \,\vec{x}_0 \,\m{R}_{-n} \,{\m{R}'}^T_{-n}\, \m{R}_{-k}\, \m{R}_{k-n}\, \m{R}'_{n-k}  \,\m{R}_k  \vec{x}_0\,\vec{x}_k^T
\right)   =   \\                                    
& =  \Tr\left(  \,\vec{x}_k\,\vec{x}_0 \,(-\m{J}) \, \m{R}_{-k}\,\, \m{J}  \, \m{R}_k  \vec{x}_0\,\vec{x}_k^T \right)    = -\left\| \vec{x}_0 \right\|^4 ~.
\end{split}
\end{equation}
The trace of the fourth  quadratic term in $\Omega'$  is $n^2 {\Omega'}^2$ times
\begin{equation}\label{eq:65}
\begin{split}
& \Tr\left(  \m{R}_{-(n-k)} \,\vec{x}_n\,\vec{x}_n^T \,{\m{R}'}^T_{-n}   \,\m{R}'_{-n} \,\vec{x}_n\,\vec{x}_n^T \, \m{R}_{n-k}  \right) = \\
& = \Tr\left(  \m{R}_{-(n-k)} \,\vec{x}_n\,\vec{x}_n^T \, \,\vec{x}_n\,\vec{x}_n^T \, \m{R}_{n-k}  \right)= \left\| \vec{x}_0 \right\|^4 ~,
\end{split}
\end{equation}
again taking into account  ${\m{R}'}^T(\alpha)\,\m{R}(\alpha)=\m{I}$.

Collecting all the four   terms   we obtain 
\begin{equation}\label{eq:66}
\begin{split}
 & \Tr\left(   (D\m{M}^{-n}(\vec{x}_n)  D\m{M}^{n-k}(\vec{x}_k))^T \,D\m{M}^{-n}(\vec{x}_n)  D\m{M}^{n-k}(\vec{x}_k)  \right)= \\
& = (\Omega')^2\,\Vert\vec{x}_0\Vert^4 \,\left(  n^2+2(n-k)+(n-k)^2)   \right)= \\
& = (\Omega')^2\,\Vert\vec{x}_0\Vert^4 \,(2n-k)^2 ~.
\end{split}
\end{equation}
Adding the contribution of equation \eqref{eq:A57} the final result for the 
\begin{equation}\label{eq:67}
\begin{split}
\mean { \vec{\Xi}_n^{(R)} \cdot \vec{\Xi}_n^{(R)} } &  =
2n +(\Omega')^2\,\left\|\vec{x}_0\right\| ^4\,  \left[ \sum_{k=1}^n (n-k)^2+ \sum_{k=1}^n (2n-k)^2 \right]= \\
&= 2n +(\Omega')^2\,\left\|\vec{x}_0\right\| ^4\, \sum_{k=1}^{2n-1} k^2 ~.
\end{split}
\end{equation}
The reversibility distance $ d_n^{(R)}$ has the following asymptotic expression 
\begin{equation}\label{eq:68}
d_n^{(R)} \sim \,\frac{\gamma}{\sqrt{3} }\, |\Omega'|\,\left\|\vec{x}_0\right\|^2 \, (2n)^{3/2} + O(\gamma^2) +O(\gamma\,n^{1/2} ) ~,
\end{equation}
which is the same as the forward error where  $n$ is  replaced by $2n$.

%
%
\bibliographystyle{mn2e}
\bibliography{ms}
\label{lastpage}
%
\end{document}